\documentclass[%
 reprint,
superscriptaddress,
 amsmath,amssymb,
 aps,prx,
 longbibliography
]{revtex4-2}

\usepackage{graphicx}
\usepackage{dcolumn}
\usepackage{bm}
\usepackage{hyperref}
\hypersetup{
	colorlinks   = true, 
	urlcolor     = blue, 
	linkcolor    = blue, 
	citecolor    = blue 
}
\usepackage{pifont}
\usepackage{multirow}
\usepackage{tabularray}
\usepackage{array}
    \newcolumntype{P}[1]{>{\centering\arraybackslash}p{#1}}
    \newcolumntype{M}[1]{>{\centering\arraybackslash}m{#1}}
\usepackage{stackengine}
    
\usepackage{romannum}
\usepackage{bigints}
\usepackage{ulem}
\usepackage{xcolor}
\usepackage{tabularx} 
\usepackage{siunitx} 
\usepackage[footnotehyper]{nicematrix}
\NiceMatrixOptions{cell-space-limits = 1.8pt}

\begin{document}

\preprint{APS/123-QED}

\title{Nanophotonic Super-dephasing in Collective Atom-Atom Interactions} 
\author{Wenbo Sun}
\affiliation{Elmore Family School of Electrical and Computer Engineering, Birck Nanotechnology Center, Purdue University, West Lafayette, Indiana 47907, USA}
\author{Adrian E. Rubio López}
\affiliation{Elmore Family School of Electrical and Computer Engineering, Birck Nanotechnology Center, Purdue University, West Lafayette, Indiana 47907, USA}
\author{Zubin Jacob}
\email{zjacob@purdue.edu}
\affiliation{Elmore Family School of Electrical and Computer Engineering, Birck Nanotechnology Center, Purdue University, West Lafayette, Indiana 47907, USA}

\date{\today}

\begin{abstract}
Pure dephasing and spontaneous emission are two non-unitary processes of atoms or spins interacting with fluctuating electromagnetic (EM) modes. The dissipative collective emission processes (e.g., superradiance) originate from interactions with EM modes in resonance with atoms and have received considerable attention. Meanwhile, the analogous non-dissipative collective dephasing phenomena mediated by EM environments remain poorly understood. Here, we introduce the nano-EM super-dephasing phenomenon arising in the photonic environments near materials. We show that collective dephasing in this nano-EM environment is enhanced by over 10 orders of magnitude compared to free space or cavities. This giant enhancement originates from long-range correlations in off-resonant, low-frequency evanescent EM fluctuations, which lead to collectively accelerated (super-) or suppressed (sub-) dephasing in many-body entangled states. We further unravel that nano-EM collective dephasing exhibits universal interaction ranges near materials with different anisotropy that can be reciprocal or non-reciprocal. This nano-EM interaction range, which is not present in free-space and cavities, leads to unique scaling laws of super-dephasing in GHZ states different from the conventional $N^2$ scaling of superradiance. Finally, we discuss how to experimentally isolate and control super-dephasing to open interesting frontiers for scalable quantum systems.
\end{abstract}

\maketitle

\normalem

\section{Introduction}
Pure dephasing and spontaneous emission are two paradigms of non-unitary processes in interactions between light and atoms or spins. While spontaneous emission involves photon radiation and energy decay, pure dephasing refers to the loss of phase coherence without energy dissipation and is the main obstacle in current quantum information technologies. Recent interest has focused on collective effects in interactions between light and ensembles of atoms or spins sharing the same fluctuating electromagnetic (EM) modes~\cite{reitz2022cooperative,chang2018colloquium,sheremet2023waveguide}. Multiple recent works have focused on collective spontaneous emission, which leads to superradiance effects, in ensembles of different sizes and dimensions in free-space, cavities, waveguides, and photonic crystals~\cite{masson2022universality,orioli2022emergent,rubies2022superradiance,sinha2020non,mok2023dicke,mlynek2014observation,pennetta2022collective,jones2020collectively,wang2020supercorrelated,pak2022long,lei2023many,zhu2024nonreciprocal}. However, for the other paradigm of non-unitary evolutions, collective effects in pure dephasing due to coupling with the EM/photonic environment are much less understood despite their importance in multiqubit decoherence~\cite{palma1996quantum,venkatesh2018cooperative,lidar1998decoherence,carnio2015robust,reina2002decoherence,doll2007incomplete,tuziemski2019reexamination}, quantum error correction~\cite{klesse2005quantum,aharonov2006fault}, and quantum metrology~\cite{kukita2021heisenberg,jeske2014quantum}. 

Superradiance originates from collective interactions between fluctuating EM modes and two-level systems (TLSs) in resonance~\cite{masson2022universality}. In conventional photonic environments, such as free-space, cavities, and photonic crystals, these resonant interactions can be made dominant by the enhancement/interference of propagating modes. In contrast, dephasing usually arises from broadband off-resonant, low-frequency ($\leq$MHz) environmental fluctuations. Here, we explore a unique regime for collective light-matter interactions which can arise in the near-field of lossy material slabs, where fluctuations of low-frequency ($\leq$MHz) EM modes are enhanced by over 20 orders of magnitude compared to free-space due to low-frequency evanescent interface mode contributions. This giant enhancement causes off-resonant collective interactions between quantum ensembles and low-frequency fluctuating EM modes to become dominant. 

In this paper, we introduce the EM-mediated super-dephasing phenomenon that emerges in this unique regime of nano-EM interactions fundamentally different from resonant cavity and photonic crystal effects. We find that nano-EM collective dephasing is enhanced by over 10 orders of magnitude compared to free-space or macroscopic cavity effects. This giant enhancement originates from the dominance of long-range correlated low-frequency evanescent EM fluctuations in the nano-EM environment. Distinct from superradiance, which accelerates energy emission and generates entanglement~\cite{orioli2022emergent}, nano-EM super-dephasing accelerates disentanglement without energy dissipation. 
Additionally, since nano-EM super-dephasing is related to many-body collective interactions with low-frequency EM fluctuations, it can exhibit behaviors distinct from other collective quantum phenomena in nanophotonic environments, including resonant dipole-dipole interactions (RDDI)~\cite{cortes2022fundamental,boddeti2021long,cortes2017super,biehs2016long,hassani2022enhancement}, Casimir-Polder frequency shifts~\cite{hanson2019non,fuchs2017casimir,block2019casimir,sinha2018collective,henkel2018influence}, and collective emission~\cite{baranov2017modifying,langsjoen2012qubit,kolkowitz2015probing,sun2023limits,bellomo2015nonequilibrium}, which hinge on high (resonant) frequency fluctuations. 

We prove that nano-EM collective dephasing exhibits universal (long-range) power law dependence $D^{-\beta}$ on the interatom distance $D$ near materials with different anisotropy that can be reciprocal or non-reciprocal. This universal nano-EM interaction range is not found in free-space, cavity, and waveguide quantum electrodynamics (QED) platforms (see Fig.~\ref{fig:figures1} in Appendix~\ref{derivation}), and is independent of EM modes' wavelengths, which is in stark contrast to superradiance effects. We further demonstrate that, due to the unique interaction range, nano-EM super-dephasing can exhibit unique scaling laws in entangled states, e.g., Greenberger–Horne–Zeilinger (GHZ) states, beyond the conventional $N^2$ scaling of superradiance in subwavelength ensembles. 

\section{Model}
\begin{figure}[!t]
    \centering
    \includegraphics[width = 3.4in]{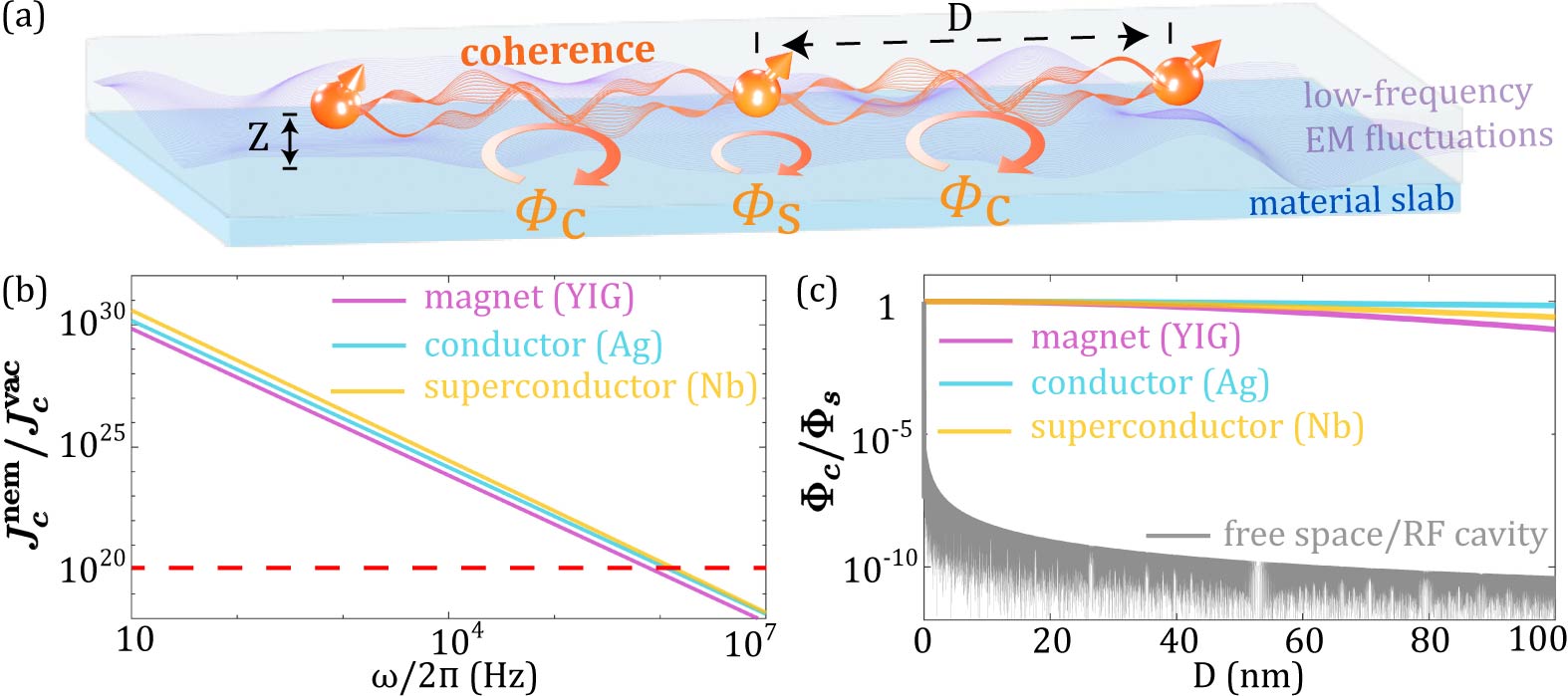}
    \caption{(a) Sketch of the magnetic TLSs ensemble interacting with off-resonant, low-frequency EM fluctuations near material slabs. TLSs are separated by interatom distance $D$ at distance $z$ from material slabs. (b) The low-frequency EM fluctuation correlations $J_c (\omega)$ are enhanced by over 20 orders of magnitude near magnets (e.g., yttrium iron garnet, YIG) and conductors (e.g., silver, niobium) compared to free space ($J_c^{nem}/J_c^{vac}>10^{20}$). (c) Collective dephasing $|\Phi_c/\Phi_s|$ is enhanced by over 10 orders of magnitude in nano-EM environments compared to free space and RF cavities. We consider TLSs perpendicular to the material slab and take $D=z=50\,\mathrm{nm}$ in (b) and $z=50\, \mathrm{nm}$ in (c). Details of material parameters are provided in Appendix~\ref{appendix_material}.} 
    \label{fig:fig1}
\end{figure}

We consider a magnetic TLSs ensemble interacting with fluctuating EM fields in the nano-EM environment near material slabs, as shown in Fig.~\ref{fig:fig1}(a). We focus on the interactions $\hat{H}_{int}=- \sum_i \mathbf{m}_{i} \hat{\sigma}_i^z \cdot \hat{\mathbf{B}}^{\mathrm{fl}}(\mathbf{r_i})$ between TLSs and longitudinal magnetic fluctuations that induce pure dephasing effects in photonic environments. $\mathbf{m}_i, \mathbf{r_i}, \hat{\sigma}_i^z$ are the spin magnetic moment, position, and Pauli-z operator of the TLSs. $\hat{\mathbf{B}}^{\mathrm{fl}}(\mathbf{r_i}) = \int d\omega \hat{\mathbf{B}}^{\mathrm{fl}}(\mathbf{r_i},\omega)+h.c.$ is the fluctuating magnetic field operator at $\mathbf{r_i}$ following the macroscopic QED quantization~\cite{buhmann2012macroscopic}. $\hat{\mathbf{B}}^{\mathrm{fl}}(\mathbf{r_i},\omega)$ is proportional to the magnetic dyadic Green's function $\overleftrightarrow{G}_m (\omega)=\overleftrightarrow{G}^0_m (\omega)+\overleftrightarrow{G}^r_m (\omega)$. $\overleftrightarrow{G}^0_m$ represents the free-space and the substrate contributions. $\overleftrightarrow{G}^r_m$ is the reflected component determined by material slabs.

Through the time-convolutionless projection operator technique~\cite{breuer2002theory}, we can find a time-local master equation from $\hat{H}_{int}$ that governs the collective dephasing dynamics in the nano-EM environment (see derivations in Appendix~\ref{derivation}): 
\begin{align}\label{ME}
\begin{aligned}
    &\frac{d \rho}{d t} =   -i[\hat{H}_{d},\rho] + \sum_{i} \frac{1}{2} \frac{d \, \Phi_{s}(\mathbf{r_i},t)}{dt}  \Big[\hat{\sigma}_{i}^z \rho \, \hat{\sigma}_{i}^z - \rho \Big] + \\ &\sum_{\substack{i \neq j}} \frac{1}{2} \frac{d \, \Phi_{c}(\mathbf{r_i},\mathbf{r_j},t)}{dt} \Big[\hat{\sigma}_{i}^z \rho \, \hat{\sigma}_{j}^z-\frac{1}{2} \hat{\sigma}_{i}^z \hat{\sigma}_{j}^z \rho - \frac{1}{2} \rho \, \hat{\sigma}_{i}^z \hat{\sigma}_{j}^z \Big],
\end{aligned}
\end{align}
where $\rho$ is the density matrix of the magnetic TLSs and $\hat{H}_{d}$ is the dipolar interaction Hamiltonian governing the unitary processes. $\Phi_{s}(\mathbf{r_i},t)$ and $\Phi_{c}(\mathbf{r_i},\mathbf{r_j},t)$ are the individual and \textit{pairwise} collective dephasing functions that govern the time evolutions of dephasing processes. The second and third terms describe the individual and collective dephasing processes induced by EM fluctuations in the nano-EM environment, respectively. We focus on the dephasing processes in this paper since they are usually the dominant sources of decoherence. Through the constructive or destructive interference of the individual and collective dephasing processes, the TLSs can undergo accelerated (super-) or suppressed (sub-) dephasing depending on their initial states. We note that the interference here is the quantum interference among individual and collective dephasing processes, and is different from the Young-type interference in wave optics. Our results for $\Phi_{c}(\mathbf{r_i},\mathbf{r_j},t)$ in Eq.~(\ref{ME}) are (see derivations in Appendix~\ref{derivation}): 
    \begin{align}
    &\begin{aligned}
          &\Phi_{c}(\mathbf{r_i},\mathbf{r_j},t) =  \int_0^{\omega_c} d\omega \, F(\omega, t) J_c(\mathbf{r_i},\mathbf{r_j},\omega)
         \label{td_cdf},
    \end{aligned}\\
    &\begin{aligned}       
          J_c(\omega) & = \frac{2\mu_0\omega^2}{\hbar \pi c^2} \coth{\frac{\hbar \omega}{2 k_B T}}  \Big[\mathbf{m}_{i} \cdot 
         \mathrm{Im}  [\overleftrightarrow{G}_m(\mathbf{r_i},\mathbf{r_j},\omega) \\ & \qquad  \qquad \qquad \qquad \quad +\overleftrightarrow{G}^\intercal_m(\mathbf{r_j},\mathbf{r_i},\omega)] \cdot \mathbf{m}_{j}\Big]
         \label{Jc},
    \end{aligned}
    \end{align} 
and $\Phi_{s}(\mathbf{r_i},t) = \Phi_{c}(\mathbf{r_i},\mathbf{r_j},t)|_{\mathbf{r_j} \rightarrow \mathbf{r_i}}$. The symmetric formalism $\Phi_c(\mathbf{r_i},\mathbf{r_j})=\Phi_c(\mathbf{r_j},\mathbf{r_i})$ ensures the Hermiticity of Eq.~(\ref{ME}) in arbitrary photonic environments, which can be reciprocal or non-reciprocal. Here, $\coth{(\hbar \omega /2 k_B T)}$ incorporates thermal fluctuation effects at temperature $T$, and $\overleftrightarrow{G}^\intercal_m$ is the transpose of $\overleftrightarrow{G}_m$. Similar to Lamb shift calculations~\cite{franke2021fermi}, we introduce a cutoff frequency $\omega_c \approx mc^2/\hbar$ to avoid the ultraviolet divergence in Eq.~(\ref{td_cdf}) due to the free-space contributions $\overleftrightarrow{G}^0_m$. $F(\omega, t)=(1-\cos{\omega t})/\omega^2$ serves as a frequency filter function usually centered around MHz frequencies. Different from collective emission induced by EM fluctuations at atom resonance frequencies~\cite{masson2022universality}, collective dephasing is determined by the frequency integral of off-resonant EM fluctuation correlations (continuum of low-frequency EM modes).

\begin{table*}[!t]
\caption{\label{tab:table1} Universal power-law dependence of nano-EM collective dephasing $|\Phi_c/\Phi_s| \sim D^{-\beta}$ near different material classes.}
    \centering
    \begin{NiceTabular}{c|c|c}
    \Hline
    \Hline
         Material class & Leading term in $r_{ss}(q\gg k_0)$ Eq.~(\ref{rss}) & Power law $\beta$\\
    \Hline
        Isotropic, scalar $\mu$ and $\varepsilon$ & $f_0(\theta)=\frac{\mu-1}{\mu+1}$, $f_2(\theta)=\frac{\varepsilon-1}{4}$ (non-magnetic) & $\beta=3$, $\beta=1$ (non-magnetic) \\ 
    \Hline
        Hyperbolic, diagonal $\overleftrightarrow{\varepsilon}$ ($\varepsilon_{xx}>0, \varepsilon_{yy}=\varepsilon_{zz}<0$)  & $f_2(\theta)=\varepsilon_{yy}(1-|\sin{\theta}|)/(4+4|\sin{\theta}|)$ & $\beta=1$, $\beta=2 \, (\varphi=0)$ \\
    \Hline
         Gyromagnetic, $\overleftrightarrow{\mu}\neq \overleftrightarrow{\mu}^\intercal $ ($\mu_{xy}= -\mu_{yx}\neq 0$) & $f_0(\theta)=\frac{\sqrt{\mu_{zz}(\mu_{xx}\cos^2{\theta}+\mu_{yy}\sin^2{\theta})}-1}{\sqrt{\mu_{zz}(\mu_{xx}\cos^2{\theta}+\mu_{yy}\sin^2{\theta})}+1} $ & $\beta=3$ \\
    \Hline
         Gyromagnetic, $\overleftrightarrow{\mu}\neq \overleftrightarrow{\mu}^\intercal $ ($\mu_{yz}= -\mu_{zy}\neq 0$) & $f_0(\theta)=\frac{\sqrt{\mu_{zz}(\mu_{xx}\cos^2{\theta}+\mu_{yy}\sin^2{\theta})}-1-i\mu_{yz}\sin{\theta}}{\sqrt{\mu_{zz}(\mu_{xx}\cos^2{\theta}+\mu_{yy}\sin^2{\theta})}+1-i\mu_{yz}\sin{\theta}} $ & $\beta=3$ \\ 
    \Hline
    \Hline
    \end{NiceTabular} 
\end{table*}

We now compare collective dephasing in different photonic environments. We focus on the ratio $\Phi_c/\Phi_s$, which determines whether the dephasing processes in EM environments are dominated by individual ($|\Phi_c/\Phi_s| \ll 1$) or collective ($|\Phi_c/\Phi_s| \sim 1$) dephasing. Unlike collective emission arising in various QED platforms (e.g., free-space), we find that EM-fluctuation-induced collective dephasing is dominant only in the nano-EM environment near material slabs or meta-surfaces. As shown in Fig.~\ref{fig:fig1}(b), at low frequencies $\omega<1\,\mathrm{MHz}$, the EM fluctuation correlations $J_c(\omega)$ in the nano-EM environment near magnets and conductors are ubiquitously broadband enhanced over 20 orders of magnitude compared to free-space ($J_c^{nem}/J_c^{vac}>10^{20}$). This reveals that low-frequency EM fluctuations exhibit strikingly different behaviors in nano-EM environments and in free-space (see further discussions in Appendix~\ref{comparison_QED}). This giant enhancement leads to distinct behaviors of the integral in Eq.~(\ref{td_cdf}). In free-space and RF cavities, Eq.~(\ref{td_cdf}) is dominated by contributions from the high-frequency components $J_c(\omega \sim \omega_c)$ that have short-range correlations. As shown in Fig.~\ref{fig:fig1}(c), $|\Phi^{vac}_c/\Phi^{vac}_s|<10^{-10}$ rapidly oscillates with the interatom distance $D=|r_i-r_j|$. Therefore, EM-fluctuation-induced dephasing of TLSs is dominated by the individual dephasing processes, and super-dephasing is negligible. In stark contrast, in the nano-EM environment, Eq.~(\ref{td_cdf}) is dominated by contributions from the low-frequency ($<$MHz) $J_c(\omega)$ with long-range correlations. As shown in Fig.~\ref{fig:fig1}(c), collective dephasing effects $|\Phi^{nem}_c/\Phi^{nem}_s| \sim 1$ are enhanced around 10 orders of magnitude compared to free space, providing a unique nano QED regime for the emergence of super-dephasing mediated by EM environments.

To elucidate the origin of this giant enhancement, we consider two magnetic TLSs perpendicular to the material slab at distance $z$ from the slab, as shown in Fig.~\ref{fig:fig2}(a). In the nano-EM environment, collective dephasing $\Phi_c$ is determined by low-frequency EM correlations $J_c (\omega) \propto \mathrm{Im} [ \overleftrightarrow{G}_m(\mathbf{r_i},\mathbf{r_j}) + \overleftrightarrow{G}_m^{\intercal}(\mathbf{r_j},\mathbf{r_i}) ]_{zz}$. In the near-field of material slabs $z\ll k_0^{-1}=c/\omega$, $ [ \overleftrightarrow{G}_m(\mathbf{r_i},\mathbf{r_j})]_{zz} \propto \int dq d\theta \, q^2 \allowbreak r_{ss}(q,\theta) e^{iqD\cos{(\theta-\varphi)}} e^{-2qz}$, where $\varphi$ is the angle between 
$\mathbf{r_i}-\mathbf{r_j}$ and x-axis (Fig.~\ref{fig:fig2}(a)) and $r_{ss}$ is the reflection coefficient. In this integral, $e^{-2qz}$ provides a momentum cutoff~\cite{machado2023quantum} $q_c \sim 1/2z \gg k_0$ and the integral is dominated by high-momentum components with $q \sim q_c \gg k_0$, which correspond to evanescent waves highly confined to the material slab interfaces~\cite{ford1984electromagnetic}. This shows that nano-EM collective dephasing is enhanced by low-frequency evanescent interface waves ($q \gg k_0$), which is fundamentally different from collective effects mediated by propagating waves ($q < k_0$) in free space/resonant structures.

\section{Nano-EM collective dephasing range}
One critical aspect of collective quantum phenomena is the underlying interaction range~\cite{mok2023dicke}. We now unravel that nano-EM collective dephasing exhibits universal (long-range) power law dependence on the interatom distance $D$ ($\Phi_c\sim D^{-\beta}$) near material slabs with local EM response, which can have different anisotropy and can be reciprocal or non-reciprocal. This is in stark contrast to other collective dephasing processes encountered in magnetic TLSs (spin qubit) systems, e.g., due to nuclear spin bath~\cite{kwiatkowski2018decoherence,bradley2019ten} or fluctuating charges~\cite{rojas2023spatial}, which are either negligible or decay exponentially with $D$. 

\begin{figure}[!b]
    \centering
    \includegraphics[width = 3.4in]{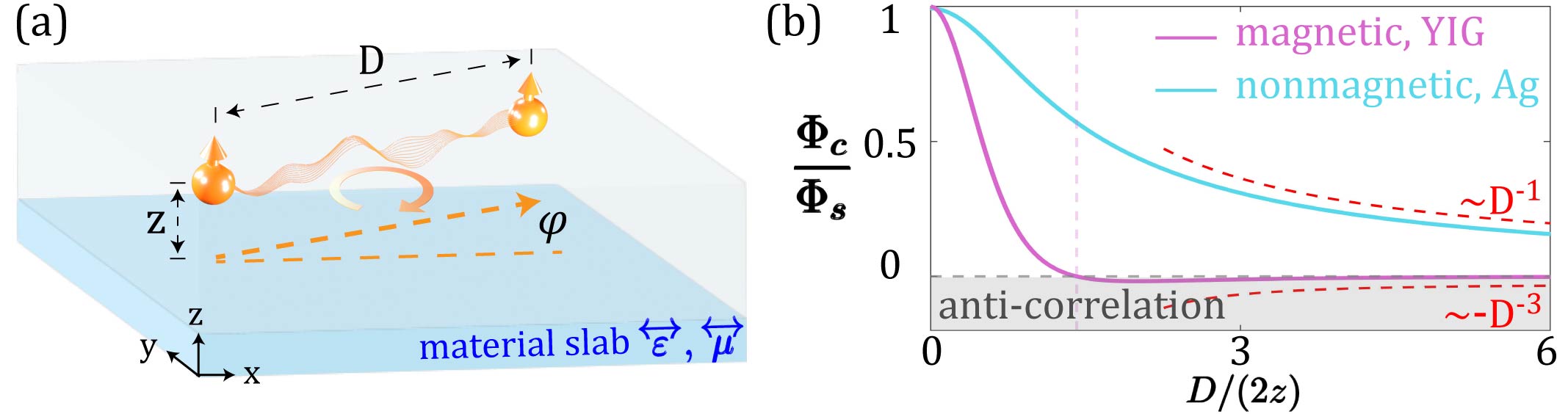}
    \caption{(a) Sketch of nano-EM collective dephasing of two magnetic TLSs separated by distance $D$ at distance $z$ from material slabs. (b) Power-law dependence of nano-EM collective dephasing on $D$ near gyromagnetic YIG and non-magnetic silver, which matches well with Eq.~(\ref{powerlaw}) at large $D$. Details of material parameters are provided in Appendix~\ref{appendix_material}.}
    \label{fig:fig2}
\end{figure}

In the following, we consider several key classes of photonic media, including isotropic media with scalar permeability $\overleftrightarrow{\mu}$ and permittivity $\overleftrightarrow{\varepsilon}$ (isotropic reciprocal), hyperbolic media with diagonal $\overleftrightarrow{\varepsilon}$ (anisotropic reciprocal), and gyrotropic media with $\overleftrightarrow{\mu}\neq \overleftrightarrow{\mu}^\intercal$ (anisotropic non-reciprocal). We focus on two magnetic TLSs with configurations shown in Fig.~\ref{fig:fig2}(a). As discussed above, nano-EM collective dephasing is dominated by evanescent interface wave contributions. For evanescent waves with momentum $q/k_0 \gg 1$, we find the reflection coefficients $r_{ss}$ can be expanded as,
\begin{equation}\label{rss}
    r_{ss}(q,\theta) \approx f_0(\theta) + f_2 (\theta)\frac{k_0^2}{q^2},
\end{equation}
where $f_0(\theta)$ and $f_2 (\theta)$ are determined by material properties $\overleftrightarrow{\mu}$ and $\overleftrightarrow{\varepsilon}$. For isotropic media, $f_0(\theta)=0, f_2(\theta)=\varepsilon/4$ (non-magnetic conductors) ~\cite{langsjoen2012qubit,kolkowitz2015probing} and $f_0(\theta)=(\mu-1)/(\mu+1)$ (magnetic materials) can be obtained from the Fresnel formula. For other materials, we find $r_{ss}(q\gg k_0)$ by solving Maxwell equations with EM boundary conditions~\cite{khandekar2019thermal} (also see Appendix~\ref{appendix_range}). We summarize our results for isotropic, hyperbolic, and gyrotropic media in Table~\ref{tab:table1} and extend the discussions to more general photonic media with local EM response in Appendix~\ref{appendix_range}. In general, we find $f_0(\theta) \neq 0$ for magnetic media and $f_0(\theta) = 0$ for non-magnetic media. Substitute Eq.~(\ref{rss}) into $[\overleftrightarrow{G}_m]_{zz}$, we have 
\begin{multline}\label{Jctheta}
    J_c(\omega) \propto \mathrm{Im} \int_0^{2\pi} \Big[ \frac{4f_0(\theta)}{k_0^3D^3} \frac{a^3-3a \cos^2(\theta-\varphi)}{[a^2+\cos^2(\theta-\varphi)]^3} \\ +\frac{2f_2(\theta)}{k_0D} \frac{a}{a^2+\cos^2(\theta-\varphi)} \Big] d\theta,
\end{multline}
where $a=2z/D$. At large $D$ (small $a$), the integral is dominated by contributions from $\theta_1=\varphi+\pi/2$ and $\theta_2=\varphi+3\pi/2$, we find (see derivations in Appendix~\ref{appendix_range})
\begin{equation}\label{powerlaw}
    J_c(\omega) \propto \sum_{i=1,2} \mathrm{Im} \Big[ -\frac{2\pi}{k_0^3D^3} f_0(\theta_i) + \frac{2 \pi}{k_0D} f_2(\theta_i) \Big],
\end{equation}
for $2z<D\ll 1/k_0$. Substitute Eq.~(\ref{powerlaw}) into Eq.~(\ref{td_cdf}), we find $\Phi_c\sim D^{-\beta}$, which proves the universal power-law dependence of nano-EM collective dephasing on $D$. We summarize our results of $\beta$ in Table~\ref{tab:table1}. In general, we find two main types of interaction range, i.e., $\Phi_c\sim -D^{-3}$ near magnetic materials ($f_0\neq 0$) and $\Phi_c\sim D^{-1}$ near non-magnetic materials ($f_0 = 0$). The minus sign indicates anti-correlations between individual and collective dephasing ($\Phi_c /\Phi_s <0$) at large $D$ near magnetic materials. We notice that this nano-EM interaction range is independent of EM fluctuation wavelengths, in stark contrast to collective emission in free-space. Meanwhile, $\beta$ can take different values when the TLSs are aligned along specific directions $\varphi$ satisfying $f_{0}(\theta_i)=f_{2}(\theta_i)=0$ (e.g., $\beta=2$ at $\varphi=0$ for hyperbolic media). We emphasize that the universality of nano-EM collective dephasing range originates from low-frequency evanescent interface wave contributions. 

In Fig.~\ref{fig:fig2}(b), we present the above power-law dependence with realistic material examples. Here, we consider two simple materials, including the gyromagnetic yttrium iron garnet (YIG, $\mu_{xy}= -\mu_{yx}\neq 0$) and isotropic conductor silver, which exhibit $\Phi_c\sim -D^{-3}$ and $\Phi_c\sim D^{-1}$ at large $D$. Additional material examples of other material classes are provided in Appendix~\ref{appendix_range}. We find that our analytic analysis in Table.~\ref{tab:table1} matches well with exact numeric calculations based on realistic material models. 

\section{Nano-EM super-dephasing}
\begin{figure}[!t]
    \centering
    \includegraphics[width=3.4 in]{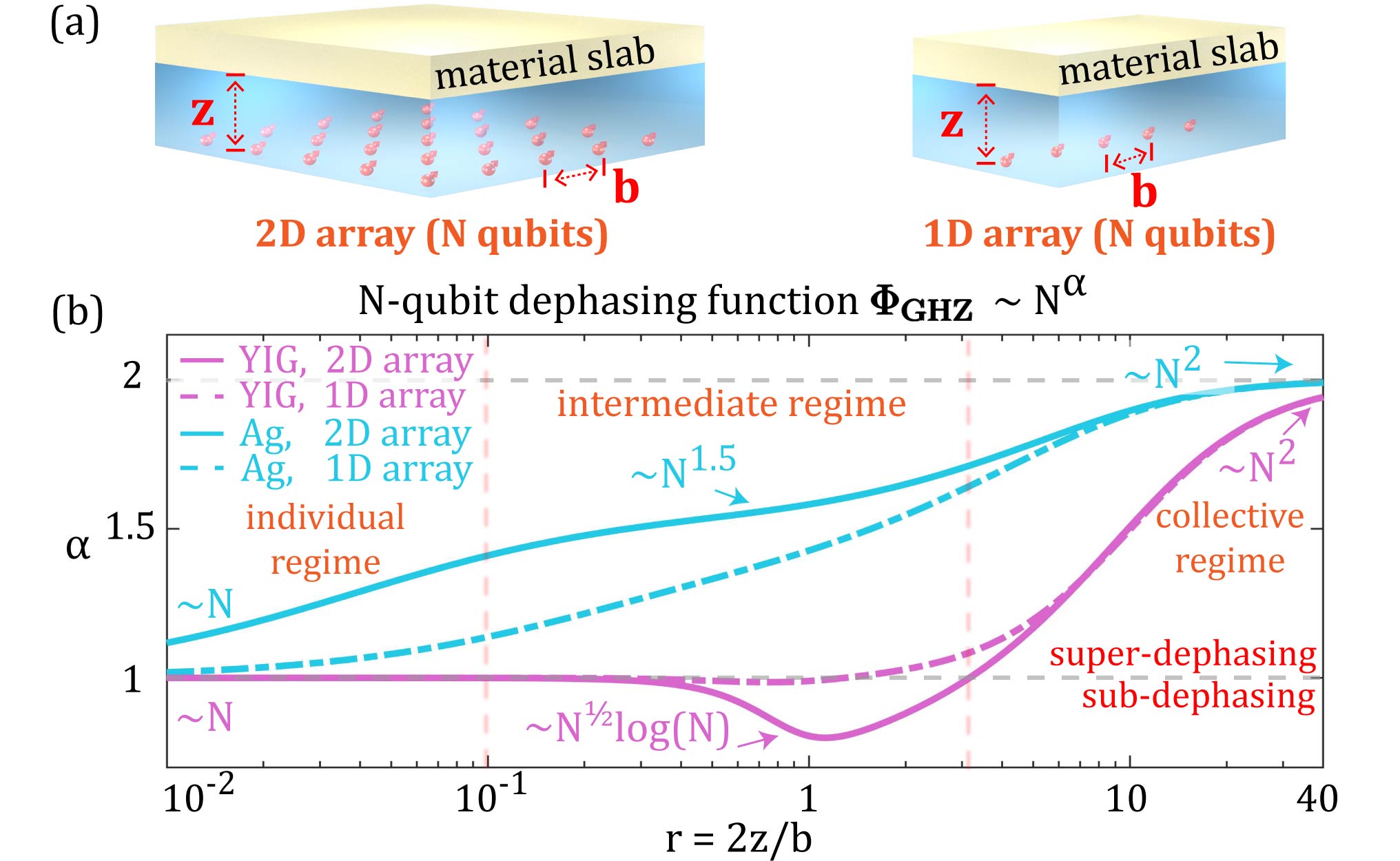}
    \caption{Scaling of nano-EM super-dephasing ($\alpha>1$) and sub-dephasing ($\alpha<1$) in $N$-qubit GHZ states $|\psi\rangle_\mathrm{GHZ}$. (a) Schematics of TLSs arranged in 1D and 2D arrays with lattice constant $b$ at distance $z$ from gyromagnetic (YIG) and non-magnetic (silver) slabs. (b) Scaling of dephasing function $\Phi_{\mathrm{GHZ}} \sim N^{\alpha} \Phi_s$ is controlled by $r=2z/b$. $\alpha$ indicates the collectively accelerated ($\alpha>1$) or suppressed ($\alpha<1$) dephasing processes of $|\psi\rangle_\mathrm{GHZ}$ due to collective dephasing effects.}
    \label{fig:fig3}
\end{figure}

The unique nano-EM interaction range, which is not present in free-space and resonant cavities (see comparison in Fig.~\ref{fig:figures1} in Appendix~\ref{comparison_QED}), leads to super-dephasing scaling laws different from conventional superradiance effects in subwavelength ensembles. We now consider the dephasing of $N$-qubit GHZ states $|\psi\rangle_\mathrm{GHZ}$ when the underlying collective dephasing range follows $\Phi_c \sim -D^{-3}$ and $D^{-1}$. As shown in Fig.~\ref{fig:fig3}(a), we consider TLSs arranged in one-dimensional (1D, $N=1 \times n$) and two-dimensional (2D, $N=n \times n$) arrays near gyromagnetic (YIG) and non-magnetic (silver) slabs. We focus on 
TLSs perpendicular to the slabs. In the following, we investigate the scaling of the dephasing function of N-qubit GHZ states $\Phi_{\mathrm{GHZ}}(t)\sim N^\alpha \Phi_s(t)$ with $N$. $\Phi_{\mathrm{GHZ}}(t)$ characterizes the decay of multiqubit coherence $e^{-\Phi_{\mathrm{GHZ}}(t)}$ of $|\psi\rangle_\mathrm{GHZ}$ and is obtained from the non-unitary part of Eq.~(\ref{ME}) (see Appendix~\ref{appendix_scalinglaws}). We characterize the super-dephasing scaling laws by $\alpha= \partial \ln{\Phi_\mathrm{GHZ}} / \partial \ln{N}$ and demonstrate the dependence of $\alpha$ on array configurations at $n=10$ in Fig.~\ref{fig:fig3}(b). 

As shown in Fig.~\ref{fig:fig3}(b), we find that the scaling of nano-EM super-dephasing with $N$ is controlled by the ratio $r=2z/b$ between lattice constant $b$ and distance from material slabs $z$. For 2D arrays, we find that nano-EM super-dephasing behaviors can be generally classified into three regimes. In the $r \ll 1$ regime, collective dephasing is negligible and individual dephasing dominates dephasing processes, leading to $\Phi_\mathrm{GHZ} \sim N$. In the $r \gtrsim n$ regime, TLSs are in close proximity and collective dephasing is comparable to individual dephasing near both materials. Therefore, collective dephasing strongly accelerates the dephasing of $|\psi\rangle_\mathrm{GHZ}$, resulting in super-dephasing behaviors with $\Phi_\mathrm{GHZ} \sim N^{2}$. Remarkably, in the intermediate regime, dephasing behaviors become sensitive to the range of collective interactions $\Phi_c/\Phi_s$. Near YIG slabs, collective dephasing suppresses the dephasing of $|\psi\rangle_\mathrm{GHZ}$ due to its anti-correlations ($\Phi_c/\Phi_s \sim -D^{-3}<0$) with individual dephasing at large $D$, leading to sub-dephasing behaviors $\Phi_\mathrm{GHZ} \sim \sqrt{N}\log{N}$. In contrast, near silver slabs, although collective dephasing is slower than individual dephasing, it is still correlated with individual dephasing ($\Phi_c/\Phi_s \sim D^{-1}>0$) and leads to moderately accelerated super-dephasing with $\Phi_\mathrm{GHZ} \sim N^{1.5}$. We derive the above scaling laws in Appendix~\ref{appendix_scalinglaws}. For comparison, we note $\Phi_\mathrm{GHZ}$ always follows $\sim N$ in free-space since individual dephasing always dominates dephasing processes in free-space.

Furthermore, we reveal that ensemble dimensionality can also affect the scaling of nano-EM super-dephasing, as shown in Fig.~\ref{fig:fig3}(b). In general, we find that collective dephasing has less prominent effects in 1D arrays due to weaker average couplings in 1D arrays. For example, anti-correlations near YIG slabs do not lead to obvious sub-dephasing of $|\psi\rangle_\mathrm{GHZ}$ in 1D arrays. 

\begin{figure}[!t]
    \centering
    \includegraphics[width = 3.4in]{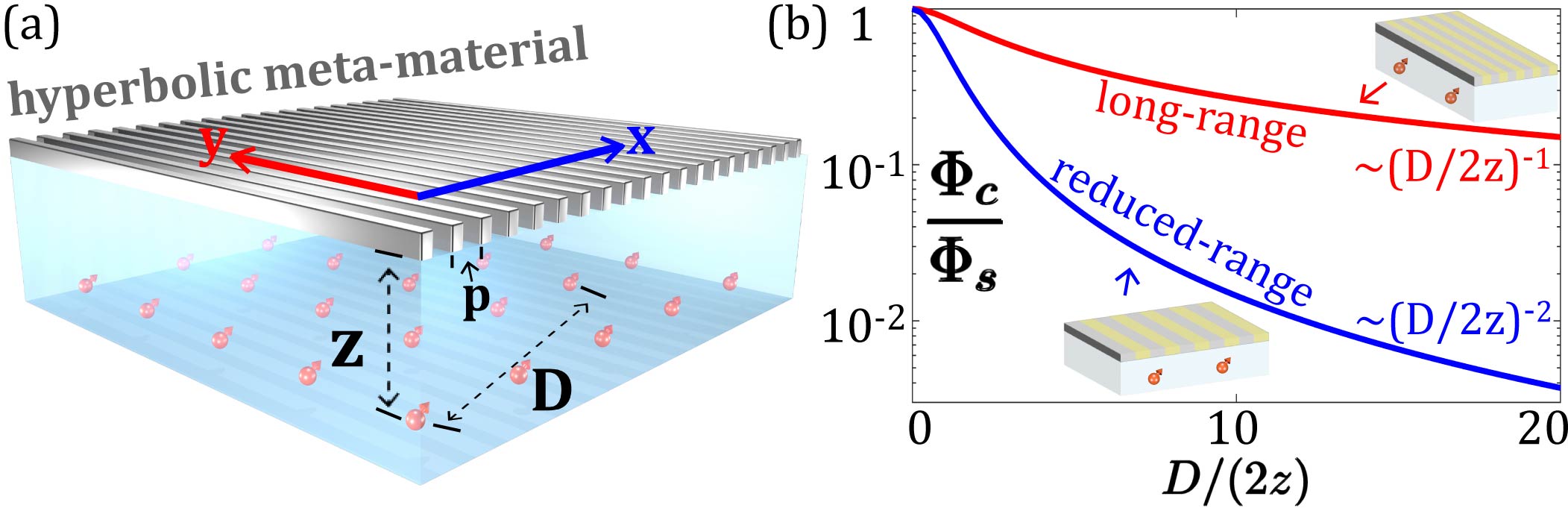} 
    \caption{Engineering nano-EM collective dephasing with hyperbolic meta-materials. (a) Sketch of a magnetic TLS array at distance $z$ from silver gratings with periodicity $p<z$. (b) Hyperbolic meta-materials reduce the range of nano-EM collective dephasing along the grating periodicity direction $\Phi_c \sim D^{-2}$.}
    \label{fig:fig4}
\end{figure}

\section{Engineering nano-EM collective dephasing}
Finally, we discuss engineering the range of nano-EM collective dephasing by exploiting controllable anisotropy in metamaterials. Table.~\ref{tab:table1} and Eq.~(\ref{powerlaw}) already indicate that the nano-EM collective dephasing exhibits reduced range along specific directions near hyperbolic media. As shown in Fig.~\ref{fig:fig4}(a), we now consider periodic silver gratings with periodicity $p$ smaller than distance $z$ from spin qubits. We employ the effective medium approximation~\cite{poddubny2013hyperbolic} valid for the low-frequency response of this structure~\cite{mcphedran1982lossy}. In Fig.~\ref{fig:fig4}(b), we demonstrate that nano-EM collective dephasing along the grating periodicity direction (x direction) is reduced with $\Phi_c \sim D^{-2}$, which matches well with our analytical analysis in Table.~\ref{tab:table1}. We note that this is in stark contrast to collective interactions at atom resonance frequencies (e.g., RDDI), which exhibit increased interaction range near hyperbolic media~\cite{cortes2017super}.

\section{Experimental considerations and discussion}
Our results can be implemented in shallow quantum impurity systems, e.g., nitrogen-vacancy (NV) centers in diamond~\cite{sangtawesin2019origins,favaro2017tailoring,rovny2022nanoscale}. Shallow spins in state-of-the-art experimental systems can have long intrinsic individual decoherence time exceeding $100\,\mathrm{\mu s}$ at depth $z>10\,\mathrm{nm}$~\cite{sangtawesin2019origins,favaro2017tailoring} at room temperatures and suffer negligible correlated noise from other sources~\cite{kwiatkowski2018decoherence,bradley2019ten}. Therefore, for shallow spins near lossy material slabs, the nano-EM environment induced dephasing can dominate over the intrinsic noise effects, enabling the isolation of nano-EM collective dephasing effects.

Nano-EM collective dephasing effects can be observed by measuring the difference in decoherence time of different entangled states. To illustrate this idea, we consider two spin qubits (e.g., two NV centers) separated by $D=20\,\mathrm{nm}$ at distance $z=20\,\mathrm{nm}$ from YIG slabs with experimentally accessible parameters. We study the decoherence of the two-qubit Bell states $|00\rangle + |11\rangle$ and $|01\rangle + |10\rangle$. From Eq.~(\ref{ME}), we find the dephasing functions $\Phi_{|00\rangle + |11\rangle}(t)$ and $\Phi_{|10\rangle + |01\rangle}(t)$ for the two Bell states are,
\begin{subequations}\label{experimental}
\begin{multline}
    \Phi_{|00\rangle + |11\rangle}(t)=\Phi_s(\mathbf{r_1},t)+\Phi_s(\mathbf{r_2},t)\\+\Phi_c(\mathbf{r_1},\mathbf{r_2},t)+\Phi_c(\mathbf{r_2},\mathbf{r_1},t), 
\end{multline}
\begin{multline}
    \Phi_{|01\rangle + |10\rangle}(t)=\Phi_s(\mathbf{r_1},t)+\Phi_s(\mathbf{r_2},t)\\-\Phi_c(\mathbf{r_1},\mathbf{r_2},t)-\Phi_c(\mathbf{r_2},\mathbf{r_1},t),
\end{multline}
\end{subequations}
where $\mathbf{r_1}$ and $\mathbf{r_2}$ are the positions of the two spin qubits. The dephasing function $\Phi_{|00\rangle + |11\rangle}(t)$ ($\Phi_{|10\rangle + |01\rangle}(t)$) determines the decay of multiqubit coherence $e^{-\Phi_{|00\rangle + |11\rangle}(t)}$ ($e^{-\Phi_{|01\rangle + |10\rangle}(t)}$) of the Bell states induced by the nano-EM environment. The decoherence time is defined as $\tau_0$ that satisfies $\Phi_{|00\rangle + |11\rangle}(\tau_0)=1$ ($\Phi_{|01\rangle + |10\rangle}(\tau_0)=1$). From Eq.~(\ref{experimental}) and experimentally accessible YIG parameters~\cite{haidar2015thickness} (see Appendix~\ref{appendix_material}), we find the nano-EM dephasing induced decoherence time for Bell states $|00\rangle+|11\rangle$ and $|01\rangle+|10\rangle$ are $12 \, \mu s$ (collectively accelerated) and $37 \, \mu s$ (collectively suppressed) at room temperatures, respectively. This indicates that the nano-EM environment induced dephasing dominates over the intrinsic noise contributions, and the differences in decoherence time of the two Bell states manifest the effects of nano-EM collective dephasing. For comparison, from Eq.~(\ref{experimental}), we note that the two Bell states would have similar decoherence time in the absence of collective dephasing (i.e., $|\Phi_c/\Phi_s| \ll 1$). 

Furthermore, our results can have broader applications in quantum sensing. Previous work exploited individual spin dephasing to probe material properties~\cite{machado2023quantum,dolgirev2024local}. Meanwhile, collective dephasing discussed in this work provides additional approaches to exploit entanglement to improve quantum sensor sensitivity and probe correlations in material response that may not have prominent effects on individual dephasing.

\begin{acknowledgments}
This work was supported by the Army Research Office under Grant No. W911NF-21-1-0287 and by the Defense Advanced Research Projects Agency under the Quantum Materials Engineering using Electromagnetic Fields (QUAMELEON) program.
\end{acknowledgments}

\appendix

\section{Derivation of the Nano-EM Collective Dephasing Master Equation}\label{derivation}
In this appendix, we provide detailed derivations of the nano-EM collective dephasing master equation (Eqs.~(\ref{ME}-\ref{Jc})). As discussed in the main text, we consider a magnetic two-level systems (TLSs) ensemble interacting with EM fluctuations. We follow the macroscopic quantum electrodynamics (QED) quantization framework~\cite{buhmann2012macroscopic} to describe fluctuating EM fields near material slabs. The total Hamiltonian $\hat{H}=\hat{H}_q+\hat{H}_f+\hat{H}_{int}$ is,
\begin{subequations}
\begin{align}
&\begin{aligned}
    \hat{H}_q=\sum_i \hbar \omega_{i} \hat{\sigma}_i^+ \hat{\sigma}_i^-,
\end{aligned}\\
&\begin{aligned}
    \hat{H}_f=\int d^3 \mathbf{r} \int_0^\infty \hbar \omega \ \hat{\mathbf{f}}^\dagger (\mathbf{r},\omega) \hat{\mathbf{f}}(\mathbf{r},\omega),
\end{aligned}\\
&\begin{aligned}\label{hint}
    \hat{H}_{int}=- \sum_i (\mathbf{m}_{i}^{eg}\hat{\sigma}_i^+ + \mathbf{m}_{i}^{ge}\hat{\sigma}_i^- + \mathbf{m}_{i}\hat{\sigma}_i^z) \cdot \hat{\mathbf{B}}(\mathbf{r_i}).
\end{aligned}
\end{align}
\end{subequations}
Here, $\hat{H}_q$ is the Hamiltonian of the magnetic TLSs, $\hat{H}_f$ is the Hamiltonian of the EM bath, and $\hat{H}_{int}$ describes the interaction between TLSs and the fluctuating magnetic fields. $\omega_{i}$ and $\mathbf{r_i}$ represent the resonance frequency and position of the i\emph{th} TLS. $\hat{\sigma}_i^{+(-)}=|1\rangle \langle0| (|0\rangle \langle1|)$ and $\hat{\sigma}_i^{z}=|1\rangle \langle1|-|0\rangle \langle0|$ are the raising (lowering) operator and the Pauli-z matrix, respectively. $\mathbf{m}^{eg}_i$, $\mathbf{m}^{ge}_i$, and $\mathbf{m}_i$ are the spin magnetic moments depending on the direction of the quantization axis of the i\emph{th} TLS (spin-$1/2$). For TLSs with quantization axes along the $z$ direction, we have $\mathbf{m}_i=[0,0,\hbar \gamma_i/2]$, where $\gamma_i$ is the gyromagnetic ratio. $\hat{\mathbf{f}}^\dagger$ and $\hat{\mathbf{f}}$ are photon creation and annihilation operators satisfying the following relation:
\begin{equation}
    [\hat{\mathrm{f}}_\alpha(\mathbf{r},\omega), \hat{\mathrm{f}}_\beta^\dagger(\mathbf{r}',\omega')]=\delta_{\alpha\beta}\delta(\mathbf{r}-\mathbf{r'})\delta(\omega-\omega'), 
\end{equation}
where $\alpha,\beta = x,y,z $. Magnetic field operator $\hat{\mathbf{B}}(\mathbf{r})$ can be expressed in terms of $\hat{\mathbf{f}}^\dagger$, $\hat{\mathbf{f}}$, and electric dyadic Green's function $\overleftrightarrow{G}$:  
\begin{subequations}
\begin{align}
&\begin{aligned}
    \hat{\mathbf{B}}(\mathbf{r}) = \int_0^\infty d\omega [\hat{\mathbf{B}}(\mathbf{r},\omega)+\hat{\mathbf{B}}^\dagger(\mathbf{r},\omega)],
\end{aligned}\\
&\begin{aligned}
    \hat{\mathbf{B}}(\mathbf{r},\omega) = (i \omega)^{-1} \int d^3 \mathbf{r}' \ \nabla_\mathbf{r} \times \overleftrightarrow{G}(\mathbf{r},\mathbf{r}',\omega) \cdot \hat{\mathbf{f}}(\mathbf{r}',\omega).
\end{aligned}
\end{align}
\end{subequations}

In the following, we focus on the interactions between longitudinal fluctuating EM fields and magnetic TLSs that induce pure dephasing effects. We note that the transverse components of EM fluctuations will induce spontaneous emission effects in the TLSs~\cite{sun2023limits}. In the interaction picture with respect to $\hat{H}_0=\hat{H}_q+\hat{H}_f$, the interaction Hamiltonian $\hat{H}_{int}^\mathrm{I}(t)$ is (superscript $\mathrm{I}$ denotes the interaction picture):
\begin{multline}\label{Hint_dep}
    \hat{H}_{int}^\mathrm{I}(t) = -\Bigg[ \int_0^\infty d\omega \ (i\omega)^{-1} 
    \int d^3\mathbf{r'} \sum_i \mathbf{m}_i \hat{\sigma}_i^z \\ \cdot [\nabla_{\mathbf{r_i}} \times \overleftrightarrow{G}(\mathbf{r_i},\mathbf{r}',\omega)] \cdot 
    \hat{\mathbf{f}}(\mathbf{r}',\omega) e^{- i\omega t} + h.c. \Bigg].
\end{multline}
In the interaction picture, the Liouville–von Neumann equation is:
\begin{equation}\label{totliouville}
    \frac{d \rho^{\mathrm{I}}_{tot}(t)}{d t}=\frac{1}{i \hbar} [\hat{H}_{int}^\mathrm{I}(t),\rho^\mathrm{I}_{tot}(t)].
\end{equation}

Substituting the total density matrix $\rho^{\mathrm{I}}_{tot}(t) = \rho^{\mathrm{I}}(t) \otimes \rho^{\mathrm{I}}_{f}(t)$ into the integral form of Eq.~(\ref{totliouville}) and tracing off the field part $\rho^{\mathrm{I}}_{f}(t)$, we have:
\begin{align}\label{liouvilleeqaftertrace}
\begin{aligned}
    &\frac{d \rho^{\mathrm{I}}(t)}{d t} =-\frac{1}{\hbar^2}\int_0^{t} d\tau \ \mathrm{Tr}_f [\hat{H}_{int}^\mathrm{I}(t),[\hat{H}_{int}^\mathrm{I}(\tau),\rho^\mathrm{I}_{tot}(\tau)]],
\end{aligned}
\end{align}
where $\rho^{\mathrm{I}}(t)$ and $\rho^{\mathrm{I}}_{f}(t)$ are the density matrices of the TLSs and EM bath, respectively.

To this end, our derivations do not involve approximation of the EM bath. Eq.~(\ref{liouvilleeqaftertrace}) relates $\rho^{\mathrm{I}}(t)$ to $\rho^{\mathrm{I}}(\tau)$ at all previous moments $\tau$ and is difficult to solve. Here, we assume the weak-coupling condition and employ the time-convolutionless projection operator (TCL) technique to obtain a time-local master equation~\cite{breuer2002theory}. The second-order TCL generator leads to:
\begin{align}\label{TCL2}
\begin{aligned}
    \frac{d \rho^{\mathrm{I}}(t)}{d t} =-\frac{1}{\hbar^2}\int_0^{t} d\tau \ \mathrm{Tr}_f [\hat{H}_{int}^\mathrm{I}(t),[\hat{H}_{int}^\mathrm{I}(\tau),\rho^\mathrm{I}_{tot}(t)]],
\end{aligned}
\end{align}
The real part of Eq.~(\ref{TCL2}) describes the dephasing processes of the TLSs ensemble, and the imaginary part of Eq.~(\ref{TCL2}) determines the unitary evolution processes. We evaluate Eq.~(\ref{TCL2}) by substituting Eq.~(\ref{Hint_dep}) into Eq.~(\ref{TCL2}). As an example, we can obtain (assuming that the EM bath is in the vacuum state):
\begin{widetext}
\begin{align}
\begin{aligned}\label{dec1}
    -\frac{1}{\hbar^2}\int_0^{t} d\tau \ \mathrm{Tr}_f [\hat{H}_{int}^\mathrm{I}(t) \hat{H}_{int}^\mathrm{I}(\tau) \rho^\mathrm{I}_{tot}(t)]
    =& \sum_{ij} \Big[ -\frac{\mu_0}{\hbar \pi} \int_0^\infty d\omega \ \frac{\omega \, \sin{\omega t}}{c^2} \, \Big[\mathbf{m}_{i} \cdot  \mathcal{I} \overleftrightarrow{G}_m(\mathbf{r_i},\mathbf{r_j},\omega) \cdot \mathbf{m}_{j}\Big] \, \hat{\sigma}^z_i \hat{\sigma}^z_j \rho^\mathrm{I}(t)\\
    & + \frac{i \mu_0}{\hbar \pi}\int_0^\infty d\omega \ \frac{\omega \, (1-\cos{\omega t})}{c^2} \Big[\mathbf{m}_{i} \cdot  \mathcal{I} \overleftrightarrow{G}_m(\mathbf{r_i},\mathbf{r_j},\omega) \cdot \mathbf{m}_{j}\Big] \,  \hat{\sigma}^z_i \hat{\sigma}^z_j \rho^\mathrm{I}(t)\Big] ,
\end{aligned}
\end{align}
\end{widetext}
and other terms in the commutator in Eq.~(\ref{TCL2}) can be calculated similarly. Here, we define the magnetic dyadic Green's function $\overleftrightarrow{G}_m(\mathbf{r_i},\mathbf{r_j},\omega)$ and $\mathcal{I} \overleftrightarrow{G}_m(\mathbf{r_i},\mathbf{r_j},\omega)$ as:
\begin{equation}\label{MGreen1}
    \overleftrightarrow{G}_m (\mathbf{r_{i},r_{j}},\omega)= \frac{1}{k_0 ^2} \nabla_i \times \overleftrightarrow{G}(\mathbf{r_{i},r_{j}},\omega) \times \nabla_j,
\end{equation}
\begin{equation}
    \mathcal{I} \overleftrightarrow{G}_m(\mathbf{r_i},\mathbf{r_j},\omega)= \frac{\overleftrightarrow{G}_m(\mathbf{r_i},\mathbf{r_j},\omega) - \overleftrightarrow{G}_m^\dagger(\mathbf{r_j},\mathbf{r_i},\omega)}{2i},
\end{equation}
and $k_0=\omega/c$. We employ the following relation~\cite{buhmann2012macroscopic} in deriving Eq.~(\ref{dec1}):
\begin{equation}
    \int d\mathbf{s} \overleftrightarrow{G}(\mathbf{r_i},s,\omega) \cdot \overleftrightarrow{G}^\dagger(\mathbf{r_j},s,\omega) = \frac{\hbar \mu_0 \omega^2}{\pi}\mathcal{I} \overleftrightarrow{G}(\mathbf{r_i},\mathbf{r_j},\omega).
\end{equation}

It is worth noting that, only in the absence of any non-reciprocal material, we have $\overleftrightarrow{G}_m(\mathbf{r_i},\mathbf{r_j},\omega)= \overleftrightarrow{G}_m^\intercal(\mathbf{r_j},\mathbf{r_i},\omega)$ and $\mathcal{I} \overleftrightarrow{G}_m(\mathbf{r_i},\mathbf{r_j},\omega) = \mathrm{Im} \overleftrightarrow{G}_m(\mathbf{r_i},\mathbf{r_j},\omega)$, where $\mathrm{Im}[\cdot]$ denotes the imaginary part of the tensor. In general, $\mathcal{I} \overleftrightarrow{G}_m(\mathbf{r_i},\mathbf{r_j},\omega) \neq \mathrm{Im} \overleftrightarrow{G}_m(\mathbf{r_i},\mathbf{r_j},\omega)$ and $\mathcal{I} \overleftrightarrow{G}_m(\mathbf{r_i},\mathbf{r_j},\omega)$ is a complex value that can be separated into a symmetric real part and anti-symmetric imaginary part under the $\mathbf{r_i}$ and $\mathbf{r_j}$ exchange:
\begin{multline}\label{complex_IM}
    \mathcal{I} \overleftrightarrow{G}_m(\mathbf{r_i},\mathbf{r_j},\omega) = \frac{\mathrm{Im}[\overleftrightarrow{G}_m(\mathbf{r_i},\mathbf{r_j},\omega)+\overleftrightarrow{G}^\intercal_m(\mathbf{r_j},\mathbf{r_i},\omega)]}{2} \\ + i \,  \frac{\mathrm{Re}[-\overleftrightarrow{G}_m(\mathbf{r_i},\mathbf{r_j},\omega)+\overleftrightarrow{G}^\intercal_m(\mathbf{r_j},\mathbf{r_i},\omega)]}{2}.
\end{multline}

Through substituting Eq.~(\ref{complex_IM}) into Eqs.~(\ref{TCL2}--\ref{dec1}), we find that \textit{the anti-symmetric part of $\mathcal{I} \overleftrightarrow{G}_m(\mathbf{r_i},\mathbf{r_j},\omega)$ does not contribute to the nano-EM collective dephasing processes (real part of Eq.~(\ref{TCL2}))}. 
We change the integration order regarding $\omega$ and $t$ in Eq.~(\ref{TCL2}) and obtain the nano-EM super-dephasing dynamics in the Schr\"{o}dinger picture (Eq.~(\ref{ME}) in the main text):
\begin{widetext}
\begin{equation}\label{me}
    \frac{d \rho(t)}{d t} = -i[\hat{H}_{d},\rho(t)] +\sum_{i} \gamma_{s}^{\phi}(\mathbf{r_i},t) \ \Big[\hat{\sigma}_{i}^z \rho(t) \hat{\sigma}_{i}^z - \rho(t) \Big] +\sum_{\substack{i \neq j}} \gamma_{c}^{\phi}(\mathbf{r_i},\mathbf{r_j},t) \ \Big[\hat{\sigma}_{i}^z \rho_{q}(t) \hat{\sigma}_{j}^z-\frac{1}{2} \hat{\sigma}_{i}^z \hat{\sigma}_{j}^z \rho(t) - \frac{1}{2} \rho(t) \hat{\sigma}_{i}^z \hat{\sigma}_{j}^z \Big],
\end{equation}
\end{widetext}
where the dipolar interaction Hamiltonian $\hat{H}_{d}$ governing the unitary evolution can be obtained from the imaginary part of Eq.~(\ref{TCL2}). The time-dependent individual and collective dephasing rates $\gamma^{s}_{\phi}(\mathbf{r_i},t)$ and $\gamma^{c}_{\phi}(\mathbf{r_i},\mathbf{r_j},t)$  at time $t$ are:
\begin{widetext}
\begin{align}
    \gamma^{c}_{\phi}(\mathbf{r_i},\mathbf{r_j},t) &= \frac{2\mu_0}{\hbar \pi} \int_0^{\omega_c} d\omega \ \frac{\sin{\omega t}}{\omega} \, (2N(\omega)+1) \,\frac{\omega^2}{c^2} \, \Big[\mathbf{m}_{i} \cdot  \frac{\mathrm{Im}[\overleftrightarrow{G}_m(\mathbf{r_i},\mathbf{r_j},\omega)+\overleftrightarrow{G}^\intercal_m(\mathbf{r_j},\mathbf{r_i},\omega)]}{2} \cdot \mathbf{m}_{j}\Big]\label{td_cd},\\
    \gamma^{s}_{\phi}(\mathbf{r_i},t) &= \frac{2\mu_0}{\hbar \pi} \int_0^{\omega_c} d\omega \ \frac{\sin{\omega t}}{\omega} \, (2N(\omega)+1) \, \frac{\omega^2}{c^2} \, \Big[\mathbf{m}_{i} \cdot  \frac{\mathrm{Im}[\overleftrightarrow{G}_m(\mathbf{r_i},\mathbf{r_i},\omega)+\overleftrightarrow{G}^\intercal_m(\mathbf{r_i},\mathbf{r_i},\omega)]}{2} \cdot \mathbf{m}_{i}\Big],\label{td_sd}
\end{align}
\end{widetext}
where we incorporate the effects of the EM bath thermal fluctuations at temperature $T$ into Eqs.~(\ref{td_cd},~\ref{td_sd}) through $2N(\omega)+1=\coth(\hbar \omega/2k_BT)$. $N(\omega)=1/(e^{\hbar \omega / k_B T}-1)$ is the mean photon number of the thermal EM bath. $\omega_c$ is the cutoff frequency that will be discussed later. In the absence of external pulse sequences, the nano-EM individual and collective dephasing functions $\Phi_s(\mathbf{r_i},t)$, $\Phi_c(\mathbf{r_i},\mathbf{r_j},t)$ can be defined by the temporal integration of Eqs.~(\ref{td_cd},~\ref{td_sd}). We obtain (Eqs.~(\ref{td_cdf},~\ref{Jc}) in the main text):
\begin{widetext}
\begin{align}
&\begin{aligned}\label{app_td_cdf}
    \Phi_{c}(\mathbf{r_i},\mathbf{r_j},t) &= 2\int_0^t dt' \, \gamma^{c}_{\phi}(\mathbf{r_i},\mathbf{r_j},t')\\
    &= \frac{4\mu_0}{\hbar \pi} \int_0^{\omega_c} d\omega \ \frac{1-\cos{\omega t}}{\omega^2} \coth{\frac{\hbar \omega}{2 k_B T}} \frac{\omega^2}{c^2} \, \Big[\mathbf{m}_{i} \cdot  \frac{\mathrm{Im}[\overleftrightarrow{G}_m(\mathbf{r_i},\mathbf{r_j},\omega)+\overleftrightarrow{G}^\intercal_m(\mathbf{r_j},\mathbf{r_i},\omega)]}{2} \cdot \mathbf{m}_{j}\Big]\\
    &=\int_0^{\omega_c} d\omega \, F(\omega, t) \, J_c(\mathbf{r_i}, \mathbf{r_j}, \omega).
\end{aligned} \\
&\quad \,  \begin{aligned}\label{td_sdf}
    \Phi_{s}(\mathbf{r_i},t) = 2\int_0^t dt' \, \gamma^{s}_{\phi}(\mathbf{r_i},t')= \int_0^{\omega_c} d\omega \, F(\omega, t) \, J_s(\mathbf{r_i}, \omega) = \Phi_{c}(\mathbf{r_i},\mathbf{r_j},t)|_{\mathbf{r_j}\rightarrow\mathbf{r_i}},
\end{aligned}   
\end{align}
\end{widetext}
Here, we decompose the integrands in Eqs.~(\ref{td_sdf},~\ref{app_td_cdf}) into the Ramsey filter function $F(\omega, t)=(1-\cos{\omega t})/\omega^2$~\cite{cywinski2008enhance} and noise (correlation) spectra $J_s(\mathbf{r_i}, \omega)$ ($J_c(\mathbf{r_i}, \mathbf{r_j}, \omega)$). The symmetric formalism $\Phi_c(\mathbf{r_i},\mathbf{r_j}) = \Phi_c(\mathbf{r_j},\mathbf{r_i})$ guarantees the Hermiticity of the master equation Eq.~(\ref{me}) in both reciprocal and non-reciprocal photonic environments. We define the timescale of nano-EM collective and individual dephasing as $t^{c}_\phi$ and $t^{s}_\phi$ satisfying $|\Phi_{c}(\mathbf{r_i},\mathbf{r_j},t^{c}_\phi)|=1$ and $|\Phi_{s}(\mathbf{r_i},t^{s}_\phi)|=1$.

    \subsection{High frequency cutoff $\omega_c$}
    Without the cutoff frequency $\omega_c$, Eqs.~(\ref{app_td_cdf},~\ref{td_sdf}) will not converge (even with the filter function $F(\omega)$) due to the free-space vacuum fluctuation contributions. At $\omega \to\infty$, we have: 
    \begin{subequations}\label{vacuumG}
    \begin{align}
    &\begin{aligned}
         \mathrm{Im} \overleftrightarrow{G}_m(\mathbf{r_i},\mathbf{r_i}, \omega) \approx  \mathrm{Im} \overleftrightarrow{G}_m^0(\mathbf{r_i},\mathbf{r_i}, \omega) = \frac{\omega}{6\pi c},
    \end{aligned} \\
    &\begin{aligned}
        \mathrm{Im} \overleftrightarrow{G}_m(\mathbf{r_i},\mathbf{r_j}, \omega) \approx  \mathrm{Im} \overleftrightarrow{G}_m^0(\mathbf{r_i},\mathbf{r_j}, \omega)  \approx \frac{\sin(\omega D/c)}{4\pi D},
    \end{aligned}
    \end{align}
    \end{subequations}
    where $\overleftrightarrow{G}_m^0(\omega)$ is the magnetic dyadic Green's function in free-space and $D=|r_i-r_j|$ is the interatom distance. Considering $\coth{\frac{\hbar \omega }{ 2k_BT}}\approx 1$ at $\omega \to +\infty$, we find that the frequency integral in Eqs.~(\ref{app_td_cdf},~\ref{td_sdf}) will not converge without the cutoff frequency $\omega_c$. This type of divergence is widely encountered in the calculations of EM-fluctuation induced effects, e.g., Lamb shift calculations and Casimir force calculations. Here, similar to calculations or other QED effects, we consider a frequency cutoff $\omega_c$ to regulate the high-frequency behavior of the integrand in Eqs.~(\ref{app_td_cdf},~\ref{td_sdf}). We select the $\omega_c=mc^2/\hbar$ to be the same as the Lamb shift calculations. We emphasize that the choice of $\omega_c$ does not affect the results presented in the paper. We note that, although the choice of $\omega_c$ can quantitatively affect the collective $\Phi_c$ and individual $\Phi_s$ dephasing functions in free-space, their ratio $|\Phi_c/\Phi_s| \ll 1$ is always small in free-space. Therefore, the dephasing processes of TLSs in free-space are always dominated by individual dephasing. Meanwhile, the choice of $\omega_c$ has negligible influence on dephasing in the nano-EM environments since $\Phi_c$ and $\Phi_s$ in the nano-EM environments are determined by low-frequency EM fluctuations, as discussed in the main text.
     
    \subsection{Completely positive and trace-preserving maps}
    Physically valid quantum dynamics must be trace-preserving and completely positive. This requires the tensor $\Psi$, defined by $[\Psi]_{ij}=\Phi_{c}(\mathbf{r_i},\mathbf{r_j},t)$ and $ [\Psi]_{ii}=\Phi_{s}(\mathbf{r_i},t)$, to be positive semi-definite~\cite{lidar2019lecture}. Therefore, we need to have $\Phi_{c}(\mathbf{r_i},\mathbf{r_j},t) = \Phi_{c}(\mathbf{r_j},\mathbf{r_i},t)$ and $\Phi_{s}(\mathbf{r_i},t) > 0$. We note that Eqs.~(\ref{td_sdf},~\ref{app_td_cdf}) satisfy both conditions in general photonic environments, which can be reciprocal or non-reciprocal. This is because the symmetric $\mathrm{Im}[\overleftrightarrow{G}_m(\mathbf{r_i},\mathbf{r_j},\omega)+\overleftrightarrow{G}^\intercal_m(\mathbf{r_j},\mathbf{r_i},\omega)]/2$ guarantees $\Phi_{c}(\mathbf{r_i},\mathbf{r_j},t) = \Phi_{c}(\mathbf{r_j},\mathbf{r_i},t)$ and positive definiteness of $\mathrm{Im}[\overleftrightarrow{G}_m(\mathbf{r_i},\mathbf{r_i},\omega)+\overleftrightarrow{G}^\intercal_m(\mathbf{r_i},\mathbf{r_i},\omega)]/2$ at all $\omega$ ensures $\Phi_{s}(\mathbf{r_i},t) > 0$. Meanwhile, we note that, in the presence of non-reciprocity, the previously discussed EM correlations $\mathrm{Im}[\overleftrightarrow{G}_m(\mathbf{r_i},\mathbf{r_j})]$~\cite{premakumar2017evanescent,kenny2021magnetic} are not good approximations of the general formalism $\mathrm{Im}[\overleftrightarrow{G}_m(\mathbf{r_i},\mathbf{r_j})+\overleftrightarrow{G}^\intercal_m(\mathbf{r_j},\mathbf{r_i})]/2$ valid in non-reciprocal environments. Neglecting the effects of non-reciprocity on EM fluctuation correlations (i.e., $\mathrm{Im}[\overleftrightarrow{G}_m(\mathbf{r_i},\mathbf{r_j})+\overleftrightarrow{G}^\intercal_m(\mathbf{r_j},\mathbf{r_i})]/2 \neq \mathrm{Im}[\overleftrightarrow{G}_m(\mathbf{r_i},\mathbf{r_j})]$) will lead to non-physical quantum dynamics in non-reciprocal photonic environments.
    
    \subsection{Non-Markovianity}
    Eq.~(\ref{me}) is not limited to the Markovian EM bath. It is worth noting that Eq.~(\ref{me}) is not a time-dependent Markovian master equation since the individual and collective dephasing rates $\gamma_{s}^{\phi}(\mathbf{r_i},t)$ and $\gamma_{c}^{\phi}(\mathbf{r_i},\mathbf{r_j},t)$ explicitly depend on the initial time $t_0$~\cite{chruscinski2010non}, which is chosen to be $t_0=0$ here. The memory effects of the EM bath can be reflected in the dependence of $\gamma_{s}^{\phi}(\mathbf{r_i},t)$ and $\gamma_{c}^{\phi}(\mathbf{r_i},\mathbf{r_j},t)$ on $t$ and $t_0$~\cite{chruscinski2010non}. However, for configurations discussed in the main text, we note that non-Markovianity does not have prominent effects on dephasing effects. For atoms/spins coupled to the EM bath, non-Markovianity is usually associated with (1) the retardation effects, which are prominent when atoms are separated by a large distance~\cite{sinha2020non}, and (2) the memory effects of the EM bath, which is usually prominent only at a time scale much shorter compared to qubit decoherence~\cite{breuer2002theory}. Here, since we are considering spin qubit ensembles much smaller than low-frequency EM fluctuation wavelengths, the retardation effects are not prominent in the configurations in this manuscript. Additionally, in this manuscript, we focus on the experimentally measurable effects of collective dephasing at time scales much longer than the memory time of the EM environment. Therefore, we neglect non-Markovianity in the main text.

\begin{figure*}[!t]
    \centering
    \includegraphics[width=5in]{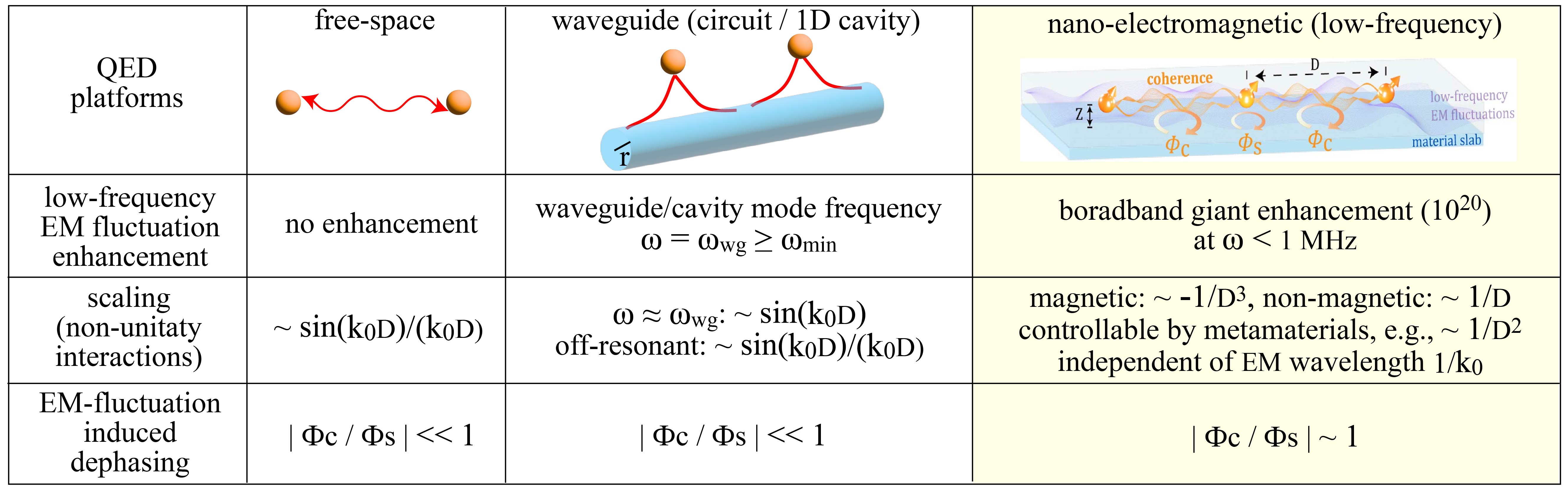}
    \caption{Comparison of collective dephasing in different QED platforms. $\omega_{wg}$ represent the waveguide resonant mode frequencies. $r$ is the radius of the circular waveguide and $D$ is the interatom distance. Waveguides and resonant cavities provide enhancement of EM fluctuations at resonance frequencies higher than the lowest mode frequency $\omega_{min}$ inversely proportional to $r$ or the cavity length. Meanwhile, nano-EM environments near magnets and conductors provide ubiquitous broadband enhancement of magnetic field fluctuations at low frequencies. In the nano-EM environments near material slabs, collective dephasing is greatly enhanced and exhibits collective interaction range not present in free-space, waveguide, or cavity QED systems.}
    \label{fig:figures1}
\end{figure*}

\section{Collective Dephasing in Different QED Platforms} \label{comparison_QED}
In this appendix, we compare collective dephasing effects in different photonic environments. As discussed in the main text, in the nano-EM environments near material slabs, collective dephasing is greatly enhanced and exhibits collective interaction range not found in free-space, waveguide, or cavity QED systems. In Fig.~\ref{fig:figures1}, we summarize our discussions in the main text and compare nano-EM environments with other commonly encountered QED platforms. We emphasize that the nano-EM environment provides a unique regime for the emergence of super-dephasing phenomena mediated by EM fluctuations.

In the following, we provide supplemental details for calculating collective dephasing $\Phi_c/\Phi_s$ and low-frequency noise correlation spectra $J_c (\omega)$ in free-space, resonant cavities, and nano-EM environments. We consider two magnetic TLSs with $\mathbf{m_i}=\mathbf{m_j}$ along the $z$ direction perpendicular to $\mathbf{r_i}-\mathbf{r_j}$ and denote $D=|\mathbf{r_i}-\mathbf{r_j}|$. 

\subsection{Free-space}
Substituting Eq.~(\ref{vacuumG}) into Eqs.~(\ref{app_td_cdf},~\ref{td_sdf}), we have dephasing functions in free-space,
\begin{widetext}
\begin{subequations}
\begin{align}
&\begin{aligned}\label{freespace_cd}
    \Phi_c(\mathbf{r_{i},r_{j}},t)=\frac{4\mu_0\mathbf{m}_i^2}{\hbar \pi c^2} \int_0^{\omega_{c}} d\omega \ \frac{1-\cos{\omega t}}{\omega^2} \, \omega^2 \coth{\frac{\hbar \omega}{2 k_B T}}  \mathrm{Im} \left[ -\frac{c^2}{4 \pi \omega^2 D^3} (1-i\frac{\omega D}{c} -\frac{\omega^2 D^2}{c^2}) \, e^{i\omega D /c} \right],
\end{aligned}\\
&\begin{aligned}\label{freespace_sd}
    \Phi_s(\mathbf{r_{i}},t)=\frac{4 \mu_0 \mathbf{m}_{i}^2}{\hbar \pi c^2} \int_0^{\omega_{c}} d\omega \ \frac{1-\cos{\omega t}}{\omega^2} \, \omega^2 \coth{\frac{\hbar \omega}{2 k_B T}} \, \frac{\omega}{6\pi c}.
\end{aligned}
\end{align}
\end{subequations}
\end{widetext}
In the main text Fig.~\ref{fig:fig1}(c), we take $\omega_{c} \approx 7.73\times 10^{20} \, \mathrm{Hz}$ and $t=100\,\mathrm{\mu s}$ in the calculations of $|\Phi_c/\Phi_s|$ in free space.

\subsection{Resonant cavity}
\begin{figure}[!h]
    \centering
    \includegraphics[width=2.3 in]{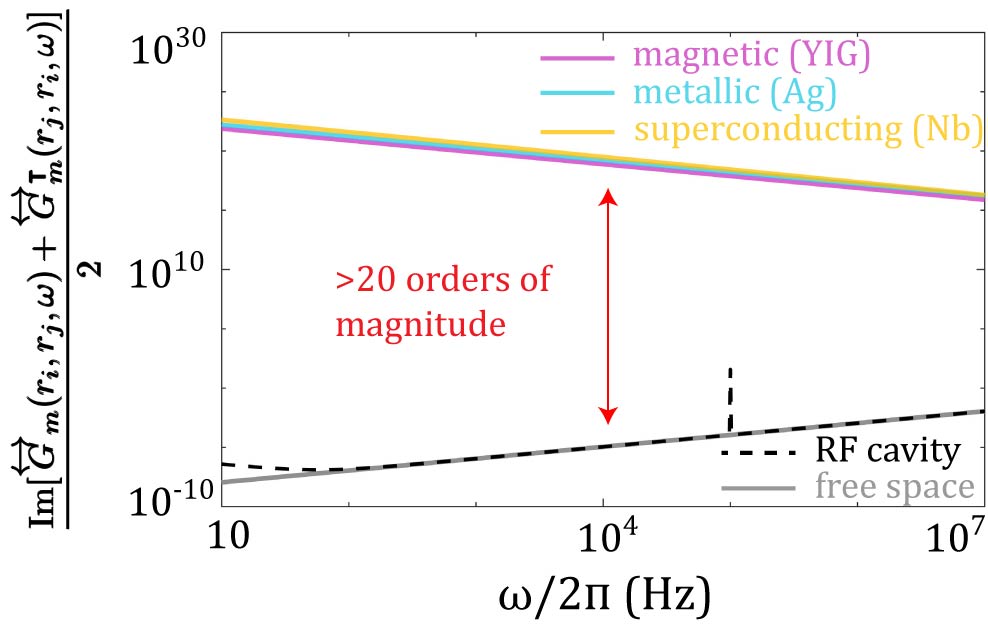}
    \caption{Low-frequency noise correlations $J_c (\omega)$ and Green’s functions $\mathrm{Im} [\overleftrightarrow{G}_m]_{zz}$ are significantly enhanced in the nano-electromagnetic environments compared to free space. We consider $D=|r_i-r_j|=50\,\mathrm{nm}$ and $z=50\,\mathrm{nm}$ in the calculations. Material parameters are provided in Appendix.~\ref{appendix_material}.}
    \label{fig:figures2}
\end{figure}

We consider an open single-mode cavity with cavity mode frequency $\omega_{cav} = 2\pi \times 0.1\,\mathrm{MHz}$. For $D=|r_i-r_j|$ much smaller than the wavelength of the cavity mode, we approximate the Green's function in cavities $\overleftrightarrow{G}^{cav}_m$ as
\begin{multline}
    \mathrm{Im} \overleftrightarrow{G}^{cav}_m(\mathbf{r_i},\mathbf{r_j}, \omega) =  \mathrm{Im} \overleftrightarrow{G}_m^0(\mathbf{r_i},\mathbf{r_j}, \omega) \\ \big[ 1+ P \frac{(\omega_{cav}/2Q)^2}{(\omega_{cav}/2Q)^2+(\omega - \omega_{cav})^2} \big],
\end{multline}
where we take the cavity Purcell factor $P=10^6$ and quality factor $Q=5\times 10^5$ in the calculations of $J_c(\omega)$ and $|\Phi_c/\Phi_s|$ in Fig.~\ref{fig:figures2}. 

\subsection{Nano-EM environment}
As discussed in the main text, the Green’s function in the nano-EM environment is given by 
\begin{multline}\label{SI_Gzz}
    \mathrm{Im} [\overleftrightarrow{G}_m(\mathbf{r_i},\mathbf{r_j})]_{zz} \approx \frac{1}{8\pi^2 k_0^2} \int dq d\theta \\ q^2 \allowbreak r_{ss}(q,\theta) e^{iqD\cos{(\theta-\varphi)}} e^{-2qz},
\end{multline}
where $k_0=\omega/c$. The reflection coefficient $r_{ss}(q,\theta)$ is determined by material properties $\overleftrightarrow{\mu}$ and $\overleftrightarrow{\varepsilon}$, $\varphi$ is the relative angle between $\mathbf{r_i}-\mathbf{r_j}$ and x-axis, and $z$ is the distance between TLSs and material slabs. Details of reflection coefficient calculations are provided in Appendix~\ref{appendix_range}. In the main text, we take $t = 100 \,\mathrm{\mu s}$ in the calculations of $|\Phi_c/\Phi_s|$ and consider gyromagnetic material yttrium iron garnet (YIG), non-magnetic conductor silver, and BCS superconductor niobium. Details of material models and experimentally accessible material parameters are provided in Appendix~\ref{appendix_material}. 

In Fig.~\ref{fig:figures2}, we plot low-frequency $\mathrm{Im} \overleftrightarrow{G}_m (\mathbf{r_i},\mathbf{r_j},\omega)$ in different photonic environments. Fig.~\ref{fig:fig1}(b) and Fig.~\ref{fig:figures2} clearly show that low-frequency $J_c (\omega) \propto \mathrm{Im} \overleftrightarrow{G}_m (\mathbf{r_i},\mathbf{r_j},\omega)$ is enhanced by about 20 orders of magnitude in the nano-EM environment compared to free-space or RF cavity. This giant enhancement makes the photonic environment-induced dephasing effects more prominent than intrinsic noise effects, thus enabling experimental observation. We further note that low-frequency EM fluctuations roughly follow $\mathrm{Im} \overleftrightarrow{G}_m \sim 1/\omega$ in the nano-EM environments considered here, while follow $\mathrm{Im} \overleftrightarrow{G}_m \sim \omega$ in free space. Therefore, this giant enhancement of $J_c(\omega)$ at low frequencies reflects the fundamental differences between low-frequency EM fluctuations in near-field nanophotonic environments and in free space. As discussed in the main text, this giant enhancement of $J_c(\omega)$ leads to distinct behaviors of collective dephasing in nano-EM environments and other QED platforms.

\subsection{Simple calculations of the enhancement magnitudes in the nano-EM environment}
In this subsection, we provide simplified toy model calculations to show that the giant enhancement of noise correlation spectra $J_c^{nem} /J_c^{vac} >10^{20}$ in Fig.~\ref{fig:fig1}(b) in the main text is a general effect near common conductors and magnetic materials. We use $J_c^{nem}$ and $J_c^{vac}$ to denote noise correlation spectra in the nano-EM environment and free-space, respectively. We focus on $J_c^{nem}/J_c^{vac} \propto \overleftrightarrow{G}_m^{nem}/\overleftrightarrow{G}_m^{vac} $. For simplicity, we consider isotropic materials with scalar $\varepsilon$ and $\mu$. As discussed in the main text, in the nano-EM environment, $\overleftrightarrow{G}_m$ is determined by evanescent interface waves with momentum $q \gg k_0$. For magnetic materials, we have the Fresnel reflection coefficients $r_{ss} (q) \approx (\mu -1)/(\mu +1)$ at $q/k_0 \to +\infty$. For non-magnetic materials, we have $r_{ss} (q) \approx (\varepsilon -1)(k_0^2)/(4q^2) $ at $q/k_0 \to +\infty$. Substituting into Eq.~(\ref{SI_Gzz}), we obtain
\begin{subequations}
\begin{align}
    &\begin{aligned}
        \mathrm{Im} [\overleftrightarrow{G}_m^{nem} (r_i,r_j,\omega)+\overleftrightarrow{G}_m^{nem,\intercal} (r_j,r_i,\omega)]_{zz}/2 \\ \approx \frac{1}{4\pi k_0^2}   \frac{8z^2-D^2 }{[4z^2+D^2 ]^{5/2}}   \mathrm{Im}  \frac{\mu -1}{\mu +1},
    \end{aligned}\\
    &\begin{aligned}
        \mathrm{Im} [\overleftrightarrow{G}_m^{nem} (r_i,r_j,\omega)+\overleftrightarrow{G}_m^{nem,\intercal} (r_j,r_i,\omega)]_{zz}/2 \\ \approx \frac{1}{16\pi \sqrt{4z^2+D^2}}   \mathrm{Im} (\varepsilon -1).
    \end{aligned}
\end{align}
\end{subequations}
Meanwhile, we can find the low-frequency Green’s function in free space at $D \ll 1/k_0$ as
\begin{equation}
    \mathrm{Im} [\overleftrightarrow{G}_m^{vac} (r_i,r_j,\omega)] \approx \frac{\omega}{6 \pi c},         \qquad  \frac{\omega D}{c} \ll 1.
\end{equation}
Since $J_c (\omega) \propto \mathrm{Im} [\overleftrightarrow{G}_m (r_i,r_j,\omega)+\overleftrightarrow{G}_m^\intercal (r_j,r_i,\omega)]$, we have
\begin{equation}
    \frac{J_c^{nem} (\omega)}{J_c^{vac} (\omega)} \approx \begin{cases}
      \frac{3}{8} \frac{\mathrm{Im} (\varepsilon -1 )}{\sqrt{4z^2+D^2}} \frac{1}{k_0}, \\
      \frac{3}{2}  \frac{8z^2-D^2 }{[(4z)^2+D^2 ]^{5/2}}   \mathrm{Im}  \frac{\mu -1}{\mu +1} \frac{1}{k_0^3}. \end{cases}
\end{equation}
We can now verify the order of $J_c^{nem} (\omega)/J_c^{vac} (\omega)$ in the nano-EM environments with simple calculations. For simplicity, we consider $D=z<100 nm$. At $\omega <10^6 \, \mathrm{s^{-1}}$, we have $k_0=\omega/c<0.003\,\mathrm{m}^{-1}$ and $k_0 D=k_0 z<3 \times 10^{-10}$. 

Near magnetic materials, we find $J_c^{nem} (\omega) / J_c^{vac} (\omega) > 6 \times 10^{27} \times \mathrm{Im}  [(\mu -1)/(\mu +1)]$. Therefore, even for magnetic materials with small magnetic damping at microwave frequencies, e.g., $\mu=2+4 \times 10^{-7} i$ (much smaller compared to common ferrites), we still have $J_c^{nem} (\omega)/J_c^{vac} (\omega)>3 \times 10^{20}$ at $\omega < 10^6 \, \mathrm{s^{-1}}$.

Near conductors, we find $J_c^{nem} (\omega) / J_c^{vac} (\omega) > 6.3 \times 10^{13} \times \sigma$, where $\sigma/\omega\varepsilon_0 = \mathrm{Im} (\varepsilon-1)$. Therefore, even for conductors with low conductivity, e.g., titanium (one of the least conductive metals) with $\sigma=2.3 \times 10^6 \, \mathrm{S/m}$, we still have $J_c^{nem} (\omega)/J_c^{vac} (\omega)>1.4 \times 10^{20}$ at $\omega <10^6  \, \mathrm{s^{-1}}$.

The above calculations demonstrate that the giant enhancement of low-frequency EM fluctuation correlations (i.e., $J_c^{nem} (\omega)/J_c^{vac} (\omega)>10^{20}$) is a general effect near common magnetic materials and conductors. 

\section{Material Models and Parameters}\label{appendix_material}
In this appendix, we provide the material models and experimentally accessible parameters for the example realistic materials considered in the main text.
\subsubsection{yttrium iron garnet (gyromagnet)}\label{YIGpara}
We consider the Landau–Lifshitz–Gilbert formula for the magnetic permeability tensor $\overleftrightarrow{\mu}_{\mathrm{YIG}}$ of the non-reciprocal gyromagnetic yttrium iron garnet (YIG). For a biasing magnetic field $H_0$ in the $\hat{\mathbf{z}}$ direction, we have~\cite{pozar2021microwave},
\begin{equation}
    \overleftrightarrow{\mu}_{\mathrm{YIG}}(\omega) = \mu_0 \begin{bmatrix}
        1+\chi_{xx} & \chi_{xy} & 0 \\
        \chi_{yx} & 1+\chi_{yy} & 0 \\
        0 & 0 & 1             
    \end{bmatrix}, 
\end{equation}
where $\chi_{xx}= \chi_{yy}, \, \chi_{xy}=-\chi_{yx}$ are the components of the susceptibility tensor given by,
\begin{align}
\begin{aligned}
    \chi_{xx} = \frac{-\omega_m (\omega_0-i\alpha\omega)}{\omega^2-(\omega_0-i\alpha\omega)^2}, \
    \chi_{xy} = \frac{i \omega_m \omega}{\omega^2-(\omega_0-i\alpha\omega)^2},
\end{aligned}
\end{align}
where the Larmor frequency $\omega_0 = \gamma_{\mathrm{YIG}} \mu_0 H_0$ depends on the gyromagnetic ratio $\gamma_{\mathrm{YIG}}$ and the bias field $H_0$. $\omega_m = \gamma_{\mathrm{YIG}} \mu_0 M_s$ depends on the saturation magnetization $M_s$. The Gilbert damping factor $\alpha_{\mathrm{YIG}}$ determines the magnetic loss. In the main text, we consider $H_0=300 \, \mathrm{G}$, $ \alpha_{\mathrm{YIG}} = 2.1\times 10^{-3}$, $ M_s=1618\,\mathrm{G}$ for the YIG slab at $T=300\,\mathrm{K}$~\cite{haidar2015thickness}. 

\subsubsection{silver (conductor)}
We consider the Drude model for the low-frequency dielectric response $\overleftrightarrow{\varepsilon}_{\mathrm{Ag}} =\varepsilon_{\mathrm{Ag}}(\omega) \overleftrightarrow{I}$ of polycrystalline silver:
\begin{equation}
    \varepsilon_{\mathrm{Ag}}(\omega)=1-\frac{\omega_p^2\tau_{\mathrm{Ag}}^2}{\omega^2\tau_{\mathrm{Ag}}^2+1} + i \frac{\sigma_{\mathrm{Ag}}}{\varepsilon_0\omega(1+\omega^2\tau_{\mathrm{Ag}}^2)},
\end{equation}
where $\omega_p$ is the plasma frequency, $1/\tau_{\mathrm{Ag}}$ is the electron mean collision rate, $\sigma_{\mathrm{Ag}}$ is the direct current (DC) conductivity. In the main text, we employ $\omega_p=2\pi \times 2.15 \times 10^{15} \, \mathrm{Hz}, \tau_{\mathrm{Ag}}=3.22 \times 10^{-14} \, \mathrm{s}^{-1}, \sigma_0=5.2 \times 10^{7} \, \mathrm{S/m}$ for silver slabs at $T=300\,\mathrm{K}$~\cite{de1988temperature}. 

\subsubsection{niobium (superconductor)}
We consider superconducting niobium with local EM response~\cite{klein1994conductivity} in Fig.~\ref{fig:fig1}(b,~c) and Figs.~\ref{fig:figures2},~\ref{fig:figs2}. In this regime, the niobium complex conductivity $\sigma_{\mathrm{Nb}}=\sigma_1+i\sigma_2$ at temperature $T$ can be described by the Mattis-Bardeen theory~\cite{klein1994conductivity}:
\begin{multline}\label{Nb_Tdep}
    \frac{\sigma_{\mathrm{Nb}} (\omega , T)}{\sigma_n}=\frac{\sigma_1 (\omega , T) + i\sigma_2 (\omega , T)}{\sigma_n}=\int_{\Delta(T)-\hbar\omega}^{\infty} \frac{dx}{\hbar\omega} \\ \tanh(\frac{x+\hbar\omega}{2k_BT})g(x) - \int_{\Delta(T)}^{\infty} \frac{dx}{\hbar\omega} \tanh(\frac{x}{2k_BT})g(x),
\end{multline}
where $\sigma_n$ is the niobium conductivity in the normal phase, $\Delta(T)$ is the temperature-dependent BCS gap. $g(x)$ is given by:
\begin{align}
    g(x)&=\frac{x^2+\Delta^2(T)+\hbar \omega x}{u_1 u_2}, \\
    u_1&=\begin{cases}
    \sqrt{x^2-\Delta^2(T)}, & \text{if } x > \Delta(T) \\
    -i\sqrt{\Delta^2(T)-x^2},               & \text{if } x < \Delta(T)
\end{cases}\\
    u_2&=\sqrt{(x+\hbar \omega)^2-\Delta^2(T)}.
\end{align}
As shown in Ref.~\cite{klein1994conductivity}, the complex conductivity of niobium predicted by the Mattis-Bardeen theory matches well with experimental measurements. At $T<T_c$ ($T_c=9.2\,\mathrm{K}$ is the critical temperature of niobium), the permittivity $\overleftrightarrow{\varepsilon}=\varepsilon_{\mathrm{Nb}}(\omega,T) \overleftrightarrow{I}$ can be written as:
\begin{equation}
        \varepsilon_{\mathrm{Nb}}(\omega,T) = 1+ \frac{i}{\varepsilon_0 \omega} \sigma_{\mathrm{Nb}}(\omega,T).
\end{equation}
In Fig.~\ref{fig:fig1}(b,~c) in the main text and Figs.~\ref{fig:figures2},~\ref{fig:figs2} in the appendices, we consider $T=7\,\mathrm{K}$, $\Delta(T=7\,\mathrm{K})\approx 1\,\mathrm{meV}$, and $\sigma_n=8.5\times 10^7 \, \mathrm{S/m}$~\cite{klein1994conductivity} for niobium slabs.

\section{Universal Power-Law Dependence of Nano-EM Collective Dephasing}\label{appendix_range}
In this appendix, we provide supplemental details of the power-law dependence of nano-EM collective dephasing $\Phi_c\sim D^{-\beta}$. In the main text, we focus on TLSs configurations with quantization axes perpendicular to the material slabs and discuss three key classes of bianisotropic media (isotropic, hyperbolic, and gyrotropic). Here, we extend our results to more general TLSs configurations and more general bianisotropic media. In addition, we provide the details of the magnetic dyadic Green's functions $\overleftrightarrow{G}_m$ calculations, reflection coefficient calculations, and additional realistic material examples to demonstrate the power-law dependence $\Phi_c \sim D^{-\beta}$ in Table.~\ref{tab:table1}. We find our analytical analysis in the main text Table.~\ref{tab:table1} matches well with exact numerical calculations based on realistic material models.

\subsection{Magnetic dyadic Green's function calculations}
In the nano-EM environment, magnetic dyadic Green's functions are dominated by contributions from the reflected component, i.e., $\overleftrightarrow{G}_m (\mathbf{r_{i},r_{j}},\omega) \approx \overleftrightarrow{G}^r_m(\mathbf{r_i},\mathbf{r_j},\omega)$. In this case, $\overleftrightarrow{G}^r_m (\mathbf{r_{i},r_{j}},\omega)$ near general material slabs can be calculated from the reflection coefficients $r_{ss}, r_{sp}, r_{pp}, r_{ps}$. Assuming the material slab is perpendicular to the $\hat{\mathbf{z}}$ direction, we have~\cite{novotny2012principles}:
\begin{widetext}
\begin{align}
\begin{aligned}\label{anacpd}
    \overleftrightarrow{G}^r_m (\mathbf{r}_i,\mathbf{r}_j,\omega)=&\frac{i}{8 \pi^2}\int \frac{d \mathbf{q}}{k_z} e^{i \mathbf{q}(\mathbf{r}_i-\mathbf{r}_j)} e^{i k_z(z_i+z_j)} \biggl( \frac{r_{pp}}{q^2}\begin{bmatrix} q_y^2&-q_x q_y&0\\-q_x q_y&q_x^2&0\\0&0&0 \end{bmatrix}+ \frac{r_{ss}}{k_0^2q^2}\begin{bmatrix} -q_x^2 k_z^2 & -q_x q_y k_z^2 & -q_x k_z q^2\\-q_x q_y k_z^2 & -q_y^2 k_z^2 & -q_y k_z q^2\\q^2 q_x k_z & q^2 q_y k_z & q^4 \end{bmatrix} \\
    & + \frac{r_{ps}}{k_0 q^2}\begin{bmatrix} q_x q_y k_z & q_y^2 k_z & q_y q^2\\-q_x^2 k_z & -q_y q_x k_z & -q_x q^2\\0 & 0 & 0 \end{bmatrix} + \frac{r_{sp}}{k_0 q^2}\begin{bmatrix} -q_x q_y k_z & q_x^2 k_z & 0\\-q_y^2 k_z & q_y q_x k_z & 0\\ q_y q^2 & - q_x q^2 & 0 \end{bmatrix} \biggr),
\end{aligned}
\end{align}
\end{widetext}
where $k_0=\omega/c$, $\mathbf{q}=q_x\,\mathbf{\hat{x}} + q_y\,\mathbf{\hat{y}}$ is the in-plane momentum, $q^2=|\mathbf{q}|^2$, $k_z=\sqrt{k_0^2-q^2}$ is the z component of momentum, and $z_i=\mathbf{r}_i \cdot \hat{\mathbf{z}}, z_j=\mathbf{r}_j \cdot \hat{\mathbf{z}}$ are the distance between TLSs and material slabs.

In the main text, we focus on $[\overleftrightarrow{G}^r_m]_{zz}$ and TLSs configurations with quantization axes perpendicular to the material slabs. From Eq.~(\ref{anacpd}), we find (Eq.~(\ref{Jctheta}) in the main text),
\begin{multline}\label{appendixD_Jc}
    J_c(\omega) \propto \mathrm{Im} [ \overleftrightarrow{G}_m(\mathbf{r_i},\mathbf{r_j}) + \overleftrightarrow{G}_m^{\intercal}(\mathbf{r_j},\mathbf{r_i}) ]_{zz} \\ \propto \mathrm{Im} \int_0^{2\pi} \Big[ \frac{4f_0(\theta)}{k_0^3D^3} \frac{a^3-3a \cos^2(\theta-\varphi)}{[a^2+\cos^2(\theta-\varphi)]^3} \\ +\frac{2f_2(\theta)}{k_0D} \frac{a}{a^2+\cos^2(\theta-\varphi)} \Big] d\theta.
\end{multline}
We notice that the integrand is proportional to $a$ except at $\cos{(\theta-\varphi)} \approx 0$. Therefore, at large $D$ (small $a<1$), the integrand is dominated by contributions from $\theta_1=\varphi+\pi/2$ and $\theta_2=\varphi+3\pi/2$. We further notice that,
\begin{subequations}\label{theta_integral}
\begin{align}
& \ \begin{aligned}
   \int_0^\pi \frac{2a^3-6a \cos^2(\theta-\varphi)}{[a^2+\cos^2(\theta-\varphi)]^3} d\theta=-\pi,
\end{aligned}\\
& \qquad \begin{aligned}
   \int_0^\pi \frac{a}{a^2+\cos^2(\theta-\varphi)} d\theta=\pi.
\end{aligned}
\end{align}
\end{subequations}
Substitute Eq.~(\ref{theta_integral}) into Eq.~(\ref{appendixD_Jc}), we obtain Eq.~(\ref{powerlaw}) in the main text.

In the following, we briefly discuss more general TLSs configurations, where the nano-EM collective dephasing is determined by other components of $\overleftrightarrow{G}^r_m$. We find that nano-EM collective dephasing generally exhibits power-law dependence on interatom distance $D$. 

As discussed in the main text, in the near-field of material slabs, low-frequency $\overleftrightarrow{G}^r_m$ is dominated by contributions from evanescent interface waves with momentum $q/k_0 \gg 1$. On the right-hand side (RHS) of Eq.~(\ref{anacpd}), we notice that the second term scales as $\sim r_{ss}(q/k_0)^2$, while other terms scale slower as $(q/k_0)^1$ or $(q/k_0)^0$. Therefore, we expect the second term on the RHS of Eq.~(\ref{anacpd}) to dominate the $q$-integral for low-frequency $\overleftrightarrow{G}^r_m$. Furthermore, at $q/k_0 \gg 1$, we have $q_x=q\cos{\theta}$, $q_y=q\sin{\theta}$, and $k_z \approx iq$. Substituting $q_x$, $q_y$, and $k_z$ into Eq.~(\ref{anacpd}), assuming that Eq.~(\ref{anacpd}) is dominated by contributions from the $r_{ss}$ term, we have
\begin{multline}\label{anacpdapprox}
    \overleftrightarrow{G}^r_m (\mathbf{r}_i,\mathbf{r}_j,\omega) \approx \frac{1}{8 \pi^2 }\int  d q d\theta e^{i \mathbf{q}(\mathbf{r}_i-\mathbf{r}_j)} e^{-q(z_i+z_j)} \frac{r_{ss}}{k_0^2} \\
    \begin{bmatrix} q^2 \cos^2{\theta} & q^2 \cos{\theta}\sin{\theta} & -iq^2\cos{\theta}\\q^2 \cos{\theta}\sin{\theta} & q^2 \sin^2{\theta} & -iq^2\sin{\theta}\\iq^2\cos{\theta} & iq^2\sin{\theta} & q^2 \end{bmatrix}.
\end{multline}
From Eq.~(\ref{anacpdapprox}), we note that the integrands of other components of the $3\times 3$ matrix $\overleftrightarrow{G}^r_m$ differ from the integrand of $[\overleftrightarrow{G}^r_m]_{zz}$ by a function of $\theta$. In Eq.~(\ref{Jctheta}) in the main text, this additional function of $\theta$ can be absorbed into $f_0(\theta)$ and $f_2(\theta)$, which are the expansion of $r_{ss}$ at $q/k_0\gg1$ (see Eq.~(\ref{rss}) in the main text). Therefore, through Eqs.~(\ref{rss}-\ref{powerlaw}) in the main text, we can similarly prove that nano-EM collective dephasing will generally exhibit power-law dependence $\Phi_c\sim D^{-\beta}$ on interatom distance $D$ when the quantization axes of TLSs are not perpendicular to the material slabs. We note that, under one specific configuration where $\mathbf{m_i}$, $\mathbf{m_j}$, and $\mathbf{r_i}-\mathbf{r_j}$ are rigorously perpendicular to each other, we have $\mathbf{m}_{i} \cdot  \mathrm{Im}[\overleftrightarrow{G}_m(\mathbf{r_i},\mathbf{r_j},\omega)+\overleftrightarrow{G}^\intercal_m(\mathbf{r_j},\mathbf{r_i},\omega)] \cdot \mathbf{m}_{j} = 0$ deviating from our discussion above. Meanwhile, the nano-EM collective dephasing behavior will recover the behaviors discussed above if there is a small deviation from this specific configuration.

\subsection{Reflection coefficient calculations}~\label{rflec}
For isotropic media, the above reflection coefficient $r_{ss}(q)$ can be obtained from the Fresnel formula. For hyperbolic and gyromagnetic media, the closed-form expressions of $r_{ss}(q,\theta)$ at all $q, \theta$ are complicated, but simple asymptotic expressions at $q\gg k_0$ can be obtained. In this paper, we find the results in Table.~\ref{tab:table1} for hyperbolic and gyrotropic media by taking $q/k_0\gg 1$ in the reflection coefficient $r_{ss}(q,\theta)$ expressions for uniaxial media (from Ref.~\cite{sosnowski1972polarization}) and gyromagnetic media (from Ref.~\cite{khandekar2019thermal}). Alternatively, we can also find $r_{ss}(q/k_0 \to \infty)$ by using the general methods described below~\cite{khandekar2019thermal}.

We now briefly describe the general approaches to calculate $r_{ss}(q,\theta)$ for general bianisotropic material slabs with local EM response, following the discussions in Ref.~\cite{khandekar2019thermal}. In general, $r_{ss}(q,\theta)$ can be solved from the Maxwell equations and EM boundary conditions. We consider a material slab perpendicular to the $z$ direction. We consider a plane wave with frequency $\omega$ and in-plane momentum $q$ incident on the slab. The x, y, and z components of momentum of the incident plane wave are $q_x=q\cos{\theta}$, $q_y=q\sin{\theta}$, and $k_z=\sqrt{k_0^2-q^2}$. For the transmitted EM fields $\mathbf{E}^t$ and $\mathbf{H}^t$ inside the material slab, we write the Maxwell equation as~\cite{khandekar2019thermal}:
\begin{align}\label{Max_material}
    & \left( \begin{bmatrix} \overleftrightarrow{\varepsilon} & 0 \\ 0 & \overleftrightarrow{\mu}  \end{bmatrix} + \begin{bmatrix} 0  & \overleftrightarrow{\kappa}/k_0 \\ -\overleftrightarrow{\kappa}/k_0 & 0 \end{bmatrix} \right) \begin{bmatrix} \mathbf{E}^t  \\ \eta_0 \mathbf{H}^t  \end{bmatrix} = 0, \\
    & \qquad \qquad \quad \overleftrightarrow{\kappa} = \begin{bmatrix} 0 & -k'_z & q_y \\ k'_z & 0 & -q_x \\ -q_y & q_x & 0 \end{bmatrix},
\end{align}
where $\overleftrightarrow{\varepsilon}$ is the permittivity tensor and $\overleftrightarrow{\mu}$ is the permeability tensor. $\eta_0=\sqrt{\mu_0/\varepsilon_0}$, $q_x$, $q_y$ are the x and y components of the transmitted fields, which are invariant at the boundary due to EM boundary conditions (in-plane momentum is continuous at the two sides of the material interface). $k_z'$ is the z component of the momentum of transmitted fields. For Eq.~(\ref{Max_material}) to have nontrivial solutions, we require the determinant:
\begin{equation}\label{kz_material}
    \mathrm{det} \left( \begin{bmatrix} \overleftrightarrow{\varepsilon} & 0 \\ 0 & \overleftrightarrow{\mu}  \end{bmatrix} + \begin{bmatrix} 0  & \overleftrightarrow{\kappa}/k_0 \\ -\overleftrightarrow{\kappa}/k_0 & 0 \end{bmatrix} \right) = \mathrm{det} \overleftrightarrow{M} = 0. 
\end{equation}
From Eq.~(\ref{kz_material}), we can solve $k'_z$. For a given pair of $(q,\theta)$ ($q_x=q\cos{\theta}, q_y=q\sin{\theta}$), there exist four $k'_z$ solutions of Eq.~(\ref{kz_material}) for anisotropic media, and the nullspace of $\overleftrightarrow{M}$ is spanned by four eigensolutions. Two of the four eigensolutions correspond to the transmitted waves $\mathbf{E}^t$ and $\mathbf{H}^t$ propagating away from the material boundary. In the following, we denote them as $(\mathbf{E}^t_1, \mathbf{H}^t_1)$ and $(\mathbf{E}^t_2, \mathbf{H}^t_2)$. The reflection coefficient $r_{ss}$ can then be solved by matching the EM boundary conditions at the interface of the material slab (e.g., see~\cite{khandekar2019thermal})
\begin{widetext}
\begin{equation}\label{EMbcs}
    \begin{bmatrix} \sin{\theta} & -\sqrt{1-(q/k_0)^2}\cos{\theta} & [\mathbf{E}^t_1]_x & [\mathbf{E}^t_2]_x \\ -\cos{\theta} & -\sqrt{1-(q/k_0)^2}\sin{\theta} & [\mathbf{E}^t_1]_y & [\mathbf{E}^t_2]_y \\ \sqrt{1-(q/k_0)^2}\cos{\theta} & \sin{\theta} & \eta_0 [\mathbf{H}^t_1]_x & \eta_0[\mathbf{H}^t_2]_x \\ \sqrt{1-(q/k_0)^2}\sin{\theta} & -\cos{\theta} & \eta_0[\mathbf{H}^t_1]_y & \eta_0[\mathbf{H}^t_2]_y \end{bmatrix} \ \begin{bmatrix} r_{ss}(q,\theta) \\ r_{ps}(q,\theta) \\ t_1 \\ t_2 \end{bmatrix}  = \begin{bmatrix} -\sin{\theta} \\ \cos{\theta} \\ \sqrt{1-(q/k_0)^2}\cos{\theta} \\ \sqrt{1-(q/k_0)^2}\sin{\theta} \end{bmatrix},
\end{equation}
\end{widetext}
where $r_{ss}(q,\theta)$, $r_{ps}(q,\theta)$ are reflection coefficients and $t_1$, $t_2$ are tranmission coefficients. In this matrix equation, the first and second rows correspond to the continuity of $[\mathbf{E}]_x$ and $[\mathbf{E}]_y$ at the material slab interface. The third and fourth lines correspond to the continuity of $[\mathbf{H}]_x$ and $[\mathbf{H}]_y$ at the material slab interface. With $(\mathbf{E}^t_1, \mathbf{H}^t_1)$ and $(\mathbf{E}^t_2, \mathbf{H}^t_2)$ solved from Eq.~(\ref{kz_material}), we can find the reflection coefficient $r_{ss}(q,\theta)$.

In the main text, we provide the expressions of $ r_{ss}(q,\theta)$ for isotropic, hyperbolic, and gyrotropic media with local EM response in Table.~\ref{tab:table1}. In the following, we extend our discussions to more general bianisotropic media with less symmetric response functions $\overleftrightarrow{\varepsilon}$ and $\overleftrightarrow{\mu}$. 

\subsubsection{more general photonic media with local EM response}
For more general bianisotropic media, it could be difficult to obtain a closed-form expression of $r_{ss}(q,\theta)$ at $q\gg k_0$. We now analyze Eqs.~(\ref{Max_material}-\ref{EMbcs}) and show that $r_{ss}$ can still be expanded as a power series of $k_0/q$ at $q/k_0\gg1$. We first note that Eq.~(\ref{kz_material}) is a quartic equation with respect to $k_z'$. For photonic media with local EM response, $\overleftrightarrow{\varepsilon}, \overleftrightarrow{\mu}$ are independent of $q$. Therefore, the coefficients of the quartic equation (\ref{kz_material}) are polynomial functions of $q/k_0$. From the general solutions of the quartic equation, we can find that the solutions $k_z'$ of Eq.~(\ref{kz_material}) are algebraic functions of $q/k_0$. Next, we consider the asymptotic behavior of $k_z'$ at large $q/k_0$. By expanding the determinant $\mathrm{det} \overleftrightarrow{M}$, we can find that $k_z'$ follows the series expansion $k_z' \sim C q + C' + C''/q \ldots$ at $q/k_0 \gg 1$, where $C, C', C''$ are constants determined by material response and $\theta$. Here, the leading-order term is $Cq$ because the left hand side of Eq.~(\ref{kz_material}) will become $k_z'^4$ or $q^4$ at large $q$ otherwise (except for very specific response functions). Substituting the solutions $k_z'$ into Eq.~(\ref{Max_material}), the nullspace of $\overleftrightarrow{M}$ can be obtained by Gaussian elimination. Since only linear operations are used in Gaussian elimination, each component of the eigensolutions $\begin{bmatrix} \mathbf{E}^t  \\ \eta_0 \mathbf{H}^t \end{bmatrix}$ is an algebraic function of $q/k_0$ and $k_z$, and can be expanded in terms of $q/k_0$ at large $q$. Therefore, we find that all the parameters in Eq.(\ref{EMbcs}) are algebraic functions of $q/k_0$. Since Eq.(\ref{EMbcs}) is a linear function, the solution $r_{ss}(q,\theta)$ is also an algebraic function of $q/k_0$. As a result, at $q/k_0 \gg 1$, we can expand $r_{ss}(q,\theta)\approx f_0(\theta) + f_1(\theta) \frac{k_0}{q} + f_2(\theta) \frac{k_0^2}{q^2}$ in terms of $k_0/q$ at large $q$.

In the main text, we demonstrate that the $f_0(\theta)$ and $f_2(\theta) k_0^2/q^2$ terms lead to the power-law dependence $\Phi_c\sim D^{-\beta}$ of nano-EM collective dephasing. Here, we discuss the contributions of the $f_1(\theta) k_0/q$ term. Substituting into Eq.~(\ref{anacpd}), we find $J_c(\omega) \propto \mathrm{Im} \int_0^{2\pi} \Big[ \frac{f_1(\theta)}{k_0^2D^2} \frac{2a^2-2\cos^2(\theta-\varphi)}{[a^2+\cos^2(\theta-\varphi)]^2} \Big] d\theta$. For $f_1(\theta)=constant$, we find $J_c(\omega) \propto D^{-3}$ at large $D$ (small $a$). Meanwhile, for general $f_1(\theta)$, we find $J_c(\omega) \propto D^{-2}$ at large $D$. This indicates that, near more general photonic media with local EM response, collective dephasing can still follow the power-law dependence $\Phi_c \sim D^{-\beta}$ on interatom distance $D$.

\subsection{Additional material examples}
\begin{figure}[h!]
    \centering
    \includegraphics[width=3 in]{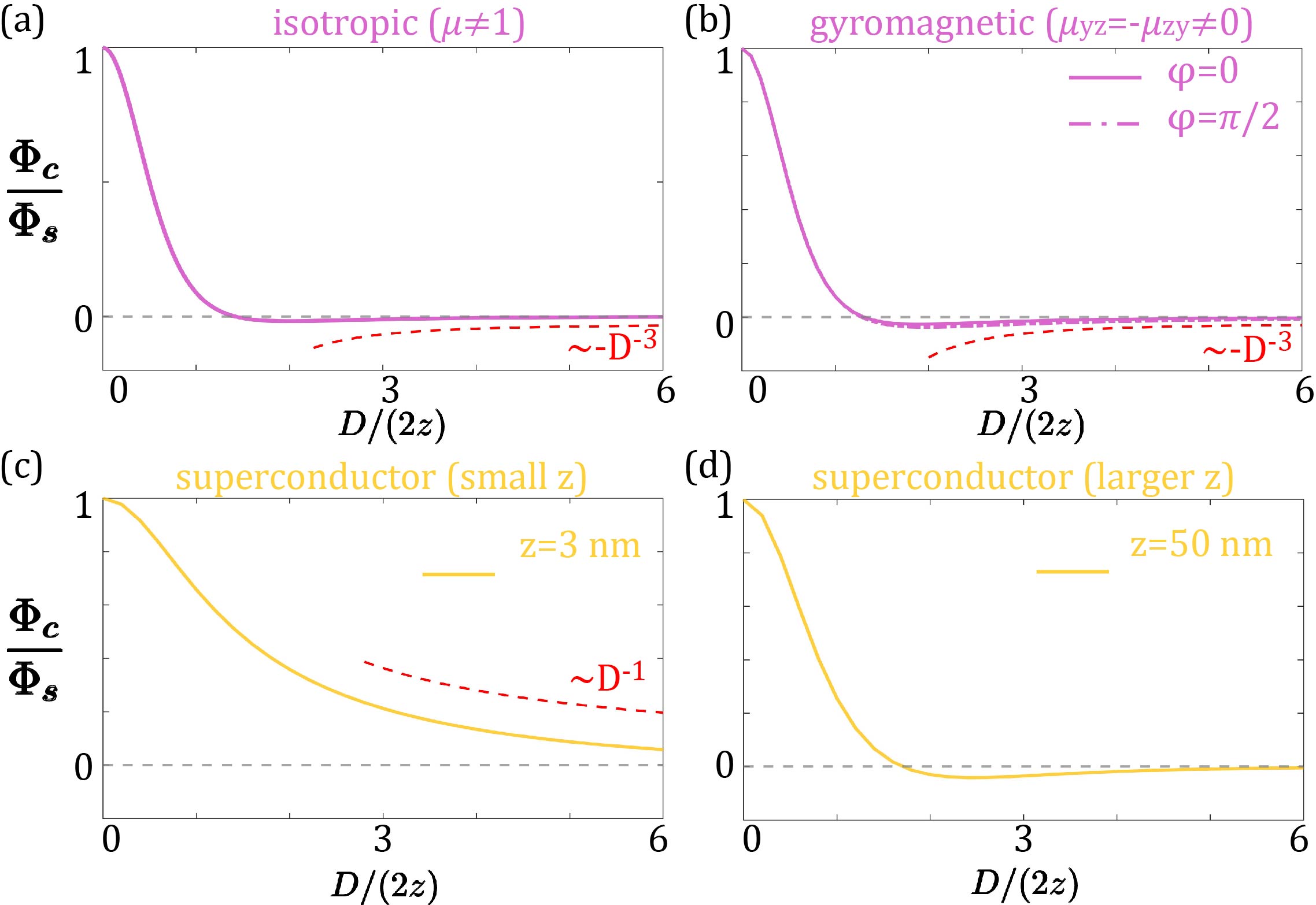}
    \caption{Additional material examples for the universal nano-EM collective dephasing range derived in the main text. Our numerical calculations match well with our analytic analysis of the nano-EM collective dephasing range in Table.~\ref{tab:table1} in the main text.}
    \label{fig:figs2}
\end{figure}

In the main text, we demonstrate the above power-law dependence of nano-EM collective dephasing $\Phi_c \sim D^{-\beta}$ using realistic material examples (gyromagnetic YIG, non-magnetic conductor silver, and hyperbolic metamaterials) in Fig.~\ref{fig:fig2}(b) and Fig.~\ref{fig:fig4}. Here, we provide additional material examples (Fig.~\ref{fig:figs2}(a,b)) that are representative of other classes of materials in Table.~\ref{tab:table1} in the main text, including isotropic magnetic media and gyromagnetic media with $\mu_{yz}= -\mu_{zy}\neq 0$.

In Fig.~\ref{fig:figs2}(a,b), we present nano-EM collective dephasing range $\Phi_c/\Phi_s$ near isotropic magnetic media and gyromagnetic media with $\mu_{yz}= -\mu_{zy}\neq 0$. We find our numerical calculations match well with our analytic analysis of the nano-EM collective dephasing range in Table.~\ref{tab:table1}.

We also briefly discuss the nano-EM collective dephasing range near BCS superconductors. Here, due to the giant screening effects from superconducting currents ((large $|\mathrm{Re} \, \varepsilon|$)), we find the approximation for Eq.~(\ref{rss}) in the main text (i.e., $r_{ss}\approx r_{ss}(q \to \infty)$) is only good when the TLSs are very close to the superconductor slabs. As shown in Fig.~\ref{fig:figs2}(c), at $z=3\,\mathrm{nm}, T=7K$, we find our numerical calculations of $\Phi_c/\Phi_s \sim D^{-1}$ match well with our analytic analysis for isotropic non-magnetic conductors. Meanwhile, at larger $z=50\,\mathrm{nm}$, $r_{ss}$ deviates from the approximation $r_{ss}(q \to \infty)$ due to screening effects (large $|\mathrm{Re} \, \varepsilon|$), the results deviate from $\Phi_c/\Phi_s \sim D^{-1}$ at large $D$ (Fig.~\ref{fig:figs2}(d)) and look similar to collective dephasing behaviors near magnetic materials. In other words, superconductor screening effects reduce the range of nano-EM collective dephasing near superconductors. These results can be helpful for controlling collective dephasing effects in hybrid quantum systems~\cite{xiang2013hybrid}.

\section{Nano-EM Super-dephasing Scaling Laws}\label{appendix_scalinglaws}
In this appendix, we provide derivations of the nano-EM super-dephasing scaling laws for GHZ states demonstrated in Fig.~(\ref{fig:fig3}) in the main text. In addition, we provide further discussions about super- and sub-dephasing in other entangled states, e.g., the decoherence-free subspace states (DFS) $|\psi\rangle_{\mathrm{DFS}}$.

As shown in Fig.~\ref{fig:fig3}(a) in the main text, we consider $N$ TLSs arranged in one-dimensional (1D) $N=1\times n$ and two-dimensional (2D) $N=n \times n$ arrays with lattice constant $b$ at distance $z$ from material slabs. We focus on $\mathbf{m}$ of all TLSs perpendicular to the material slab. We consider the dephasing function for N-qubit entangled states, e.g., $\Phi_\mathrm{GHZ}(t)$ for GHZ states. Here, the dephasing functions characterize the decay of multiqubit coherence defined through the $l_1$ norm measure $C_{l_1}(\rho)=\sum_{\substack{n \neq k}}|\rho_{nk}|$~\cite{baumgratz2014quantifying}, where $\rho$ is the density matrix of the magnetic TLSs ensemble and $\rho_{nk}$ are off-diagonal elements of $\rho$. For example, the decay of $C_{l_1}(\rho)$ induced by nano-EM dephasing for GHZ states follows $C_{l_1}(\rho(t=\tau))= e^{-\Phi_\mathrm{GHZ}(\tau)} C_{l_1}(\rho(t=0))$. In the following, we focus on the scaling of $\Phi_\mathrm{GHZ} \sim N^\alpha \Phi_s$ with $N$. We characterize the super- or sub-dephasing scaling laws by $\alpha=\partial \ln{\Phi_\mathrm{GHZ}} / \partial \ln{N}$.

\subsection{Greenberger–Horne–Zeilinger (GHZ) state}
The $N$-qubit GHZ state is $|\psi\rangle_\mathrm{GHZ}=\frac{1}{\sqrt{2}}(|0\rangle^{\otimes N} + |1\rangle^{\otimes N})$. From non-unitary processes in Eq.~(\ref{me}) (Eq.~(\ref{ME}) in the main text), we find the dephasing function $\Phi^{\mathrm{2D}}_{\mathrm{GHZ}}(n,t)$ for $|\psi\rangle_{\mathrm{GHZ}}$ in 2D $n\times n$ TLSs arrays:
\begin{multline}\label{DecGHZ2D}
    \Phi^{\mathrm{2D}}_{\mathrm{GHZ}}(n,t) =  2 \bigintssss_0^t \Big[ \sum_{x_i, \, y_i=1}^n \gamma_\phi^s (\mathbf{r_i},t') \\ + \sum_{\substack{x_i,\, x_j,\, y_i,\, y_j=1\\ (x_i, \, y_i)\neq (x_j, \, y_j)}}^n \gamma_\phi^c (\mathbf{r_i},\mathbf{r_j},t') \Big] dt',
\end{multline}
where $\mathbf{r_i}=b x_i \, \hat{\mathbf{x}} + b y_i \, \hat{\mathbf{y}}$, $\mathbf{r_j}=b x_j \, \hat{\mathbf{x}} + b y_j \, \hat{\mathbf{y}}$. For 1D $1\times n$ TLSs arrays, the dephasing function $\Phi^{\mathrm{1D}}_{\mathrm{GHZ}}(n,t)$ is:
\begin{equation}\label{DecGHZ1D}
    \Phi^{\mathrm{1D}}_{\mathrm{GHZ}}(n,t) =  2 \bigintssss_0^t \Big[ \sum_{x_i=1}^n \gamma_\phi^s (\mathbf{r_i},t') + \sum_{\substack{x_i, x_j=1\\ x_i \neq x_j }}^n \gamma_\phi^c (\mathbf{r_i},\mathbf{r_j},t') \Big] dt',
\end{equation}
where $\mathbf{r_i}=bx_i \, \hat{\mathbf{x}}$, $\mathbf{r_j}=bx_j \, \hat{\mathbf{x}}$. 

\subsubsection{Collective dephasing with $\Phi_c/\Phi_s \sim D^{-1}$}
We first derive the scaling of super-dephasing when the underlying collective dephasing range follows $\Phi_c/\Phi_s \sim D^{-1}$ at large interatom distance $D$. Here, we consider super-dephasing near non-magnetic conductors. From Table.~\ref{tab:table1} in the main text and Eqs.~(\ref{SI_Gzz},~\ref{app_td_cdf},~\ref{td_sdf}), we find
\begin{equation}\label{Agratio}
    \frac{\Phi_c(\mathbf{r_i}, \, \mathbf{r_j})}{\Phi_s(\mathbf{r_i})} \approx \frac{2z}{\sqrt{(2z)^2 + (\mathbf{r_i} - \mathbf{r_j})^2}}.
\end{equation}
Substituting Eq.~(\ref{Agratio}) into Eqs.~(\ref{DecGHZ2D},~\ref{DecGHZ1D}), we obtain:
\begin{align}\label{DecGHZAg}
\begin{aligned}
    \frac{\Phi^{\mathrm{2D}}_{\mathrm{GHZ}}(n,t)}{\Phi_{s}(t)}& \approx  \sum_{\substack{x_i, x_j, \\ y_i, y_j =1}}^n \frac{r}{\sqrt{r^2+(x_i-x_j)^2+(y_i-y_j)^2}}, \\
    \frac{\Phi^{\mathrm{1D}}_{\mathrm{GHZ}}(n,t)}{\Phi_{s}(t)}& \approx  \sum_{x_i, x_j =1}^n \frac{r}{\sqrt{r^2+(x_i-x_j)^2}},
\end{aligned}
\end{align}
where $r=2z/b$. For 2D arrays, we have --

\textit{In the regime $r \ll 1$}, the summation terms corresponding to $(x_i,y_i)=(x_j,y_j)$ are much larger than other terms in Eq.~(\ref{DecGHZAg}). Therefore, we have:
\begin{equation}
    \Phi^{\mathrm{2D}}_{\mathrm{GHZ}} / \Phi_{s} \sim n^2.
\end{equation}

\textit{In the regime $r \gtrsim n$}, we can convert the summation in Eq.~(\ref{DecGHZAg}) into an integral through the Euler–Maclaurin formula:
\begin{align}
\begin{aligned}
    &\Phi^{\mathrm{2D}}_{\mathrm{GHZ}}/ \Phi_{s} \\ &\sim n^3 \int_0^1 \int_0^1 \int_0^1 \int_0^1 \frac{dx_1 dx_2 dy_1 dy_2}{\sqrt{(\frac{r}{n})^2+(x_1-x_2)^2+(y_1-y_2)^2}}, \\
    &=4 n^3 \int_0^1 \int_0^1 \frac{(1-x)(1-y)}{\sqrt{(\frac{r}{n})^2+x^2+y^2}}dx dy,  
\end{aligned}
\end{align}
where, in the last step, we consider that the distribution of the absolute difference of two uniform random variables is $(1-x)$. Converting the integral into the polar coordinates $(\mathcal{\varrho},\theta)$, we have:
\begin{align}
\begin{aligned}
    &\Phi^{\mathrm{2D}}_{\mathrm{GHZ}}/ \Phi_{s} \\ &\sim 8n^3 \int_0^{\pi/4} d\theta \int_0^{1/\cos{\theta}} d\varrho \, \frac{\varrho(1-\varrho\cos{\theta})(1-\varrho\sin{\theta})}{\sqrt{\varrho^2+(r/n)^2}}  \\
    &\approx 8n^3 \int_0^{\pi/4} d\theta \int_0^{1/\cos{\theta}} d\varrho \, \varrho (1-\varrho\cos{\theta})(1-\varrho\sin{\theta}) \frac{n}{r},\label{DecGHZAg2Drg}
\end{aligned}
\end{align}
where we employ $r/n \gtrsim 1$ for the last equation. From Eq.~(\ref{DecGHZAg2Drg}), we have:
\begin{equation}
    \Phi^{\mathrm{2D}}_{\mathrm{GHZ}}/ \Phi_{s} \sim n^4,
\end{equation}

\textit{In the intermediate regime}, from Eq.~(\ref{DecGHZAg2Drg}), we can obtain for $r/n \ll 1$:    
\begin{align}
\begin{aligned}\label{DecGHZAg2Dri}
    &\Phi^{\mathrm{2D}}_{\mathrm{GHZ}}/ \Phi_{s} \\ &\sim 8n^3 \int_0^{\pi/4} d\theta \int_0^{1/\cos{\theta}} d\varrho \, \frac{\varrho(1-\varrho\cos{\theta})(1-\varrho\sin{\theta})}{\sqrt{\varrho^2+(r/n)^2}}  \\
    &\approx 8n^3 \int_0^{\pi/4} d\theta \int_0^{1/\cos{\theta}} d\varrho \, (1-\varrho\cos{\theta})(1-\varrho\sin{\theta}).
\end{aligned}
\end{align}
Therefore, we have:
\begin{equation}
    \Phi^{\mathrm{2D}}_{\mathrm{GHZ}}/ \Phi_{s} \sim n^3.
\end{equation}  

For 1D arrays, from Eq.~(\ref{DecGHZAg}), we can similarly find $\Phi^{\mathrm{1D}}_{\mathrm{GHZ}} / \Phi_{s} \sim n$ and $\Phi^{\mathrm{1D}}_{\mathrm{GHZ}} / \Phi_{s} \sim n^2$ in the $r \ll 1$ and $r \gtrsim n$ regimes, respectively. In the intermediate regime, through the Euler–Maclaurin formula, we obtain $\Phi^{\mathrm{1D}}_{\mathrm{GHZ}} / \Phi_{s} \sim n \log{n}$. 

The above analytical derivations match well with our numerical results in Fig.~(\ref{fig:fig4}) in the main text. We have,
\begin{equation}
\Phi_{\mathrm{GHZ}}^{2D}/\Phi_s \sim
    \begin{cases}
N & \text{$r\ll 1$,} \\
N^{1.5} & \text{intermediate regime,} \\
N^{2} & \text{$r\gtrsim n$,}
\end{cases}
\end{equation}
for 2D arrays ($N=n^2$). Meanwhile, differences between $\Phi^{\mathrm{2D}}_{\mathrm{GHZ}}/ \Phi_{s}$ and $\Phi^{\mathrm{1D}}_{\mathrm{GHZ}}/ \Phi_{s}$ in the intermediate regime shows the dimensionality effects on nano-EM super-dephasing phenomena.

\subsubsection{Collective dephasing with $\Phi_c/\Phi_s \sim -D^{-3}$}
We now derive the scaling of super-dephasing when the underlying collective dephasing range follows $\Phi_c/\Phi_s \sim -D^{-3}$ at large interatom distance $D$. Here, we consider super-dephasing near gyromagnetic media, e.g., YIG. From Table.~\ref{tab:table1} in the main text and Eqs.~(\ref{SI_Gzz},~\ref{app_td_cdf},~\ref{td_sdf}), we find 
\begin{equation}\label{YIGratio}
    \frac{\Phi_c(\mathbf{r_i}, \, \mathbf{r_j})}{\Phi_s(\mathbf{r_i})} \approx \frac{2(2z)^5-(2z)^3 (\mathbf{r_i} - \mathbf{r_j})^2}{[(2z)^2 + (\mathbf{r_i} - \mathbf{r_j})^2]^{\frac{5}{2}}}.
\end{equation}
Substituting Eq.~(\ref{YIGratio}) into Eqs.~(\ref{DecGHZ2D},~\ref{DecGHZ1D}), we have:
\begin{align}\label{DecGHZYIG2D}
\begin{aligned}
    &\frac{\Phi^{\mathrm{2D}}_{\mathrm{GHZ}}(n,t)}{\Phi_{s}(t)} \approx \sum_{\substack{x_i, x_j, \\ y_i, y_j =1}}^n \frac{2r^5-r^3[(x_i-x_j)^2+(y_i-y_j)^2]}{2[r^2 + (x_i-x_j)^2+(y_i-y_j)^2]^{\frac{5}{2}}}, \\
    &\frac{\Phi^{\mathrm{1D}}_{\mathrm{GHZ}}(n,t)}{\Phi_{s}(t)} \approx \sum_{x_i, x_j =1}^n \frac{2r^5-r^3(x_i-x_j)^2}{2[r^2 + (x_i-x_j)^2]^{\frac{5}{2}}},
\end{aligned}
\end{align}
where $r=2z/b$. For 2D arrays, we have --

\textit{In the regime $r \ll 1$}, Eq.~(\ref{DecGHZYIG2D}) is dominated by terms corresponding to $(x_i,y_i)=(x_j,y_j)$. Therefore, we have:
\begin{equation}
    \Phi^{\mathrm{2D}}_{\mathrm{GHZ}} / \Phi_{s} \sim n^2.
\end{equation}

\textit{In the regime $r \gtrsim n$}, with the Euler–Maclaurin formula, we convert Eq.~(\ref{DecGHZYIG2D}) into the integral:
\begin{multline}\label{DecGHZMrg}
    \Phi^{\mathrm{2D}}_{\mathrm{GHZ}} / \Phi_{s}  \sim n \int_0^1 \int_0^1 \int_0^1 \int_0^1 dx_1 dx_2 dy_1 dy_2 \\ \qquad \qquad \qquad \qquad \frac{2(\frac{r}{n})^2-(x_1-x_2)^2-(y_1-y_2)^2}{[(\frac{r}{n})^2+(x_1-x_2)^2+(y_1-y_2)^2]^{\frac{5}{2}}} \\
    =4 n \int_0^1 \int_0^1 \frac{2(\frac{r}{n})^2-x^2-y^2}{[(\frac{r}{n})^2+x^2+y^2]^{\frac{5}{2}}} (1-x)(1-y)dx dy.
\end{multline}
In polar coordinates $(\mathcal{\varrho},\theta)$, we have:
\begin{multline}
    \Phi^{\mathrm{2D}}_{\mathrm{GHZ}}/ \Phi_{s} \sim 8n \int_0^{\pi/4} d\theta \int_0^{1/\cos{\theta}} d\varrho \\ \qquad \qquad \varrho(1-\varrho\cos{\theta})(1-\varrho\sin{\theta}) \frac{2(r/n)^2-\varrho^2}{(\varrho^2+(r/n)^2)^{5/2}} \\
    \approx 8n \int_0^{\pi/4} d\theta \int_0^{1/\cos{\theta}} d\varrho \, \varrho (1-\varrho\cos{\theta})(1-\varrho\sin{\theta}) (\frac{n}{r})^3, \label{DecGHZYIG2Drg}
\end{multline}
where we employ $r/n \gtrsim 1$ for the last equation. From Eq.~(\ref{DecGHZYIG2Drg}), we have:
\begin{equation}
    \Phi^{\mathrm{2D}}_{\mathrm{GHZ}}/ \Phi_{s} \sim n^4
\end{equation}

\textit{In the intermediate regime}, taking $r/n \ll 1$ for the integral in Eq.~(\ref{DecGHZYIG2Drg}), we can obtain:
\begin{align}
    \Phi^{\mathrm{2D}}_{\mathrm{GHZ}}/ \Phi_{s} & \sim n \log{n}.
\end{align}

For 1D arrays, from Eq.~(\ref{DecGHZYIG2D}), we can similarly find $\Phi^{\mathrm{1D}}_{\mathrm{GHZ}} / \Phi_{s} \sim n$ and $\Phi^{\mathrm{1D}}_{\mathrm{GHZ}} / \Phi_{s} \sim n^2$ in the $r \ll 1$ and $r \gtrsim n$ regimes, respectively. We obtain $\Phi^{\mathrm{1D}}_{\mathrm{GHZ}} / \Phi_{s} \sim n$ through the Euler–Maclaurin formula in the intermediate regime. 

The above analytical analysis matches well with our numerical results in Fig.~(\ref{fig:fig4}) in the main text. We have,
\begin{equation}
\Phi_{\mathrm{GHZ}}^{2D}/\Phi_s \sim
    \begin{cases}
N & \text{$r\ll 1$,} \\
N^{0.5} \log{N} & \text{intermediate regime,} \\
N^{2} & \text{$r\gtrsim n$,}
\end{cases}
\end{equation}
for 2D arrays ($N=n^2$). Differences between $\Phi^{\mathrm{2D}}_{\mathrm{GHZ}} / \Phi_{s} \sim n \log{n}$ and $\Phi^{\mathrm{1D}}_{\mathrm{GHZ}} / \Phi_{s} \sim n$ in the intermediate regime reveals the dimensionality effects on nano-EM sub-dephasing in 1D and 2D arrays. 

\begin{figure}[!t]
    \centering
    \includegraphics[width = 3in]{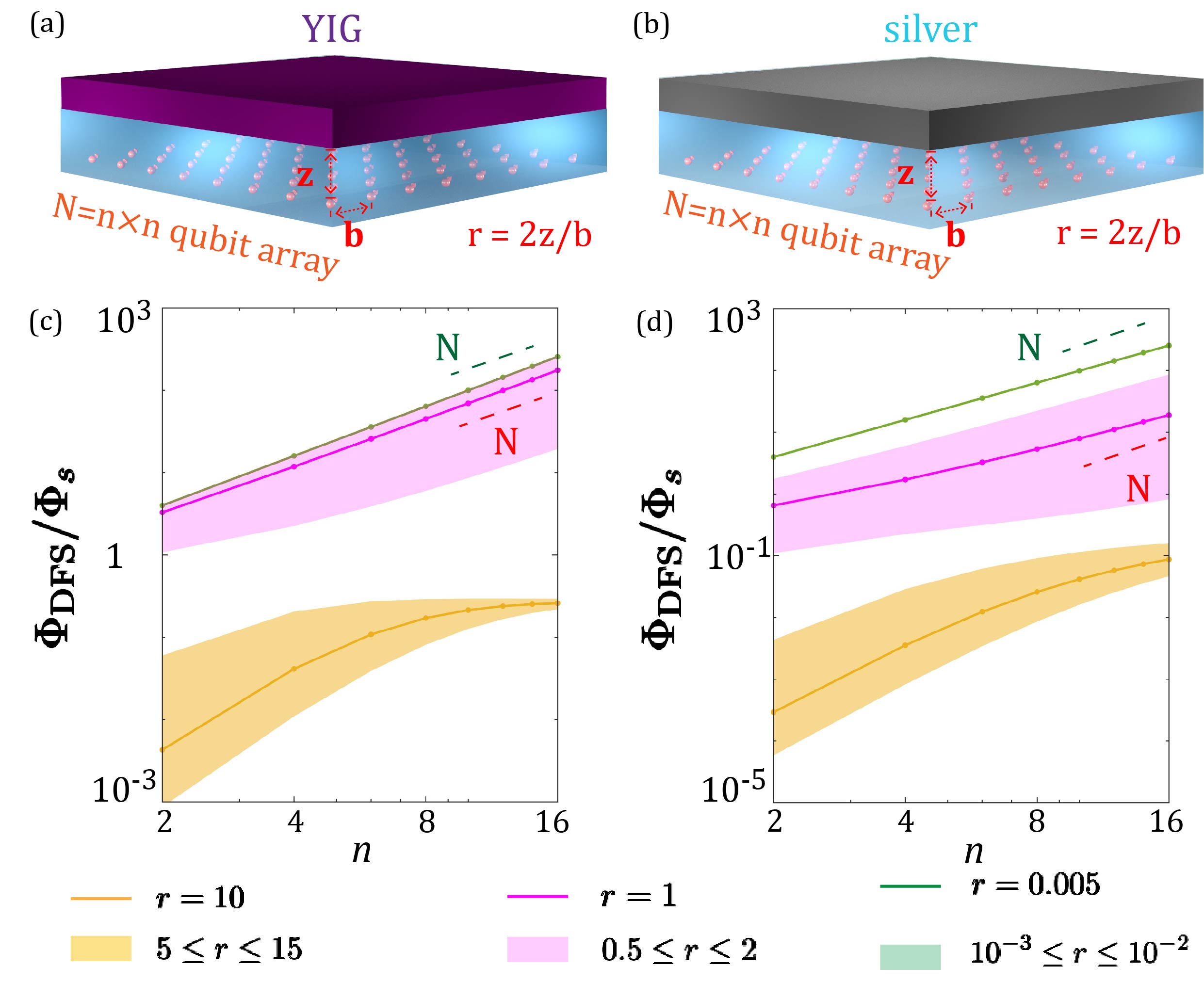}
    \caption{Nano-EM sub-dephasing of $|\psi\rangle_{\mathrm{DFS}}$ near gyromagnetic (YIG) and non-magnetic (silver) slabs. (a,b) Schematics of 2D $n\times n$ TLSs arrays with lattice constant $b$ at a distance $z$ from (a) YIG and (b) silver slabs. (c,d) Scaling of nano-EM sub-dephasing in $|\psi\rangle_{\mathrm{DFS}}$ in the three regimes corresponding to $r \ll1$, $r \gtrsim n$, and the intermediate regime.}
    \label{fig:figs3}
\end{figure}

\subsection{Decoherence-free subspace (DFS) state}
To this end, we extend our discussions in the main text to states in the decoherence-free subspace (DFS). We consider $|\psi\rangle_\mathrm{DFS}=\frac{1}{\sqrt{2}}(|0101\cdots\rangle + |1010\cdots\rangle)$ with arrangements of $|0101\cdots\rangle$ and $|1010\cdots\rangle$ satisfying $\hat{\sigma}^z_m\hat{\sigma}^z_k|0101\cdots\rangle<0$ and  $\hat{\sigma}^z_m\hat{\sigma}^z_k|1010\cdots\rangle<0$ for all pairs of nearest-neighbor qubits $m, k$ in the 2D array.

From Eq.~(\ref{me}), the dephasing function $\Phi^{\mathrm{2D}}_{\mathrm{DFS}}(n,t)$ for $|\psi\rangle_{\mathrm{DFS}}$ in 2D arrays is:
\begin{multline}\label{DecDFS2D}
    \Phi^{\mathrm{2D}}_{\mathrm{DFS}}(n,t) =  2 \bigintsss_0^t \Big[ \sum_{x_i, \, y_i=1}^n \gamma_s^\phi (\mathbf{r_i},t') + \\ \sum_{\substack{x_i,\, x_j,\, y_i,\, y_j=1\\ (x_i, \, y_i)\neq (x_j, \, y_j)}}^n (-1)^{|x_i-x_j|+|y_i-y_j|} \ \gamma_c^\phi (\mathbf{r_i},\mathbf{r_j},t') \Big] dt',
\end{multline}
and $\Phi^{\mathrm{2D}}_{\mathrm{DFS}}$ characterizes the decay of multiqubit coherence $e^{-\Phi^{\mathrm{2D}}_{\mathrm{DFS}}(t)}$ of the DFS state.

We demonstrate nano-EM sub-dephasing behaviors $\Phi_{\mathrm{DFS}}/\Phi_s$ in 2D arrays near gyromagnetic and non-magnetic slabs, as shown in Fig.~\ref{fig:figs3}. We focus on collective effects on multiqubit decoherence behaviors in the three regimes distinguished by $r$. In the $r \ll 1$ regime near both material interfaces, $\Phi_{\mathrm{DFS}} \sim N$ as shown by the green line in Fig.~\ref{fig:figs3}. This is because the dephasing processes are dominated by individual dephasing. In the $r \gtrsim n$ regime near both interfaces, nano-EM collective and individual dephasing have comparable rates. Therefore, collective dephasing strongly suppresses the dephasing of $|\psi\rangle_{\mathrm{DFS}}$ through destructive interference with individual dephasing, leading to nano-EM sub-dephasing behaviors. In the intermediate regime, nano-EM collective dephasing can moderately suppress multiqubit decoherence of $|\psi\rangle_{\mathrm{DFS}}$ near both materials. Meanwhile, $\Phi_{\mathrm{DFS}} \sim N$ follows qualitatively similar behaviors as in the $r \ll 1$ regime.

\nocite{*}
\bibliography{reference}

\begin{thebibliography}{69}%
\makeatletter
\providecommand \@ifxundefined [1]{%
 \@ifx{#1\undefined}
}%
\providecommand \@ifnum [1]{%
 \ifnum #1\expandafter \@firstoftwo
 \else \expandafter \@secondoftwo
 \fi
}%
\providecommand \@ifx [1]{%
 \ifx #1\expandafter \@firstoftwo
 \else \expandafter \@secondoftwo
 \fi
}%
\providecommand \natexlab [1]{#1}%
\providecommand \enquote  [1]{``#1''}%
\providecommand \bibnamefont  [1]{#1}%
\providecommand \bibfnamefont [1]{#1}%
\providecommand \citenamefont [1]{#1}%
\providecommand \href@noop [0]{\@secondoftwo}%
\providecommand \href [0]{\begingroup \@sanitize@url \@href}%
\providecommand \@href[1]{\@@startlink{#1}\@@href}%
\providecommand \@@href[1]{\endgroup#1\@@endlink}%
\providecommand \@sanitize@url [0]{\catcode `\\12\catcode `\$12\catcode `\&12\catcode `\#12\catcode `\^12\catcode `\_12\catcode `\%12\relax}%
\providecommand \@@startlink[1]{}%
\providecommand \@@endlink[0]{}%
\providecommand \url  [0]{\begingroup\@sanitize@url \@url }%
\providecommand \@url [1]{\endgroup\@href {#1}{\urlprefix }}%
\providecommand \urlprefix  [0]{URL }%
\providecommand \Eprint [0]{\href }%
\providecommand \doibase [0]{https://doi.org/}%
\providecommand \selectlanguage [0]{\@gobble}%
\providecommand \bibinfo  [0]{\@secondoftwo}%
\providecommand \bibfield  [0]{\@secondoftwo}%
\providecommand \translation [1]{[#1]}%
\providecommand \BibitemOpen [0]{}%
\providecommand \bibitemStop [0]{}%
\providecommand \bibitemNoStop [0]{.\EOS\space}%
\providecommand \EOS [0]{\spacefactor3000\relax}%
\providecommand \BibitemShut  [1]{\csname bibitem#1\endcsname}%
\let\auto@bib@innerbib\@empty
\bibitem [{\citenamefont {Reitz}\ \emph {et~al.}(2022)\citenamefont {Reitz}, \citenamefont {Sommer},\ and\ \citenamefont {Genes}}]{reitz2022cooperative}%
  \BibitemOpen
  \bibfield  {author} {\bibinfo {author} {\bibfnamefont {M.}~\bibnamefont {Reitz}}, \bibinfo {author} {\bibfnamefont {C.}~\bibnamefont {Sommer}},\ and\ \bibinfo {author} {\bibfnamefont {C.}~\bibnamefont {Genes}},\ }\bibfield  {title} {\bibinfo {title} {Cooperative quantum phenomena in light-matter platforms},\ }\href {https://link.aps.org/doi/10.1103/PRXQuantum.3.010201} {\bibfield  {journal} {\bibinfo  {journal} {PRX Quantum}\ }\textbf {\bibinfo {volume} {3}},\ \bibinfo {pages} {010201} (\bibinfo {year} {2022})}\BibitemShut {NoStop}%
\bibitem [{\citenamefont {Chang}\ \emph {et~al.}(2018)\citenamefont {Chang}, \citenamefont {Douglas}, \citenamefont {Gonz{\'a}lez-Tudela}, \citenamefont {Hung},\ and\ \citenamefont {Kimble}}]{chang2018colloquium}%
  \BibitemOpen
  \bibfield  {author} {\bibinfo {author} {\bibfnamefont {D.}~\bibnamefont {Chang}}, \bibinfo {author} {\bibfnamefont {J.}~\bibnamefont {Douglas}}, \bibinfo {author} {\bibfnamefont {A.}~\bibnamefont {Gonz{\'a}lez-Tudela}}, \bibinfo {author} {\bibfnamefont {C.-L.}\ \bibnamefont {Hung}},\ and\ \bibinfo {author} {\bibfnamefont {H.}~\bibnamefont {Kimble}},\ }\bibfield  {title} {\bibinfo {title} {Colloquium: Quantum matter built from nanoscopic lattices of atoms and photons},\ }\href {https://link.aps.org/doi/10.1103/RevModPhys.90.031002} {\bibfield  {journal} {\bibinfo  {journal} {Reviews of Modern Physics}\ }\textbf {\bibinfo {volume} {90}},\ \bibinfo {pages} {031002} (\bibinfo {year} {2018})}\BibitemShut {NoStop}%
\bibitem [{\citenamefont {Sheremet}\ \emph {et~al.}(2023)\citenamefont {Sheremet}, \citenamefont {Petrov}, \citenamefont {Iorsh}, \citenamefont {Poshakinskiy},\ and\ \citenamefont {Poddubny}}]{sheremet2023waveguide}%
  \BibitemOpen
  \bibfield  {author} {\bibinfo {author} {\bibfnamefont {A.~S.}\ \bibnamefont {Sheremet}}, \bibinfo {author} {\bibfnamefont {M.~I.}\ \bibnamefont {Petrov}}, \bibinfo {author} {\bibfnamefont {I.~V.}\ \bibnamefont {Iorsh}}, \bibinfo {author} {\bibfnamefont {A.~V.}\ \bibnamefont {Poshakinskiy}},\ and\ \bibinfo {author} {\bibfnamefont {A.~N.}\ \bibnamefont {Poddubny}},\ }\bibfield  {title} {\bibinfo {title} {Waveguide quantum electrodynamics: collective radiance and photon-photon correlations},\ }\href {https://link.aps.org/doi/10.1103/RevModPhys.95.015002} {\bibfield  {journal} {\bibinfo  {journal} {Reviews of Modern Physics}\ }\textbf {\bibinfo {volume} {95}},\ \bibinfo {pages} {015002} (\bibinfo {year} {2023})}\BibitemShut {NoStop}%
\bibitem [{\citenamefont {Masson}\ and\ \citenamefont {Asenjo-Garcia}(2022)}]{masson2022universality}%
  \BibitemOpen
  \bibfield  {author} {\bibinfo {author} {\bibfnamefont {S.~J.}\ \bibnamefont {Masson}}\ and\ \bibinfo {author} {\bibfnamefont {A.}~\bibnamefont {Asenjo-Garcia}},\ }\bibfield  {title} {\bibinfo {title} {Universality of dicke superradiance in arrays of quantum emitters},\ }\href {https://doi.org/10.1038/s41467-022-29805-4} {\bibfield  {journal} {\bibinfo  {journal} {Nature Communications}\ }\textbf {\bibinfo {volume} {13}},\ \bibinfo {pages} {2285} (\bibinfo {year} {2022})}\BibitemShut {NoStop}%
\bibitem [{\citenamefont {Orioli}\ \emph {et~al.}(2022)\citenamefont {Orioli}, \citenamefont {Thompson},\ and\ \citenamefont {Rey}}]{orioli2022emergent}%
  \BibitemOpen
  \bibfield  {author} {\bibinfo {author} {\bibfnamefont {A.~P.}\ \bibnamefont {Orioli}}, \bibinfo {author} {\bibfnamefont {J.~K.}\ \bibnamefont {Thompson}},\ and\ \bibinfo {author} {\bibfnamefont {A.~M.}\ \bibnamefont {Rey}},\ }\bibfield  {title} {\bibinfo {title} {Emergent dark states from superradiant dynamics in multilevel atoms in a cavity},\ }\href {https://link.aps.org/doi/10.1103/PhysRevX.12.011054} {\bibfield  {journal} {\bibinfo  {journal} {Physical Review X}\ }\textbf {\bibinfo {volume} {12}},\ \bibinfo {pages} {011054} (\bibinfo {year} {2022})}\BibitemShut {NoStop}%
\bibitem [{\citenamefont {Rubies-Bigorda}\ and\ \citenamefont {Yelin}(2022)}]{rubies2022superradiance}%
  \BibitemOpen
  \bibfield  {author} {\bibinfo {author} {\bibfnamefont {O.}~\bibnamefont {Rubies-Bigorda}}\ and\ \bibinfo {author} {\bibfnamefont {S.~F.}\ \bibnamefont {Yelin}},\ }\bibfield  {title} {\bibinfo {title} {Superradiance and subradiance in inverted atomic arrays},\ }\href {https://link.aps.org/doi/10.1103/PhysRevA.106.053717} {\bibfield  {journal} {\bibinfo  {journal} {Physical Review A}\ }\textbf {\bibinfo {volume} {106}},\ \bibinfo {pages} {053717} (\bibinfo {year} {2022})}\BibitemShut {NoStop}%
\bibitem [{\citenamefont {Sinha}\ \emph {et~al.}(2020)\citenamefont {Sinha}, \citenamefont {Meystre}, \citenamefont {Goldschmidt}, \citenamefont {Fatemi}, \citenamefont {Rolston},\ and\ \citenamefont {Solano}}]{sinha2020non}%
  \BibitemOpen
  \bibfield  {author} {\bibinfo {author} {\bibfnamefont {K.}~\bibnamefont {Sinha}}, \bibinfo {author} {\bibfnamefont {P.}~\bibnamefont {Meystre}}, \bibinfo {author} {\bibfnamefont {E.~A.}\ \bibnamefont {Goldschmidt}}, \bibinfo {author} {\bibfnamefont {F.~K.}\ \bibnamefont {Fatemi}}, \bibinfo {author} {\bibfnamefont {S.~L.}\ \bibnamefont {Rolston}},\ and\ \bibinfo {author} {\bibfnamefont {P.}~\bibnamefont {Solano}},\ }\bibfield  {title} {\bibinfo {title} {Non-markovian collective emission from macroscopically separated emitters},\ }\href {https://link.aps.org/doi/10.1103/PhysRevLett.124.043603} {\bibfield  {journal} {\bibinfo  {journal} {Physical review letters}\ }\textbf {\bibinfo {volume} {124}},\ \bibinfo {pages} {043603} (\bibinfo {year} {2020})}\BibitemShut {NoStop}%
\bibitem [{\citenamefont {Mok}\ \emph {et~al.}(2023)\citenamefont {Mok}, \citenamefont {Asenjo-Garcia}, \citenamefont {Sum},\ and\ \citenamefont {Kwek}}]{mok2023dicke}%
  \BibitemOpen
  \bibfield  {author} {\bibinfo {author} {\bibfnamefont {W.-K.}\ \bibnamefont {Mok}}, \bibinfo {author} {\bibfnamefont {A.}~\bibnamefont {Asenjo-Garcia}}, \bibinfo {author} {\bibfnamefont {T.~C.}\ \bibnamefont {Sum}},\ and\ \bibinfo {author} {\bibfnamefont {L.-C.}\ \bibnamefont {Kwek}},\ }\bibfield  {title} {\bibinfo {title} {Dicke superradiance requires interactions beyond nearest neighbors},\ }\href {https://link.aps.org/doi/10.1103/PhysRevLett.130.213605} {\bibfield  {journal} {\bibinfo  {journal} {Physical Review Letters}\ }\textbf {\bibinfo {volume} {130}},\ \bibinfo {pages} {213605} (\bibinfo {year} {2023})}\BibitemShut {NoStop}%
\bibitem [{\citenamefont {Mlynek}\ \emph {et~al.}(2014)\citenamefont {Mlynek}, \citenamefont {Abdumalikov}, \citenamefont {Eichler},\ and\ \citenamefont {Wallraff}}]{mlynek2014observation}%
  \BibitemOpen
  \bibfield  {author} {\bibinfo {author} {\bibfnamefont {J.~A.}\ \bibnamefont {Mlynek}}, \bibinfo {author} {\bibfnamefont {A.~A.}\ \bibnamefont {Abdumalikov}}, \bibinfo {author} {\bibfnamefont {C.}~\bibnamefont {Eichler}},\ and\ \bibinfo {author} {\bibfnamefont {A.}~\bibnamefont {Wallraff}},\ }\bibfield  {title} {\bibinfo {title} {Observation of dicke superradiance for two artificial atoms in a cavity with high decay rate},\ }\href {https://doi.org/10.1038/ncomms6186} {\bibfield  {journal} {\bibinfo  {journal} {Nature communications}\ }\textbf {\bibinfo {volume} {5}},\ \bibinfo {pages} {5186} (\bibinfo {year} {2014})}\BibitemShut {NoStop}%
\bibitem [{\citenamefont {Pennetta}\ \emph {et~al.}(2022)\citenamefont {Pennetta}, \citenamefont {Blaha}, \citenamefont {Johnson}, \citenamefont {Lechner}, \citenamefont {Schneeweiss}, \citenamefont {Volz},\ and\ \citenamefont {Rauschenbeutel}}]{pennetta2022collective}%
  \BibitemOpen
  \bibfield  {author} {\bibinfo {author} {\bibfnamefont {R.}~\bibnamefont {Pennetta}}, \bibinfo {author} {\bibfnamefont {M.}~\bibnamefont {Blaha}}, \bibinfo {author} {\bibfnamefont {A.}~\bibnamefont {Johnson}}, \bibinfo {author} {\bibfnamefont {D.}~\bibnamefont {Lechner}}, \bibinfo {author} {\bibfnamefont {P.}~\bibnamefont {Schneeweiss}}, \bibinfo {author} {\bibfnamefont {J.}~\bibnamefont {Volz}},\ and\ \bibinfo {author} {\bibfnamefont {A.}~\bibnamefont {Rauschenbeutel}},\ }\bibfield  {title} {\bibinfo {title} {Collective radiative dynamics of an ensemble of cold atoms coupled to an optical waveguide},\ }\href {https://link.aps.org/doi/10.1103/PhysRevLett.128.073601} {\bibfield  {journal} {\bibinfo  {journal} {Physical Review Letters}\ }\textbf {\bibinfo {volume} {128}},\ \bibinfo {pages} {073601} (\bibinfo {year} {2022})}\BibitemShut {NoStop}%
\bibitem [{\citenamefont {Jones}\ \emph {et~al.}(2020)\citenamefont {Jones}, \citenamefont {Buonaiuto}, \citenamefont {Lang}, \citenamefont {Lesanovsky},\ and\ \citenamefont {Olmos}}]{jones2020collectively}%
  \BibitemOpen
  \bibfield  {author} {\bibinfo {author} {\bibfnamefont {R.}~\bibnamefont {Jones}}, \bibinfo {author} {\bibfnamefont {G.}~\bibnamefont {Buonaiuto}}, \bibinfo {author} {\bibfnamefont {B.}~\bibnamefont {Lang}}, \bibinfo {author} {\bibfnamefont {I.}~\bibnamefont {Lesanovsky}},\ and\ \bibinfo {author} {\bibfnamefont {B.}~\bibnamefont {Olmos}},\ }\bibfield  {title} {\bibinfo {title} {Collectively enhanced chiral photon emission from an atomic array near a nanofiber},\ }\href {https://link.aps.org/doi/10.1103/PhysRevLett.124.093601} {\bibfield  {journal} {\bibinfo  {journal} {Physical review letters}\ }\textbf {\bibinfo {volume} {124}},\ \bibinfo {pages} {093601} (\bibinfo {year} {2020})}\BibitemShut {NoStop}%
\bibitem [{\citenamefont {Wang}\ \emph {et~al.}(2020)\citenamefont {Wang}, \citenamefont {Jaako}, \citenamefont {Kirton},\ and\ \citenamefont {Rabl}}]{wang2020supercorrelated}%
  \BibitemOpen
  \bibfield  {author} {\bibinfo {author} {\bibfnamefont {Z.}~\bibnamefont {Wang}}, \bibinfo {author} {\bibfnamefont {T.}~\bibnamefont {Jaako}}, \bibinfo {author} {\bibfnamefont {P.}~\bibnamefont {Kirton}},\ and\ \bibinfo {author} {\bibfnamefont {P.}~\bibnamefont {Rabl}},\ }\bibfield  {title} {\bibinfo {title} {Supercorrelated radiance in nonlinear photonic waveguides},\ }\href {https://link.aps.org/doi/10.1103/PhysRevLett.124.213601} {\bibfield  {journal} {\bibinfo  {journal} {Physical Review Letters}\ }\textbf {\bibinfo {volume} {124}},\ \bibinfo {pages} {213601} (\bibinfo {year} {2020})}\BibitemShut {NoStop}%
\bibitem [{\citenamefont {Pak}\ \emph {et~al.}(2022)\citenamefont {Pak}, \citenamefont {Nandi}, \citenamefont {Titze}, \citenamefont {Bielejec}, \citenamefont {Alaeian},\ and\ \citenamefont {Hosseini}}]{pak2022long}%
  \BibitemOpen
  \bibfield  {author} {\bibinfo {author} {\bibfnamefont {D.}~\bibnamefont {Pak}}, \bibinfo {author} {\bibfnamefont {A.}~\bibnamefont {Nandi}}, \bibinfo {author} {\bibfnamefont {M.}~\bibnamefont {Titze}}, \bibinfo {author} {\bibfnamefont {E.~S.}\ \bibnamefont {Bielejec}}, \bibinfo {author} {\bibfnamefont {H.}~\bibnamefont {Alaeian}},\ and\ \bibinfo {author} {\bibfnamefont {M.}~\bibnamefont {Hosseini}},\ }\bibfield  {title} {\bibinfo {title} {Long-range cooperative resonances in rare-earth ion arrays inside photonic resonators},\ }\href {https://doi.org/10.1038/s42005-022-00871-w} {\bibfield  {journal} {\bibinfo  {journal} {Communications Physics}\ }\textbf {\bibinfo {volume} {5}},\ \bibinfo {pages} {89} (\bibinfo {year} {2022})}\BibitemShut {NoStop}%
\bibitem [{\citenamefont {Lei}\ \emph {et~al.}(2023)\citenamefont {Lei}, \citenamefont {Fukumori}, \citenamefont {Rochman}, \citenamefont {Zhu}, \citenamefont {Endres}, \citenamefont {Choi},\ and\ \citenamefont {Faraon}}]{lei2023many}%
  \BibitemOpen
  \bibfield  {author} {\bibinfo {author} {\bibfnamefont {M.}~\bibnamefont {Lei}}, \bibinfo {author} {\bibfnamefont {R.}~\bibnamefont {Fukumori}}, \bibinfo {author} {\bibfnamefont {J.}~\bibnamefont {Rochman}}, \bibinfo {author} {\bibfnamefont {B.}~\bibnamefont {Zhu}}, \bibinfo {author} {\bibfnamefont {M.}~\bibnamefont {Endres}}, \bibinfo {author} {\bibfnamefont {J.}~\bibnamefont {Choi}},\ and\ \bibinfo {author} {\bibfnamefont {A.}~\bibnamefont {Faraon}},\ }\bibfield  {title} {\bibinfo {title} {Many-body cavity quantum electrodynamics with driven inhomogeneous emitters},\ }\href {https://doi.org/10.1038/s41586-023-05884-1} {\bibfield  {journal} {\bibinfo  {journal} {Nature}\ }\textbf {\bibinfo {volume} {617}},\ \bibinfo {pages} {271} (\bibinfo {year} {2023})}\BibitemShut {NoStop}%
\bibitem [{\citenamefont {Zhu}\ \emph {et~al.}(2024)\citenamefont {Zhu}, \citenamefont {Hu}, \citenamefont {Wang}, \citenamefont {Qin}, \citenamefont {L{\"u}},\ and\ \citenamefont {Nori}}]{zhu2024nonreciprocal}%
  \BibitemOpen
  \bibfield  {author} {\bibinfo {author} {\bibfnamefont {G.-L.}\ \bibnamefont {Zhu}}, \bibinfo {author} {\bibfnamefont {C.-S.}\ \bibnamefont {Hu}}, \bibinfo {author} {\bibfnamefont {H.}~\bibnamefont {Wang}}, \bibinfo {author} {\bibfnamefont {W.}~\bibnamefont {Qin}}, \bibinfo {author} {\bibfnamefont {X.-Y.}\ \bibnamefont {L{\"u}}},\ and\ \bibinfo {author} {\bibfnamefont {F.}~\bibnamefont {Nori}},\ }\bibfield  {title} {\bibinfo {title} {Nonreciprocal superradiant phase transitions and multicriticality in a cavity qed system},\ }\href {https://doi.org/10.1103/PhysRevLett.132.193602} {\bibfield  {journal} {\bibinfo  {journal} {Physical Review Letters}\ }\textbf {\bibinfo {volume} {132}},\ \bibinfo {pages} {193602} (\bibinfo {year} {2024})}\BibitemShut {NoStop}%
\bibitem [{\citenamefont {Palma}\ \emph {et~al.}(1996)\citenamefont {Palma}, \citenamefont {Suominen},\ and\ \citenamefont {Ekert}}]{palma1996quantum}%
  \BibitemOpen
  \bibfield  {author} {\bibinfo {author} {\bibfnamefont {G.~M.}\ \bibnamefont {Palma}}, \bibinfo {author} {\bibfnamefont {K.-A.}\ \bibnamefont {Suominen}},\ and\ \bibinfo {author} {\bibfnamefont {A.}~\bibnamefont {Ekert}},\ }\bibfield  {title} {\bibinfo {title} {Quantum computers and dissipation},\ }\href {http://www.jstor.org/stable/52838} {\bibfield  {journal} {\bibinfo  {journal} {Proceedings of the Royal Society of London. Series A: Mathematical, Physical and Engineering Sciences}\ }\textbf {\bibinfo {volume} {452}},\ \bibinfo {pages} {567} (\bibinfo {year} {1996})}\BibitemShut {NoStop}%
\bibitem [{\citenamefont {Venkatesh}\ \emph {et~al.}(2018)\citenamefont {Venkatesh}, \citenamefont {Juan},\ and\ \citenamefont {Romero-Isart}}]{venkatesh2018cooperative}%
  \BibitemOpen
  \bibfield  {author} {\bibinfo {author} {\bibfnamefont {B.~P.}\ \bibnamefont {Venkatesh}}, \bibinfo {author} {\bibfnamefont {M.}~\bibnamefont {Juan}},\ and\ \bibinfo {author} {\bibfnamefont {O.}~\bibnamefont {Romero-Isart}},\ }\bibfield  {title} {\bibinfo {title} {Cooperative effects in closely packed quantum emitters with collective dephasing},\ }\href {https://link.aps.org/doi/10.1103/PhysRevLett.120.033602} {\bibfield  {journal} {\bibinfo  {journal} {Physical Review Letters}\ }\textbf {\bibinfo {volume} {120}},\ \bibinfo {pages} {033602} (\bibinfo {year} {2018})}\BibitemShut {NoStop}%
\bibitem [{\citenamefont {Lidar}\ \emph {et~al.}(1998)\citenamefont {Lidar}, \citenamefont {Chuang},\ and\ \citenamefont {Whaley}}]{lidar1998decoherence}%
  \BibitemOpen
  \bibfield  {author} {\bibinfo {author} {\bibfnamefont {D.~A.}\ \bibnamefont {Lidar}}, \bibinfo {author} {\bibfnamefont {I.~L.}\ \bibnamefont {Chuang}},\ and\ \bibinfo {author} {\bibfnamefont {K.~B.}\ \bibnamefont {Whaley}},\ }\bibfield  {title} {\bibinfo {title} {Decoherence-free subspaces for quantum computation},\ }\href {https://doi.org/10.1103/PhysRevLett.81.2594} {\bibfield  {journal} {\bibinfo  {journal} {Physical Review Letters}\ }\textbf {\bibinfo {volume} {81}},\ \bibinfo {pages} {2594} (\bibinfo {year} {1998})}\BibitemShut {NoStop}%
\bibitem [{\citenamefont {Carnio}\ \emph {et~al.}(2015)\citenamefont {Carnio}, \citenamefont {Buchleitner},\ and\ \citenamefont {Gessner}}]{carnio2015robust}%
  \BibitemOpen
  \bibfield  {author} {\bibinfo {author} {\bibfnamefont {E.~G.}\ \bibnamefont {Carnio}}, \bibinfo {author} {\bibfnamefont {A.}~\bibnamefont {Buchleitner}},\ and\ \bibinfo {author} {\bibfnamefont {M.}~\bibnamefont {Gessner}},\ }\bibfield  {title} {\bibinfo {title} {Robust asymptotic entanglement under multipartite collective dephasing},\ }\href {https://link.aps.org/doi/10.1103/PhysRevLett.115.010404} {\bibfield  {journal} {\bibinfo  {journal} {Physical Review Letters}\ }\textbf {\bibinfo {volume} {115}},\ \bibinfo {pages} {010404} (\bibinfo {year} {2015})}\BibitemShut {NoStop}%
\bibitem [{\citenamefont {Reina}\ \emph {et~al.}(2002)\citenamefont {Reina}, \citenamefont {Quiroga},\ and\ \citenamefont {Johnson}}]{reina2002decoherence}%
  \BibitemOpen
  \bibfield  {author} {\bibinfo {author} {\bibfnamefont {J.~H.}\ \bibnamefont {Reina}}, \bibinfo {author} {\bibfnamefont {L.}~\bibnamefont {Quiroga}},\ and\ \bibinfo {author} {\bibfnamefont {N.~F.}\ \bibnamefont {Johnson}},\ }\bibfield  {title} {\bibinfo {title} {Decoherence of quantum registers},\ }\href {https://link.aps.org/doi/10.1103/PhysRevA.65.032326} {\bibfield  {journal} {\bibinfo  {journal} {Physical Review A}\ }\textbf {\bibinfo {volume} {65}},\ \bibinfo {pages} {032326} (\bibinfo {year} {2002})}\BibitemShut {NoStop}%
\bibitem [{\citenamefont {Doll}\ \emph {et~al.}(2007)\citenamefont {Doll}, \citenamefont {Wubs}, \citenamefont {H{\"a}nggi},\ and\ \citenamefont {Kohler}}]{doll2007incomplete}%
  \BibitemOpen
  \bibfield  {author} {\bibinfo {author} {\bibfnamefont {R.}~\bibnamefont {Doll}}, \bibinfo {author} {\bibfnamefont {M.}~\bibnamefont {Wubs}}, \bibinfo {author} {\bibfnamefont {P.}~\bibnamefont {H{\"a}nggi}},\ and\ \bibinfo {author} {\bibfnamefont {S.}~\bibnamefont {Kohler}},\ }\bibfield  {title} {\bibinfo {title} {Incomplete pure dephasing of n-qubit entangled w states},\ }\href {https://link.aps.org/doi/10.1103/PhysRevB.76.045317} {\bibfield  {journal} {\bibinfo  {journal} {Physical Review B}\ }\textbf {\bibinfo {volume} {76}},\ \bibinfo {pages} {045317} (\bibinfo {year} {2007})}\BibitemShut {NoStop}%
\bibitem [{\citenamefont {Tuziemski}\ \emph {et~al.}(2019)\citenamefont {Tuziemski}, \citenamefont {Lampo}, \citenamefont {Lewenstein},\ and\ \citenamefont {Korbicz}}]{tuziemski2019reexamination}%
  \BibitemOpen
  \bibfield  {author} {\bibinfo {author} {\bibfnamefont {J.}~\bibnamefont {Tuziemski}}, \bibinfo {author} {\bibfnamefont {A.}~\bibnamefont {Lampo}}, \bibinfo {author} {\bibfnamefont {M.}~\bibnamefont {Lewenstein}},\ and\ \bibinfo {author} {\bibfnamefont {J.~K.}\ \bibnamefont {Korbicz}},\ }\bibfield  {title} {\bibinfo {title} {Reexamination of the decoherence of spin registers},\ }\href {https://link.aps.org/doi/10.1103/PhysRevA.99.022122} {\bibfield  {journal} {\bibinfo  {journal} {Physical Review A}\ }\textbf {\bibinfo {volume} {99}},\ \bibinfo {pages} {022122} (\bibinfo {year} {2019})}\BibitemShut {NoStop}%
\bibitem [{\citenamefont {Klesse}\ and\ \citenamefont {Frank}(2005)}]{klesse2005quantum}%
  \BibitemOpen
  \bibfield  {author} {\bibinfo {author} {\bibfnamefont {R.}~\bibnamefont {Klesse}}\ and\ \bibinfo {author} {\bibfnamefont {S.}~\bibnamefont {Frank}},\ }\bibfield  {title} {\bibinfo {title} {Quantum error correction in spatially correlated quantum noise},\ }\href {https://link.aps.org/doi/10.1103/PhysRevLett.95.230503} {\bibfield  {journal} {\bibinfo  {journal} {Physical review letters}\ }\textbf {\bibinfo {volume} {95}},\ \bibinfo {pages} {230503} (\bibinfo {year} {2005})}\BibitemShut {NoStop}%
\bibitem [{\citenamefont {Aharonov}\ \emph {et~al.}(2006)\citenamefont {Aharonov}, \citenamefont {Kitaev},\ and\ \citenamefont {Preskill}}]{aharonov2006fault}%
  \BibitemOpen
  \bibfield  {author} {\bibinfo {author} {\bibfnamefont {D.}~\bibnamefont {Aharonov}}, \bibinfo {author} {\bibfnamefont {A.}~\bibnamefont {Kitaev}},\ and\ \bibinfo {author} {\bibfnamefont {J.}~\bibnamefont {Preskill}},\ }\bibfield  {title} {\bibinfo {title} {Fault-tolerant quantum computation with long-range correlated noise},\ }\href {https://link.aps.org/doi/10.1103/PhysRevLett.96.050504} {\bibfield  {journal} {\bibinfo  {journal} {Physical review letters}\ }\textbf {\bibinfo {volume} {96}},\ \bibinfo {pages} {050504} (\bibinfo {year} {2006})}\BibitemShut {NoStop}%
\bibitem [{\citenamefont {Kukita}\ \emph {et~al.}(2021)\citenamefont {Kukita}, \citenamefont {Matsuzaki},\ and\ \citenamefont {Kondo}}]{kukita2021heisenberg}%
  \BibitemOpen
  \bibfield  {author} {\bibinfo {author} {\bibfnamefont {S.}~\bibnamefont {Kukita}}, \bibinfo {author} {\bibfnamefont {Y.}~\bibnamefont {Matsuzaki}},\ and\ \bibinfo {author} {\bibfnamefont {Y.}~\bibnamefont {Kondo}},\ }\bibfield  {title} {\bibinfo {title} {Heisenberg-limited quantum metrology using collective dephasing},\ }\href {https://doi.org/10.1103/PhysRevApplied.16.064026} {\bibfield  {journal} {\bibinfo  {journal} {Physical Review Applied}\ }\textbf {\bibinfo {volume} {16}},\ \bibinfo {pages} {064026} (\bibinfo {year} {2021})}\BibitemShut {NoStop}%
\bibitem [{\citenamefont {Jeske}\ \emph {et~al.}(2014)\citenamefont {Jeske}, \citenamefont {Cole},\ and\ \citenamefont {Huelga}}]{jeske2014quantum}%
  \BibitemOpen
  \bibfield  {author} {\bibinfo {author} {\bibfnamefont {J.}~\bibnamefont {Jeske}}, \bibinfo {author} {\bibfnamefont {J.~H.}\ \bibnamefont {Cole}},\ and\ \bibinfo {author} {\bibfnamefont {S.~F.}\ \bibnamefont {Huelga}},\ }\bibfield  {title} {\bibinfo {title} {Quantum metrology subject to spatially correlated markovian noise: restoring the heisenberg limit},\ }\href {https://doi.org/10.1088/1367-2630/16/7/073039} {\bibfield  {journal} {\bibinfo  {journal} {New Journal of Physics}\ }\textbf {\bibinfo {volume} {16}},\ \bibinfo {pages} {073039} (\bibinfo {year} {2014})}\BibitemShut {NoStop}%
\bibitem [{\citenamefont {Cortes}\ \emph {et~al.}(2022)\citenamefont {Cortes}, \citenamefont {Sun},\ and\ \citenamefont {Jacob}}]{cortes2022fundamental}%
  \BibitemOpen
  \bibfield  {author} {\bibinfo {author} {\bibfnamefont {C.~L.}\ \bibnamefont {Cortes}}, \bibinfo {author} {\bibfnamefont {W.}~\bibnamefont {Sun}},\ and\ \bibinfo {author} {\bibfnamefont {Z.}~\bibnamefont {Jacob}},\ }\bibfield  {title} {\bibinfo {title} {Fundamental efficiency bound for quantum coherent energy transfer in nanophotonics},\ }\href {https://doi.org/10.1364/OE.465703} {\bibfield  {journal} {\bibinfo  {journal} {Optics Express}\ }\textbf {\bibinfo {volume} {30}},\ \bibinfo {pages} {34725} (\bibinfo {year} {2022})}\BibitemShut {NoStop}%
\bibitem [{\citenamefont {Boddeti}\ \emph {et~al.}(2021)\citenamefont {Boddeti}, \citenamefont {Guan}, \citenamefont {Sentz}, \citenamefont {Juarez}, \citenamefont {Newman}, \citenamefont {Cortes}, \citenamefont {Odom},\ and\ \citenamefont {Jacob}}]{boddeti2021long}%
  \BibitemOpen
  \bibfield  {author} {\bibinfo {author} {\bibfnamefont {A.~K.}\ \bibnamefont {Boddeti}}, \bibinfo {author} {\bibfnamefont {J.}~\bibnamefont {Guan}}, \bibinfo {author} {\bibfnamefont {T.}~\bibnamefont {Sentz}}, \bibinfo {author} {\bibfnamefont {X.}~\bibnamefont {Juarez}}, \bibinfo {author} {\bibfnamefont {W.}~\bibnamefont {Newman}}, \bibinfo {author} {\bibfnamefont {C.}~\bibnamefont {Cortes}}, \bibinfo {author} {\bibfnamefont {T.~W.}\ \bibnamefont {Odom}},\ and\ \bibinfo {author} {\bibfnamefont {Z.}~\bibnamefont {Jacob}},\ }\bibfield  {title} {\bibinfo {title} {Long-range dipole--dipole interactions in a plasmonic lattice},\ }\href {https://doi.org/10.1021/acs.nanolett.1c02835} {\bibfield  {journal} {\bibinfo  {journal} {Nano letters}\ }\textbf {\bibinfo {volume} {22}},\ \bibinfo {pages} {22} (\bibinfo {year} {2021})}\BibitemShut {NoStop}%
\bibitem [{\citenamefont {Cortes}\ and\ \citenamefont {Jacob}(2017)}]{cortes2017super}%
  \BibitemOpen
  \bibfield  {author} {\bibinfo {author} {\bibfnamefont {C.~L.}\ \bibnamefont {Cortes}}\ and\ \bibinfo {author} {\bibfnamefont {Z.}~\bibnamefont {Jacob}},\ }\bibfield  {title} {\bibinfo {title} {Super-coulombic atom--atom interactions in hyperbolic media},\ }\href {https://doi.org/10.1038/ncomms14144} {\bibfield  {journal} {\bibinfo  {journal} {Nature communications}\ }\textbf {\bibinfo {volume} {8}},\ \bibinfo {pages} {14144} (\bibinfo {year} {2017})}\BibitemShut {NoStop}%
\bibitem [{\citenamefont {Biehs}\ \emph {et~al.}(2016)\citenamefont {Biehs}, \citenamefont {Menon},\ and\ \citenamefont {Agarwal}}]{biehs2016long}%
  \BibitemOpen
  \bibfield  {author} {\bibinfo {author} {\bibfnamefont {S.-A.}\ \bibnamefont {Biehs}}, \bibinfo {author} {\bibfnamefont {V.~M.}\ \bibnamefont {Menon}},\ and\ \bibinfo {author} {\bibfnamefont {G.~S.}\ \bibnamefont {Agarwal}},\ }\bibfield  {title} {\bibinfo {title} {Long-range dipole-dipole interaction and anomalous f\"orster energy transfer across a hyperbolic metamaterial},\ }\href {https://doi.org/10.1103/PhysRevB.93.245439} {\bibfield  {journal} {\bibinfo  {journal} {Phys. Rev. B}\ }\textbf {\bibinfo {volume} {93}},\ \bibinfo {pages} {245439} (\bibinfo {year} {2016})}\BibitemShut {NoStop}%
\bibitem [{\citenamefont {Hassani~Gangaraj}\ \emph {et~al.}(2022)\citenamefont {Hassani~Gangaraj}, \citenamefont {Ying}, \citenamefont {Monticone},\ and\ \citenamefont {Yu}}]{hassani2022enhancement}%
  \BibitemOpen
  \bibfield  {author} {\bibinfo {author} {\bibfnamefont {S.~A.}\ \bibnamefont {Hassani~Gangaraj}}, \bibinfo {author} {\bibfnamefont {L.}~\bibnamefont {Ying}}, \bibinfo {author} {\bibfnamefont {F.}~\bibnamefont {Monticone}},\ and\ \bibinfo {author} {\bibfnamefont {Z.}~\bibnamefont {Yu}},\ }\bibfield  {title} {\bibinfo {title} {Enhancement of quantum excitation transport by photonic nonreciprocity},\ }\href {https://doi.org/10.1103/PhysRevA.106.033501} {\bibfield  {journal} {\bibinfo  {journal} {Physical Review A}\ }\textbf {\bibinfo {volume} {106}},\ \bibinfo {pages} {033501} (\bibinfo {year} {2022})}\BibitemShut {NoStop}%
\bibitem [{\citenamefont {Hanson}\ \emph {et~al.}(2019)\citenamefont {Hanson}, \citenamefont {Gangaraj}, \citenamefont {Silveirinha}, \citenamefont {Antezza},\ and\ \citenamefont {Monticone}}]{hanson2019non}%
  \BibitemOpen
  \bibfield  {author} {\bibinfo {author} {\bibfnamefont {G.~W.}\ \bibnamefont {Hanson}}, \bibinfo {author} {\bibfnamefont {S.~A.~H.}\ \bibnamefont {Gangaraj}}, \bibinfo {author} {\bibfnamefont {M.~G.}\ \bibnamefont {Silveirinha}}, \bibinfo {author} {\bibfnamefont {M.}~\bibnamefont {Antezza}},\ and\ \bibinfo {author} {\bibfnamefont {F.}~\bibnamefont {Monticone}},\ }\bibfield  {title} {\bibinfo {title} {Non-markovian transient casimir-polder force and population dynamics on excited-and ground-state atoms: Weak-and strong-coupling regimes in generally nonreciprocal environments},\ }\href {https://doi.org/10.1103/PhysRevA.99.042508} {\bibfield  {journal} {\bibinfo  {journal} {Physical Review A}\ }\textbf {\bibinfo {volume} {99}},\ \bibinfo {pages} {042508} (\bibinfo {year} {2019})}\BibitemShut {NoStop}%
\bibitem [{\citenamefont {Fuchs}\ \emph {et~al.}(2017)\citenamefont {Fuchs}, \citenamefont {Crosse},\ and\ \citenamefont {Buhmann}}]{fuchs2017casimir}%
  \BibitemOpen
  \bibfield  {author} {\bibinfo {author} {\bibfnamefont {S.}~\bibnamefont {Fuchs}}, \bibinfo {author} {\bibfnamefont {J.}~\bibnamefont {Crosse}},\ and\ \bibinfo {author} {\bibfnamefont {S.~Y.}\ \bibnamefont {Buhmann}},\ }\bibfield  {title} {\bibinfo {title} {Casimir-polder shift and decay rate in the presence of nonreciprocal media},\ }\href {https://doi.org/10.1103/PhysRevA.95.023805} {\bibfield  {journal} {\bibinfo  {journal} {Physical Review A}\ }\textbf {\bibinfo {volume} {95}},\ \bibinfo {pages} {023805} (\bibinfo {year} {2017})}\BibitemShut {NoStop}%
\bibitem [{\citenamefont {Block}\ and\ \citenamefont {Scheel}(2019)}]{block2019casimir}%
  \BibitemOpen
  \bibfield  {author} {\bibinfo {author} {\bibfnamefont {J.}~\bibnamefont {Block}}\ and\ \bibinfo {author} {\bibfnamefont {S.}~\bibnamefont {Scheel}},\ }\bibfield  {title} {\bibinfo {title} {Casimir-polder-induced rydberg macrodimers},\ }\href {https://doi.org/10.1103/PhysRevA.100.062508} {\bibfield  {journal} {\bibinfo  {journal} {Physical Review A}\ }\textbf {\bibinfo {volume} {100}},\ \bibinfo {pages} {062508} (\bibinfo {year} {2019})}\BibitemShut {NoStop}%
\bibitem [{\citenamefont {Sinha}\ \emph {et~al.}(2018)\citenamefont {Sinha}, \citenamefont {Venkatesh},\ and\ \citenamefont {Meystre}}]{sinha2018collective}%
  \BibitemOpen
  \bibfield  {author} {\bibinfo {author} {\bibfnamefont {K.}~\bibnamefont {Sinha}}, \bibinfo {author} {\bibfnamefont {B.~P.}\ \bibnamefont {Venkatesh}},\ and\ \bibinfo {author} {\bibfnamefont {P.}~\bibnamefont {Meystre}},\ }\bibfield  {title} {\bibinfo {title} {Collective effects in casimir-polder forces},\ }\href {https://doi.org/10.1103/PhysRevLett.121.183605} {\bibfield  {journal} {\bibinfo  {journal} {Physical Review Letters}\ }\textbf {\bibinfo {volume} {121}},\ \bibinfo {pages} {183605} (\bibinfo {year} {2018})}\BibitemShut {NoStop}%
\bibitem [{\citenamefont {Henkel}\ \emph {et~al.}(2018)\citenamefont {Henkel}, \citenamefont {Klimchitskaya},\ and\ \citenamefont {Mostepanenko}}]{henkel2018influence}%
  \BibitemOpen
  \bibfield  {author} {\bibinfo {author} {\bibfnamefont {C.}~\bibnamefont {Henkel}}, \bibinfo {author} {\bibfnamefont {G.}~\bibnamefont {Klimchitskaya}},\ and\ \bibinfo {author} {\bibfnamefont {V.}~\bibnamefont {Mostepanenko}},\ }\bibfield  {title} {\bibinfo {title} {Influence of the chemical potential on the casimir-polder interaction between an atom and gapped graphene or a graphene-coated substrate},\ }\href {https://doi.org/10.1103/PhysRevA.97.032504} {\bibfield  {journal} {\bibinfo  {journal} {Physical Review A}\ }\textbf {\bibinfo {volume} {97}},\ \bibinfo {pages} {032504} (\bibinfo {year} {2018})}\BibitemShut {NoStop}%
\bibitem [{\citenamefont {Baranov}\ \emph {et~al.}(2017)\citenamefont {Baranov}, \citenamefont {Savelev}, \citenamefont {Li}, \citenamefont {Krasnok},\ and\ \citenamefont {Al{\`u}}}]{baranov2017modifying}%
  \BibitemOpen
  \bibfield  {author} {\bibinfo {author} {\bibfnamefont {D.~G.}\ \bibnamefont {Baranov}}, \bibinfo {author} {\bibfnamefont {R.~S.}\ \bibnamefont {Savelev}}, \bibinfo {author} {\bibfnamefont {S.~V.}\ \bibnamefont {Li}}, \bibinfo {author} {\bibfnamefont {A.~E.}\ \bibnamefont {Krasnok}},\ and\ \bibinfo {author} {\bibfnamefont {A.}~\bibnamefont {Al{\`u}}},\ }\bibfield  {title} {\bibinfo {title} {Modifying magnetic dipole spontaneous emission with nanophotonic structures},\ }\href {https://doi.org/10.1002/lpor.201600268} {\bibfield  {journal} {\bibinfo  {journal} {Laser \& Photonics Reviews}\ }\textbf {\bibinfo {volume} {11}},\ \bibinfo {pages} {1600268} (\bibinfo {year} {2017})}\BibitemShut {NoStop}%
\bibitem [{\citenamefont {Langsjoen}\ \emph {et~al.}(2012)\citenamefont {Langsjoen}, \citenamefont {Poudel}, \citenamefont {Vavilov},\ and\ \citenamefont {Joynt}}]{langsjoen2012qubit}%
  \BibitemOpen
  \bibfield  {author} {\bibinfo {author} {\bibfnamefont {L.~S.}\ \bibnamefont {Langsjoen}}, \bibinfo {author} {\bibfnamefont {A.}~\bibnamefont {Poudel}}, \bibinfo {author} {\bibfnamefont {M.~G.}\ \bibnamefont {Vavilov}},\ and\ \bibinfo {author} {\bibfnamefont {R.}~\bibnamefont {Joynt}},\ }\bibfield  {title} {\bibinfo {title} {Qubit relaxation from evanescent-wave johnson noise},\ }\href {https://doi.org/10.1103/PhysRevA.86.010301} {\bibfield  {journal} {\bibinfo  {journal} {Physical Review A}\ }\textbf {\bibinfo {volume} {86}},\ \bibinfo {pages} {010301} (\bibinfo {year} {2012})}\BibitemShut {NoStop}%
\bibitem [{\citenamefont {Kolkowitz}\ \emph {et~al.}(2015)\citenamefont {Kolkowitz}, \citenamefont {Safira}, \citenamefont {High}, \citenamefont {Devlin}, \citenamefont {Choi}, \citenamefont {Unterreithmeier}, \citenamefont {Patterson}, \citenamefont {Zibrov}, \citenamefont {Manucharyan}, \citenamefont {Park} \emph {et~al.}}]{kolkowitz2015probing}%
  \BibitemOpen
  \bibfield  {author} {\bibinfo {author} {\bibfnamefont {S.}~\bibnamefont {Kolkowitz}}, \bibinfo {author} {\bibfnamefont {A.}~\bibnamefont {Safira}}, \bibinfo {author} {\bibfnamefont {A.}~\bibnamefont {High}}, \bibinfo {author} {\bibfnamefont {R.}~\bibnamefont {Devlin}}, \bibinfo {author} {\bibfnamefont {S.}~\bibnamefont {Choi}}, \bibinfo {author} {\bibfnamefont {Q.}~\bibnamefont {Unterreithmeier}}, \bibinfo {author} {\bibfnamefont {D.}~\bibnamefont {Patterson}}, \bibinfo {author} {\bibfnamefont {A.}~\bibnamefont {Zibrov}}, \bibinfo {author} {\bibfnamefont {V.}~\bibnamefont {Manucharyan}}, \bibinfo {author} {\bibfnamefont {H.}~\bibnamefont {Park}}, \emph {et~al.},\ }\bibfield  {title} {\bibinfo {title} {Probing johnson noise and ballistic transport in normal metals with a single-spin qubit},\ }\href {https://doi.org/10.1126/science.aaa4298} {\bibfield  {journal} {\bibinfo  {journal} {Science}\ }\textbf {\bibinfo {volume} {347}},\ \bibinfo {pages} {1129} (\bibinfo {year} {2015})}\BibitemShut {NoStop}%
\bibitem [{\citenamefont {Sun}\ \emph {et~al.}(2023)\citenamefont {Sun}, \citenamefont {Bharadwaj}, \citenamefont {Yang}, \citenamefont {Hsueh}, \citenamefont {Wang}, \citenamefont {Jiao}, \citenamefont {Rahman},\ and\ \citenamefont {Jacob}}]{sun2023limits}%
  \BibitemOpen
  \bibfield  {author} {\bibinfo {author} {\bibfnamefont {W.}~\bibnamefont {Sun}}, \bibinfo {author} {\bibfnamefont {S.}~\bibnamefont {Bharadwaj}}, \bibinfo {author} {\bibfnamefont {L.-P.}\ \bibnamefont {Yang}}, \bibinfo {author} {\bibfnamefont {Y.-L.}\ \bibnamefont {Hsueh}}, \bibinfo {author} {\bibfnamefont {Y.}~\bibnamefont {Wang}}, \bibinfo {author} {\bibfnamefont {D.}~\bibnamefont {Jiao}}, \bibinfo {author} {\bibfnamefont {R.}~\bibnamefont {Rahman}},\ and\ \bibinfo {author} {\bibfnamefont {Z.}~\bibnamefont {Jacob}},\ }\bibfield  {title} {\bibinfo {title} {Limits to quantum gate fidelity from near-field thermal and vacuum fluctuations},\ }\href {https://doi.org/10.1103/PhysRevApplied.19.064038} {\bibfield  {journal} {\bibinfo  {journal} {Physical Review Applied}\ }\textbf {\bibinfo {volume} {19}},\ \bibinfo {pages} {064038} (\bibinfo {year} {2023})}\BibitemShut {NoStop}%
\bibitem [{\citenamefont {Bellomo}\ and\ \citenamefont {Antezza}(2015)}]{bellomo2015nonequilibrium}%
  \BibitemOpen
  \bibfield  {author} {\bibinfo {author} {\bibfnamefont {B.}~\bibnamefont {Bellomo}}\ and\ \bibinfo {author} {\bibfnamefont {M.}~\bibnamefont {Antezza}},\ }\bibfield  {title} {\bibinfo {title} {Nonequilibrium dissipation-driven steady many-body entanglement},\ }\href {https://doi.org/10.1103/PhysRevA.91.042124} {\bibfield  {journal} {\bibinfo  {journal} {Physical Review A}\ }\textbf {\bibinfo {volume} {91}},\ \bibinfo {pages} {042124} (\bibinfo {year} {2015})}\BibitemShut {NoStop}%
\bibitem [{\citenamefont {Buhmann}\ \emph {et~al.}(2012)\citenamefont {Buhmann}, \citenamefont {Butcher},\ and\ \citenamefont {Scheel}}]{buhmann2012macroscopic}%
  \BibitemOpen
  \bibfield  {author} {\bibinfo {author} {\bibfnamefont {S.~Y.}\ \bibnamefont {Buhmann}}, \bibinfo {author} {\bibfnamefont {D.~T.}\ \bibnamefont {Butcher}},\ and\ \bibinfo {author} {\bibfnamefont {S.}~\bibnamefont {Scheel}},\ }\bibfield  {title} {\bibinfo {title} {Macroscopic quantum electrodynamics in nonlocal and nonreciprocal media},\ }\href {https://doi.org/10.1088/1367-2630/14/8/083034} {\bibfield  {journal} {\bibinfo  {journal} {New Journal of Physics}\ }\textbf {\bibinfo {volume} {14}},\ \bibinfo {pages} {083034} (\bibinfo {year} {2012})}\BibitemShut {NoStop}%
\bibitem [{\citenamefont {Breuer}\ and\ \citenamefont {Petruccione}(2002)}]{breuer2002theory}%
  \BibitemOpen
  \bibfield  {author} {\bibinfo {author} {\bibfnamefont {H.-P.}\ \bibnamefont {Breuer}}\ and\ \bibinfo {author} {\bibfnamefont {F.}~\bibnamefont {Petruccione}},\ }\href {https://doi.org/10.1093/acprof:oso/9780199213900.001.0001} {\emph {\bibinfo {title} {The theory of open quantum systems}}}\ (\bibinfo  {publisher} {Oxford University Press, USA},\ \bibinfo {year} {2002})\BibitemShut {NoStop}%
\bibitem [{\citenamefont {Franke}\ \emph {et~al.}(2021)\citenamefont {Franke}, \citenamefont {Ren}, \citenamefont {Richter}, \citenamefont {Knorr},\ and\ \citenamefont {Hughes}}]{franke2021fermi}%
  \BibitemOpen
  \bibfield  {author} {\bibinfo {author} {\bibfnamefont {S.}~\bibnamefont {Franke}}, \bibinfo {author} {\bibfnamefont {J.}~\bibnamefont {Ren}}, \bibinfo {author} {\bibfnamefont {M.}~\bibnamefont {Richter}}, \bibinfo {author} {\bibfnamefont {A.}~\bibnamefont {Knorr}},\ and\ \bibinfo {author} {\bibfnamefont {S.}~\bibnamefont {Hughes}},\ }\bibfield  {title} {\bibinfo {title} {Fermi’s golden rule for spontaneous emission in absorptive and amplifying media},\ }\href {https://doi.org/10.1103/PhysRevLett.127.013602} {\bibfield  {journal} {\bibinfo  {journal} {Physical Review Letters}\ }\textbf {\bibinfo {volume} {127}},\ \bibinfo {pages} {013602} (\bibinfo {year} {2021})}\BibitemShut {NoStop}%
\bibitem [{\citenamefont {Machado}\ \emph {et~al.}(2023)\citenamefont {Machado}, \citenamefont {Demler}, \citenamefont {Yao},\ and\ \citenamefont {Chatterjee}}]{machado2023quantum}%
  \BibitemOpen
  \bibfield  {author} {\bibinfo {author} {\bibfnamefont {F.}~\bibnamefont {Machado}}, \bibinfo {author} {\bibfnamefont {E.~A.}\ \bibnamefont {Demler}}, \bibinfo {author} {\bibfnamefont {N.~Y.}\ \bibnamefont {Yao}},\ and\ \bibinfo {author} {\bibfnamefont {S.}~\bibnamefont {Chatterjee}},\ }\bibfield  {title} {\bibinfo {title} {Quantum noise spectroscopy of dynamical critical phenomena},\ }\href {https://doi.org/10.1103/PhysRevLett.131.070801} {\bibfield  {journal} {\bibinfo  {journal} {Physical Review Letters}\ }\textbf {\bibinfo {volume} {131}},\ \bibinfo {pages} {070801} (\bibinfo {year} {2023})}\BibitemShut {NoStop}%
\bibitem [{\citenamefont {Ford}\ and\ \citenamefont {Weber}(1984)}]{ford1984electromagnetic}%
  \BibitemOpen
  \bibfield  {author} {\bibinfo {author} {\bibfnamefont {G.~W.}\ \bibnamefont {Ford}}\ and\ \bibinfo {author} {\bibfnamefont {W.~H.}\ \bibnamefont {Weber}},\ }\bibfield  {title} {\bibinfo {title} {Electromagnetic interactions of molecules with metal surfaces},\ }\href {https://doi.org/https://doi.org/10.1016/0370-1573(84)90098-X} {\bibfield  {journal} {\bibinfo  {journal} {Physics Reports}\ }\textbf {\bibinfo {volume} {113}},\ \bibinfo {pages} {195} (\bibinfo {year} {1984})}\BibitemShut {NoStop}%
\bibitem [{\citenamefont {Kwiatkowski}\ and\ \citenamefont {Cywi{\'n}ski}(2018)}]{kwiatkowski2018decoherence}%
  \BibitemOpen
  \bibfield  {author} {\bibinfo {author} {\bibfnamefont {D.}~\bibnamefont {Kwiatkowski}}\ and\ \bibinfo {author} {\bibfnamefont {{\L}.}~\bibnamefont {Cywi{\'n}ski}},\ }\bibfield  {title} {\bibinfo {title} {Decoherence of two entangled spin qubits coupled to an interacting sparse nuclear spin bath: Application to nitrogen vacancy centers},\ }\href {https://doi.org/10.1103/PhysRevB.98.155202} {\bibfield  {journal} {\bibinfo  {journal} {Physical Review B}\ }\textbf {\bibinfo {volume} {98}},\ \bibinfo {pages} {155202} (\bibinfo {year} {2018})}\BibitemShut {NoStop}%
\bibitem [{\citenamefont {Bradley}\ \emph {et~al.}(2019)\citenamefont {Bradley}, \citenamefont {Randall}, \citenamefont {Abobeih}, \citenamefont {Berrevoets}, \citenamefont {Degen}, \citenamefont {Bakker}, \citenamefont {Markham}, \citenamefont {Twitchen},\ and\ \citenamefont {Taminiau}}]{bradley2019ten}%
  \BibitemOpen
  \bibfield  {author} {\bibinfo {author} {\bibfnamefont {C.~E.}\ \bibnamefont {Bradley}}, \bibinfo {author} {\bibfnamefont {J.}~\bibnamefont {Randall}}, \bibinfo {author} {\bibfnamefont {M.~H.}\ \bibnamefont {Abobeih}}, \bibinfo {author} {\bibfnamefont {R.}~\bibnamefont {Berrevoets}}, \bibinfo {author} {\bibfnamefont {M.}~\bibnamefont {Degen}}, \bibinfo {author} {\bibfnamefont {M.~A.}\ \bibnamefont {Bakker}}, \bibinfo {author} {\bibfnamefont {M.}~\bibnamefont {Markham}}, \bibinfo {author} {\bibfnamefont {D.}~\bibnamefont {Twitchen}},\ and\ \bibinfo {author} {\bibfnamefont {T.~H.}\ \bibnamefont {Taminiau}},\ }\bibfield  {title} {\bibinfo {title} {A ten-qubit solid-state spin register with quantum memory up to one minute},\ }\href {https://doi.org/10.1103/PhysRevX.9.031045} {\bibfield  {journal} {\bibinfo  {journal} {Physical Review X}\ }\textbf {\bibinfo {volume} {9}},\ \bibinfo {pages} {031045} (\bibinfo {year} {2019})}\BibitemShut {NoStop}%
\bibitem [{\citenamefont {Rojas-Arias}\ \emph {et~al.}(2023)\citenamefont {Rojas-Arias}, \citenamefont {Noiri}, \citenamefont {Stano}, \citenamefont {Nakajima}, \citenamefont {Yoneda}, \citenamefont {Takeda}, \citenamefont {Kobayashi}, \citenamefont {Sammak}, \citenamefont {Scappucci}, \citenamefont {Loss} \emph {et~al.}}]{rojas2023spatial}%
  \BibitemOpen
  \bibfield  {author} {\bibinfo {author} {\bibfnamefont {J.~S.}\ \bibnamefont {Rojas-Arias}}, \bibinfo {author} {\bibfnamefont {A.}~\bibnamefont {Noiri}}, \bibinfo {author} {\bibfnamefont {P.}~\bibnamefont {Stano}}, \bibinfo {author} {\bibfnamefont {T.}~\bibnamefont {Nakajima}}, \bibinfo {author} {\bibfnamefont {J.}~\bibnamefont {Yoneda}}, \bibinfo {author} {\bibfnamefont {K.}~\bibnamefont {Takeda}}, \bibinfo {author} {\bibfnamefont {T.}~\bibnamefont {Kobayashi}}, \bibinfo {author} {\bibfnamefont {A.}~\bibnamefont {Sammak}}, \bibinfo {author} {\bibfnamefont {G.}~\bibnamefont {Scappucci}}, \bibinfo {author} {\bibfnamefont {D.}~\bibnamefont {Loss}}, \emph {et~al.},\ }\bibfield  {title} {\bibinfo {title} {Spatial noise correlations beyond nearest neighbors in 28 si/si-ge spin qubits},\ }\href {https://doi.org/10.1103/PhysRevApplied.20.054024} {\bibfield  {journal} {\bibinfo  {journal} {Physical Review Applied}\ }\textbf {\bibinfo {volume} {20}},\ \bibinfo {pages} {054024} (\bibinfo {year} {2023})}\BibitemShut
  {NoStop}%
\bibitem [{\citenamefont {Khandekar}\ and\ \citenamefont {Jacob}(2019)}]{khandekar2019thermal}%
  \BibitemOpen
  \bibfield  {author} {\bibinfo {author} {\bibfnamefont {C.}~\bibnamefont {Khandekar}}\ and\ \bibinfo {author} {\bibfnamefont {Z.}~\bibnamefont {Jacob}},\ }\bibfield  {title} {\bibinfo {title} {Thermal spin photonics in the near-field of nonreciprocal media},\ }\href {https://dx.doi.org/10.1088/1367-2630/ab494d} {\bibfield  {journal} {\bibinfo  {journal} {New Journal of Physics}\ }\textbf {\bibinfo {volume} {21}},\ \bibinfo {pages} {103030} (\bibinfo {year} {2019})}\BibitemShut {NoStop}%
\bibitem [{\citenamefont {Poddubny}\ \emph {et~al.}(2013)\citenamefont {Poddubny}, \citenamefont {Iorsh}, \citenamefont {Belov},\ and\ \citenamefont {Kivshar}}]{poddubny2013hyperbolic}%
  \BibitemOpen
  \bibfield  {author} {\bibinfo {author} {\bibfnamefont {A.}~\bibnamefont {Poddubny}}, \bibinfo {author} {\bibfnamefont {I.}~\bibnamefont {Iorsh}}, \bibinfo {author} {\bibfnamefont {P.}~\bibnamefont {Belov}},\ and\ \bibinfo {author} {\bibfnamefont {Y.}~\bibnamefont {Kivshar}},\ }\bibfield  {title} {\bibinfo {title} {Hyperbolic metamaterials},\ }\href {https://doi.org/10.1038/nphoton.2013.243} {\bibfield  {journal} {\bibinfo  {journal} {Nature photonics}\ }\textbf {\bibinfo {volume} {7}},\ \bibinfo {pages} {948} (\bibinfo {year} {2013})}\BibitemShut {NoStop}%
\bibitem [{\citenamefont {McPhedran}\ \emph {et~al.}(1982)\citenamefont {McPhedran}, \citenamefont {Botten}, \citenamefont {Craig}, \citenamefont {Nevi{\`e}re},\ and\ \citenamefont {Maystre}}]{mcphedran1982lossy}%
  \BibitemOpen
  \bibfield  {author} {\bibinfo {author} {\bibfnamefont {R.}~\bibnamefont {McPhedran}}, \bibinfo {author} {\bibfnamefont {L.}~\bibnamefont {Botten}}, \bibinfo {author} {\bibfnamefont {M.}~\bibnamefont {Craig}}, \bibinfo {author} {\bibfnamefont {M.}~\bibnamefont {Nevi{\`e}re}},\ and\ \bibinfo {author} {\bibfnamefont {D.}~\bibnamefont {Maystre}},\ }\bibfield  {title} {\bibinfo {title} {Lossy lamellar gratings in the quasistatic limit},\ }\href {https://doi.org/10.1080/713820844} {\bibfield  {journal} {\bibinfo  {journal} {Optica Acta: International Journal of Optics}\ }\textbf {\bibinfo {volume} {29}},\ \bibinfo {pages} {289} (\bibinfo {year} {1982})}\BibitemShut {NoStop}%
\bibitem [{\citenamefont {Sangtawesin}\ \emph {et~al.}(2019)\citenamefont {Sangtawesin}, \citenamefont {Dwyer}, \citenamefont {Srinivasan}, \citenamefont {Allred}, \citenamefont {Rodgers}, \citenamefont {De~Greve}, \citenamefont {Stacey}, \citenamefont {Dontschuk}, \citenamefont {O’Donnell}, \citenamefont {Hu} \emph {et~al.}}]{sangtawesin2019origins}%
  \BibitemOpen
  \bibfield  {author} {\bibinfo {author} {\bibfnamefont {S.}~\bibnamefont {Sangtawesin}}, \bibinfo {author} {\bibfnamefont {B.~L.}\ \bibnamefont {Dwyer}}, \bibinfo {author} {\bibfnamefont {S.}~\bibnamefont {Srinivasan}}, \bibinfo {author} {\bibfnamefont {J.~J.}\ \bibnamefont {Allred}}, \bibinfo {author} {\bibfnamefont {L.~V.}\ \bibnamefont {Rodgers}}, \bibinfo {author} {\bibfnamefont {K.}~\bibnamefont {De~Greve}}, \bibinfo {author} {\bibfnamefont {A.}~\bibnamefont {Stacey}}, \bibinfo {author} {\bibfnamefont {N.}~\bibnamefont {Dontschuk}}, \bibinfo {author} {\bibfnamefont {K.~M.}\ \bibnamefont {O’Donnell}}, \bibinfo {author} {\bibfnamefont {D.}~\bibnamefont {Hu}}, \emph {et~al.},\ }\bibfield  {title} {\bibinfo {title} {Origins of diamond surface noise probed by correlating single-spin measurements with surface spectroscopy},\ }\href {https://doi.org/10.1103/PhysRevX.9.031052} {\bibfield  {journal} {\bibinfo  {journal} {Physical Review X}\ }\textbf {\bibinfo {volume} {9}},\ \bibinfo {pages} {031052} (\bibinfo
  {year} {2019})}\BibitemShut {NoStop}%
\bibitem [{\citenamefont {F{\'a}varo~de Oliveira}\ \emph {et~al.}(2017)\citenamefont {F{\'a}varo~de Oliveira}, \citenamefont {Antonov}, \citenamefont {Wang}, \citenamefont {Neumann}, \citenamefont {Momenzadeh}, \citenamefont {H{\"a}u{\ss}ermann}, \citenamefont {Pasquarelli}, \citenamefont {Denisenko},\ and\ \citenamefont {Wrachtrup}}]{favaro2017tailoring}%
  \BibitemOpen
  \bibfield  {author} {\bibinfo {author} {\bibfnamefont {F.}~\bibnamefont {F{\'a}varo~de Oliveira}}, \bibinfo {author} {\bibfnamefont {D.}~\bibnamefont {Antonov}}, \bibinfo {author} {\bibfnamefont {Y.}~\bibnamefont {Wang}}, \bibinfo {author} {\bibfnamefont {P.}~\bibnamefont {Neumann}}, \bibinfo {author} {\bibfnamefont {S.~A.}\ \bibnamefont {Momenzadeh}}, \bibinfo {author} {\bibfnamefont {T.}~\bibnamefont {H{\"a}u{\ss}ermann}}, \bibinfo {author} {\bibfnamefont {A.}~\bibnamefont {Pasquarelli}}, \bibinfo {author} {\bibfnamefont {A.}~\bibnamefont {Denisenko}},\ and\ \bibinfo {author} {\bibfnamefont {J.}~\bibnamefont {Wrachtrup}},\ }\bibfield  {title} {\bibinfo {title} {Tailoring spin defects in diamond by lattice charging},\ }\href {https://doi.org/10.1038/ncomms15409} {\bibfield  {journal} {\bibinfo  {journal} {Nature communications}\ }\textbf {\bibinfo {volume} {8}},\ \bibinfo {pages} {15409} (\bibinfo {year} {2017})}\BibitemShut {NoStop}%
\bibitem [{\citenamefont {Rovny}\ \emph {et~al.}(2022)\citenamefont {Rovny}, \citenamefont {Yuan}, \citenamefont {Fitzpatrick}, \citenamefont {Abdalla}, \citenamefont {Futamura}, \citenamefont {Fox}, \citenamefont {Cambria}, \citenamefont {Kolkowitz},\ and\ \citenamefont {de~Leon}}]{rovny2022nanoscale}%
  \BibitemOpen
  \bibfield  {author} {\bibinfo {author} {\bibfnamefont {J.}~\bibnamefont {Rovny}}, \bibinfo {author} {\bibfnamefont {Z.}~\bibnamefont {Yuan}}, \bibinfo {author} {\bibfnamefont {M.}~\bibnamefont {Fitzpatrick}}, \bibinfo {author} {\bibfnamefont {A.~I.}\ \bibnamefont {Abdalla}}, \bibinfo {author} {\bibfnamefont {L.}~\bibnamefont {Futamura}}, \bibinfo {author} {\bibfnamefont {C.}~\bibnamefont {Fox}}, \bibinfo {author} {\bibfnamefont {M.~C.}\ \bibnamefont {Cambria}}, \bibinfo {author} {\bibfnamefont {S.}~\bibnamefont {Kolkowitz}},\ and\ \bibinfo {author} {\bibfnamefont {N.~P.}\ \bibnamefont {de~Leon}},\ }\bibfield  {title} {\bibinfo {title} {Nanoscale covariance magnetometry with diamond quantum sensors},\ }\href {https://doi.org/10.1126/science.ade9858} {\bibfield  {journal} {\bibinfo  {journal} {Science}\ }\textbf {\bibinfo {volume} {378}},\ \bibinfo {pages} {1301} (\bibinfo {year} {2022})}\BibitemShut {NoStop}%
\bibitem [{\citenamefont {Haidar}\ \emph {et~al.}(2015)\citenamefont {Haidar}, \citenamefont {Ranjbar}, \citenamefont {Balinsky}, \citenamefont {Dumas}, \citenamefont {Khartsev},\ and\ \citenamefont {{\AA}kerman}}]{haidar2015thickness}%
  \BibitemOpen
  \bibfield  {author} {\bibinfo {author} {\bibfnamefont {M.}~\bibnamefont {Haidar}}, \bibinfo {author} {\bibfnamefont {M.}~\bibnamefont {Ranjbar}}, \bibinfo {author} {\bibfnamefont {M.}~\bibnamefont {Balinsky}}, \bibinfo {author} {\bibfnamefont {R.}~\bibnamefont {Dumas}}, \bibinfo {author} {\bibfnamefont {S.}~\bibnamefont {Khartsev}},\ and\ \bibinfo {author} {\bibfnamefont {J.}~\bibnamefont {{\AA}kerman}},\ }\bibfield  {title} {\bibinfo {title} {Thickness-and temperature-dependent magnetodynamic properties of yttrium iron garnet thin films},\ }\href {https://doi.org/10.1063/1.4914363} {\bibfield  {journal} {\bibinfo  {journal} {Journal of Applied Physics}\ }\textbf {\bibinfo {volume} {117}} (\bibinfo {year} {2015})}\BibitemShut {NoStop}%
\bibitem [{\citenamefont {Dolgirev}\ \emph {et~al.}(2024)\citenamefont {Dolgirev}, \citenamefont {Esterlis}, \citenamefont {Zibrov}, \citenamefont {Lukin}, \citenamefont {Giamarchi},\ and\ \citenamefont {Demler}}]{dolgirev2024local}%
  \BibitemOpen
  \bibfield  {author} {\bibinfo {author} {\bibfnamefont {P.~E.}\ \bibnamefont {Dolgirev}}, \bibinfo {author} {\bibfnamefont {I.}~\bibnamefont {Esterlis}}, \bibinfo {author} {\bibfnamefont {A.~A.}\ \bibnamefont {Zibrov}}, \bibinfo {author} {\bibfnamefont {M.~D.}\ \bibnamefont {Lukin}}, \bibinfo {author} {\bibfnamefont {T.}~\bibnamefont {Giamarchi}},\ and\ \bibinfo {author} {\bibfnamefont {E.}~\bibnamefont {Demler}},\ }\bibfield  {title} {\bibinfo {title} {Local noise spectroscopy of wigner crystals in two-dimensional materials},\ }\href {https://doi.org/10.1103/PhysRevLett.132.246504} {\bibfield  {journal} {\bibinfo  {journal} {Physical Review Letters}\ }\textbf {\bibinfo {volume} {132}},\ \bibinfo {pages} {246504} (\bibinfo {year} {2024})}\BibitemShut {NoStop}%
\bibitem [{\citenamefont {Cywi{\'n}ski}\ \emph {et~al.}(2008)\citenamefont {Cywi{\'n}ski}, \citenamefont {Lutchyn}, \citenamefont {Nave},\ and\ \citenamefont {Sarma}}]{cywinski2008enhance}%
  \BibitemOpen
  \bibfield  {author} {\bibinfo {author} {\bibfnamefont {{\L}.}~\bibnamefont {Cywi{\'n}ski}}, \bibinfo {author} {\bibfnamefont {R.~M.}\ \bibnamefont {Lutchyn}}, \bibinfo {author} {\bibfnamefont {C.~P.}\ \bibnamefont {Nave}},\ and\ \bibinfo {author} {\bibfnamefont {S.~D.}\ \bibnamefont {Sarma}},\ }\bibfield  {title} {\bibinfo {title} {How to enhance dephasing time in superconducting qubits},\ }\href {https://doi.org/10.1103/PhysRevB.77.174509} {\bibfield  {journal} {\bibinfo  {journal} {Physical Review B}\ }\textbf {\bibinfo {volume} {77}},\ \bibinfo {pages} {174509} (\bibinfo {year} {2008})}\BibitemShut {NoStop}%
\bibitem [{\citenamefont {Lidar}(2019)}]{lidar2019lecture}%
  \BibitemOpen
  \bibfield  {author} {\bibinfo {author} {\bibfnamefont {D.~A.}\ \bibnamefont {Lidar}},\ }\bibfield  {title} {\bibinfo {title} {Lecture notes on the theory of open quantum systems},\ }\href@noop {} {\bibfield  {journal} {\bibinfo  {journal} {arXiv preprint arXiv:1902.00967}\ } (\bibinfo {year} {2019})}\BibitemShut {NoStop}%
\bibitem [{\citenamefont {Premakumar}\ \emph {et~al.}(2017)\citenamefont {Premakumar}, \citenamefont {Vavilov},\ and\ \citenamefont {Joynt}}]{premakumar2017evanescent}%
  \BibitemOpen
  \bibfield  {author} {\bibinfo {author} {\bibfnamefont {V.~N.}\ \bibnamefont {Premakumar}}, \bibinfo {author} {\bibfnamefont {M.~G.}\ \bibnamefont {Vavilov}},\ and\ \bibinfo {author} {\bibfnamefont {R.}~\bibnamefont {Joynt}},\ }\bibfield  {title} {\bibinfo {title} {Evanescent-wave johnson noise in small devices},\ }\href {https://doi.org/10.1088/2058-9565/aa8e15} {\bibfield  {journal} {\bibinfo  {journal} {Quantum Science and Technology}\ }\textbf {\bibinfo {volume} {3}},\ \bibinfo {pages} {015001} (\bibinfo {year} {2017})}\BibitemShut {NoStop}%
\bibitem [{\citenamefont {Kenny}\ \emph {et~al.}(2021)\citenamefont {Kenny}, \citenamefont {Mallubhotla},\ and\ \citenamefont {Joynt}}]{kenny2021magnetic}%
  \BibitemOpen
  \bibfield  {author} {\bibinfo {author} {\bibfnamefont {J.}~\bibnamefont {Kenny}}, \bibinfo {author} {\bibfnamefont {H.}~\bibnamefont {Mallubhotla}},\ and\ \bibinfo {author} {\bibfnamefont {R.}~\bibnamefont {Joynt}},\ }\bibfield  {title} {\bibinfo {title} {Magnetic noise from metal objects near qubit arrays},\ }\href {https://doi.org/10.1103/PhysRevA.103.062401} {\bibfield  {journal} {\bibinfo  {journal} {Physical Review A}\ }\textbf {\bibinfo {volume} {103}},\ \bibinfo {pages} {062401} (\bibinfo {year} {2021})}\BibitemShut {NoStop}%
\bibitem [{\citenamefont {Chru{\'s}ci{\'n}ski}\ and\ \citenamefont {Kossakowski}(2010)}]{chruscinski2010non}%
  \BibitemOpen
  \bibfield  {author} {\bibinfo {author} {\bibfnamefont {D.}~\bibnamefont {Chru{\'s}ci{\'n}ski}}\ and\ \bibinfo {author} {\bibfnamefont {A.}~\bibnamefont {Kossakowski}},\ }\bibfield  {title} {\bibinfo {title} {Non-markovian quantum dynamics: local versus nonlocal},\ }\href {https://doi.org/10.1103/PhysRevLett.104.070406} {\bibfield  {journal} {\bibinfo  {journal} {Physical review letters}\ }\textbf {\bibinfo {volume} {104}},\ \bibinfo {pages} {070406} (\bibinfo {year} {2010})}\BibitemShut {NoStop}%
\bibitem [{\citenamefont {Pozar}(2021)}]{pozar2021microwave}%
  \BibitemOpen
  \bibfield  {author} {\bibinfo {author} {\bibfnamefont {D.~M.}\ \bibnamefont {Pozar}},\ }\href@noop {} {\emph {\bibinfo {title} {Microwave engineering: theory and techniques}}}\ (\bibinfo  {publisher} {John wiley \& sons},\ \bibinfo {year} {2021})\BibitemShut {NoStop}%
\bibitem [{\citenamefont {De~Vries}(1988)}]{de1988temperature}%
  \BibitemOpen
  \bibfield  {author} {\bibinfo {author} {\bibfnamefont {J.}~\bibnamefont {De~Vries}},\ }\bibfield  {title} {\bibinfo {title} {Temperature and thickness dependence of the resistivity of thin polycrystalline aluminium, cobalt, nickel, palladium, silver and gold films},\ }\href {https://doi.org/10.1016/0040-6090(88)90478-6} {\bibfield  {journal} {\bibinfo  {journal} {Thin Solid Films}\ }\textbf {\bibinfo {volume} {167}},\ \bibinfo {pages} {25} (\bibinfo {year} {1988})}\BibitemShut {NoStop}%
\bibitem [{\citenamefont {Klein}\ \emph {et~al.}(1994)\citenamefont {Klein}, \citenamefont {Nicol}, \citenamefont {Holczer},\ and\ \citenamefont {Gr{\"u}ner}}]{klein1994conductivity}%
  \BibitemOpen
  \bibfield  {author} {\bibinfo {author} {\bibfnamefont {O.}~\bibnamefont {Klein}}, \bibinfo {author} {\bibfnamefont {E.}~\bibnamefont {Nicol}}, \bibinfo {author} {\bibfnamefont {K.}~\bibnamefont {Holczer}},\ and\ \bibinfo {author} {\bibfnamefont {G.}~\bibnamefont {Gr{\"u}ner}},\ }\bibfield  {title} {\bibinfo {title} {Conductivity coherence factors in the conventional superconductors nb and pb},\ }\href {https://doi.org/10.1103/PhysRevB.50.6307} {\bibfield  {journal} {\bibinfo  {journal} {Physical Review B}\ }\textbf {\bibinfo {volume} {50}},\ \bibinfo {pages} {6307} (\bibinfo {year} {1994})}\BibitemShut {NoStop}%
\bibitem [{\citenamefont {Novotny}\ and\ \citenamefont {Hecht}(2012)}]{novotny2012principles}%
  \BibitemOpen
  \bibfield  {author} {\bibinfo {author} {\bibfnamefont {L.}~\bibnamefont {Novotny}}\ and\ \bibinfo {author} {\bibfnamefont {B.}~\bibnamefont {Hecht}},\ }\href@noop {} {\emph {\bibinfo {title} {Principles of nano-optics}}}\ (\bibinfo  {publisher} {Cambridge university press},\ \bibinfo {year} {2012})\BibitemShut {NoStop}%
\bibitem [{\citenamefont {Sosnowski}(1972)}]{sosnowski1972polarization}%
  \BibitemOpen
  \bibfield  {author} {\bibinfo {author} {\bibfnamefont {T.}~\bibnamefont {Sosnowski}},\ }\bibfield  {title} {\bibinfo {title} {Polarization mode filters for integrated optics},\ }\href {https://www.sciencedirect.com/science/article/pii/0030401872901125} {\bibfield  {journal} {\bibinfo  {journal} {Optics Communications}\ }\textbf {\bibinfo {volume} {4}},\ \bibinfo {pages} {408} (\bibinfo {year} {1972})}\BibitemShut {NoStop}%
\bibitem [{\citenamefont {Xiang}\ \emph {et~al.}(2013)\citenamefont {Xiang}, \citenamefont {Ashhab}, \citenamefont {You},\ and\ \citenamefont {Nori}}]{xiang2013hybrid}%
  \BibitemOpen
  \bibfield  {author} {\bibinfo {author} {\bibfnamefont {Z.-L.}\ \bibnamefont {Xiang}}, \bibinfo {author} {\bibfnamefont {S.}~\bibnamefont {Ashhab}}, \bibinfo {author} {\bibfnamefont {J.}~\bibnamefont {You}},\ and\ \bibinfo {author} {\bibfnamefont {F.}~\bibnamefont {Nori}},\ }\bibfield  {title} {\bibinfo {title} {Hybrid quantum circuits: Superconducting circuits interacting with other quantum systems},\ }\href {https://doi.org/10.1103/RevModPhys.85.623} {\bibfield  {journal} {\bibinfo  {journal} {Reviews of Modern Physics}\ }\textbf {\bibinfo {volume} {85}},\ \bibinfo {pages} {623} (\bibinfo {year} {2013})}\BibitemShut {NoStop}%
\bibitem [{\citenamefont {Baumgratz}\ \emph {et~al.}(2014)\citenamefont {Baumgratz}, \citenamefont {Cramer},\ and\ \citenamefont {Plenio}}]{baumgratz2014quantifying}%
  \BibitemOpen
  \bibfield  {author} {\bibinfo {author} {\bibfnamefont {T.}~\bibnamefont {Baumgratz}}, \bibinfo {author} {\bibfnamefont {M.}~\bibnamefont {Cramer}},\ and\ \bibinfo {author} {\bibfnamefont {M.~B.}\ \bibnamefont {Plenio}},\ }\bibfield  {title} {\bibinfo {title} {Quantifying coherence},\ }\href {https://doi.org/10.1103/PhysRevLett.113.140401} {\bibfield  {journal} {\bibinfo  {journal} {Physical review letters}\ }\textbf {\bibinfo {volume} {113}},\ \bibinfo {pages} {140401} (\bibinfo {year} {2014})}\BibitemShut {NoStop}%
\end{thebibliography}%

\end{document}